\RequirePackage{fix-cm}
\documentclass[smallextended]{svjour3}       
\smartqed  

\usepackage{etex} 

\usepackage{latexsym,amsbsy,amsmath,epsfig,subfigure,fancybox}
\usepackage{amstext,mathrsfs}
\usepackage{amsfonts,amssymb}
\usepackage{mathrsfs}
\usepackage{stmaryrd,dsfont}
\usepackage{bbm}
\usepackage{url}
\usepackage{enumerate}
\usepackage{wrapfig,bm}
\usepackage{color,colordvi}
\usepackage[numbers,sort&compress]{natbib}
\usepackage{hypernat}
\usepackage[colorlinks=true,citecolor=blue]{hyperref}
\usepackage{tikz}
\usepackage{citesort}
\usepackage{pstricks}
\usepackage{caption}
\usepackage{upgreek}
\usepackage{isomath}
\usepackage[dvips]{geometry}
\usepackage{latexsym}
\usepackage{graphicx}
\usepackage{array}
\usepackage{multirow}
\usepackage{booktabs}
\usepackage{colortbl}
\usepackage{verbatim}
\usepackage{framed}
\usepackage[normalem]{ulem}
\usepackage[titletoc,title]{appendix}

\usepackage[algoruled]{algorithm2e}

\usepackage{listings}
\usepackage{color}

\usepackage{caption} 

\usepackage[misc,geometry]{ifsym}

\usepackage[dont-mess-around]{fnpct}

\usepackage{etoolbox}
\newcommand*{\affaddr}[1]{#1} 
\newcommand*{\affmark}[1][*]{\textsuperscript{#1}}

\definecolor{dkgreen}{rgb}{0,0.6,0}
\definecolor{gray}{rgb}{0.5,0.5,0.5}
\definecolor{mauve}{rgb}{0.58,0,0.82}
\definecolor{backcolour}{rgb}{0.95,0.95,0.92}

\usepackage{float}
\newfloat{Code}{H}{myc}

\newcommand{\eref}[1]{(\ref{#1})}
\newcommand{\fref}[1]{Fig.~\ref{#1}}

\renewcommand{\Re}{\mathrm{I\!R}}
\newcommand{\transpose}{\mathsf{T}}
\newcommand{\vm}[1]{\bm{#1}} 
\newcommand{\vx}{\vm{x}}
\newcommand{\bsym}{\boldsymbol}
\newcommand{\mat}[1]{\bm{#1}} 
\newcommand{\diffx}{\mathrm{d}\vm{x}}
\newcommand{\diffs}{\mathrm{d}s}

\newcommand{\smat}[2][ccccccccccccccccccccccccccccccccccccccccccccccccccc]{\left
[\arraycolsep=4.0pt\def\arraystretch{0.7}\begin{array}{#1}#2 \\ \end{array} \right]}

\tikzstyle{nicebox}=[draw=black!100, fill=white!10, rectangle, inner sep=4pt, inner ysep=16pt]
\tikzstyle{niceboxtitle}=[draw=black!100, fill=white, text=black, rectangle]

\definecolor{forestgreen}{RGB}{34, 139, 34}
\definecolor{lightgray}{gray}{0.92}

\usepackage[symbol]{footmisc}

\journalname{Numerical Algorithms}

\begin{document}

\captionsetup[figure]{labelfont={bf},labelformat={default},labelsep=quad,name={Fig.},justification=raggedright,singlelinecheck=false}
\captionsetup[lstlisting]{labelfont={bf},labelsep=quad,justification=raggedright,singlelinecheck=false}
\captionsetup[table]{labelfont={bf},labelsep=quad,justification=raggedright,singlelinecheck=false}

\title{\texttt{Veamy}: an extensible object-oriented C++ library for the virtual element method
}


\author{A.~Ortiz-Bernardin\protect\affmark[1,2] \and C.~Alvarez\affmark[1,3] \and N.~Hitschfeld-Kahler\affmark[3,4] \and A.~Russo\affmark[5,6] \and R.~Silva-Valenzuela\affmark[1,2] \and E.~Olate-Sanzana\affmark[1,2]
}

\authorrunning{Ortiz-Bernardin et al.} 

\institute{\Letter \, A.~Ortiz-Bernardin\\ \email{aortizb@uchile.cl}  \\ \\       
           C.~Alvarez \\ \email{catalin@uchile.cl} \\ \\
           N.~Hitschfeld-Kahler \\ \email{nancy@dcc.uchile.cl} \\ \\
           A.~Russo \\ \email{alessandro.russo@unimib.it} \\ \\
           R.~Silva-Valenzuela \\ \email{rosilva@ug.uchile.cl} \\ \\
           E.~Olate-Sanzana \\ \email{edgardo.olate@ing.uchile.cl} \\ \\
           \affaddr{\affmark[1]Department of Mechanical Engineering, Universidad de Chile, Av. Beauchef 851, Santiago 8370456, Chile.}\\ \\
           \affaddr{\affmark[2]Computational and Applied Mechanics Laboratory, Center for Modern Computational Engineering, Facultad de Ciencias F\'isicas y Matem\'aticas, Universidad de Chile, Av. Beauchef 851, Santiago 8370456, Chile.}\\ \\
           \affaddr{\affmark[3]Department of Computer Science, Universidad de Chile, Av. Beauchef 851, Santiago 8370456, Chile.}\\ \\
           \affaddr{\affmark[4]Meshing for Applied Science Laboratory, Center for Modern Computational Engineering, Facultad de Ciencias F\'isicas y Matem\'aticas, Universidad de Chile, Av. Beauchef 851, Santiago 8370456, Chile.}\\ \\
           \affaddr{\affmark[5]Dipartimento di Matematica e Applicazioni, Universit\`a di Milano-Bicocca, 20153 Milano, Italy.}\\ \\
           \affaddr{\affmark[6]Istituto di Matematica Applicata e Tecnologie Informatiche del CNR, via Ferrata 1, 27100 Pavia, Italy.}
}

\date{Received: date / Accepted: date}

\maketitle

\begin{abstract}
This paper summarizes the development of \texttt{Veamy}, an object-oriented C++ library for the virtual element method (VEM) on general polygonal meshes, whose modular design is focused on its extensibility. The linear elastostatic and Poisson problems in two dimensions have been chosen as the starting stage for the development of this library. The theory of the VEM, upon which \texttt{Veamy} is built, is presented using a notation and a terminology that resemble the language of the finite element method (FEM) in engineering analysis. Several examples are provided to demonstrate the usage of \texttt{Veamy}, and in particular, one of them features the interaction between \texttt{Veamy} and the polygonal mesh generator \texttt{PolyMesher}. A computational performance comparison between VEM and FEM is also conducted. \texttt{Veamy} is free and open source software.
\end{abstract}

\section{Introduction}
\label{sec:intro}

When the Galerkin weak formulation of a boundary-value problem such as the linear elastostatic problem is solved numerically, the trial and test displacements are replaced by their discrete representations using basis functions. Herein, we consider basis functions that span the space of functions of degree 1 (i.e., affine functions). Due to the nature of some basis functions, the discrete trial and test displacement fields may represent linear fields plus some additional functions that are non-polynomials or high-order monomials. Such additional terms cause inhomogeneous deformations, and when present, integration errors appear in the numerical integration of the stiffness matrix leading to stability issues that affect the convergence of the approximation method. This is the case of polygonal and polyhedral finite element methods~\cite{Talischi-Paulino:2014,talischi:2015,francis:LSP:2016}, and meshfree Galerkin methods~\cite{dolbow:1999:NIO,chen:2001:ASC,babuska:2008:QMM,babuska:2009:NIM,ortiz:2010:MEM,ortiz:2011:MEI,duan:2012:SOI,duan:2014:CEF,duan:2014:FPI,ortiz:VAN:2015,ortiz:IRN:2015}.

The virtual element
method~\cite{BeiraoDaVeiga-Brezzi-Cangiani-Manzini-Marini-Russo:2013}
(VEM) has been presented to deal with these integration issues. In short, the method consists in the construction of an algebraic (exact) representation of
the stiffness matrix without the explicit evaluation of
basis functions (basis functions are {\em virtual}). In the VEM, the stiffness matrix is decomposed into two parts: a consistent matrix that guarantees the exact reproduction of a linear displacement field and a correction matrix that provides stability. Such a decomposition is formulated in the spirit of the Lax equivalence theorem (consistency $+$ stability
$\to$ convergence) for finite-difference schemes and is sufficient
for the method to pass the patch test~\cite{cangiani:2015}. Recently, the virtual element framework has been used to correct integration errors in polygonal finite element methods~\cite{Gain-Talischi-Paulino:2014,Manzini-Russo-Sukumar:2014,BeiraodaVeiga-Lovadina-Mora:2015} and in meshfree Galerkin methods~\cite{ortiz:CSMGMVEM:2017}.

Some of the advantages that the VEM exhibits over the standard finite element method (FEM) are:
\begin{itemize}
\item Ability to perform simulations using meshes formed by elements with arbitrary number of edges, not necessarily convex, having coplanar edges and collapsing nodes, while retaining the same approximation properties of the FEM.
\item Possibility of formulating high-order approximations with arbitrary order of global regularity~\cite{daVeiga:VEMAR:2013}.
\item Adaptive mesh refinement techniques are greatly facilitated since hanging nodes become automatically included as elements with coplanar edges are accepted~\cite{cangiani:PEE:2017}.
\end{itemize}

In this paper, object-oriented programming concepts are adopted to develop a C++ library, named \texttt{Veamy}, that implements the VEM on general polygonal meshes. The current status of this library has a focus on the linear elastostatic and Poisson problems in two dimensions, but its design is geared towards its extensibility. \texttt{Veamy} uses Eigen library~\cite{eigenweb} for linear algebra, and Triangle~\cite{shewchuk96b} and Clipper~\cite{clipperweb} are used for the implementation of its polygonal mesh generator, \texttt{Delynoi}~\cite{delynoiweb}, which is based on the constrained Voronoi diagram. Despite this built-in polygonal mesh generator, \texttt{Veamy} is capable of interacting straightforwardly with \texttt{PolyMesher}~\cite{Talischi:POLYM:2012}, a polygonal mesh generator that is widely used in the VEM and polygonal finite elements communities.

In presenting the theory of the VEM, upon which \texttt{Veamy} is built, we adopt a notation and a terminology that resemble the language of the FEM in engineering analysis. The work of Gain et al.~\cite{Gain-Talischi-Paulino:2014} is in line with this aim and has inspired most of the notation and terminology used in this paper.

In \texttt{Veamy}'s programming philosophy entities commonly found in the VEM and FEM literature such as mesh, degree of freedom, element, element stiffness matrix and element force vector, are represented by objects. In contrast to some of the well-established free and open source object-oriented FEM codes such as FreeFEM++~\cite{hecht:FFEM:2012}, FEniCS~\cite{alnaes:FENI:2015} and Feel++~\cite{prud:FEEL:2012}, \texttt{Veamy} does not generate code from the variational form of a particular problem, since that kind of software design tends to hide the implementation details that are fundamental to understand the method. On the contrary, since \texttt{Veamy}'s scope is research and teaching, in its design we wanted a direct and balanced correspondence between theory and implementation. In this sense, \texttt{Veamy} is very similar in its spirit to the 50-line MATLAB implementation of the VEM~\cite{Sutton:VEM:2017}. However, compared to this MATLAB implementation, \texttt{Veamy} is an improvement in the following aspects:

\begin{itemize}
\item Its core VEM numerical implementation is entirely built on free and open source libraries.
\item It offers the possibility of using a built-in polygonal mesh generator, whose implementation is also entirely built on free and open source libraries. In addition, it allows a straightforward interaction with \texttt{PolyMesher}~\cite{Talischi:POLYM:2012}, a popular and widely used MATLAB-based polygonal mesh generator.
\item It is designed using the object-oriented paradigm, which allows a safer and better code design, facilitates code reuse and recycling, code maintenance, and therefore code extension.
\item Its initial release implements both the two-dimensional linear elastostatic problem and the two-dimensional Poisson problem.
\end{itemize}

We are also aware of the MATLAB Reservoir Simulation Toolbox~\cite{lie:MRS:2016}, which provides a module for first- and second-order virtual element methods for Poisson-type flow equations that was developed as part of a master thesis~\cite{klemetsdal:VEM:2016}. The toolbox also implements a module dedicated to the VEM in linear elasticity for geomechanics simulations.

\texttt{Veamy} is free and open source software, and to the best of our knowledge is the first object-oriented C++ implementation of the VEM.

The remainder of this paper is structured as follows. The model problem for two-dimensional linear elastostatics is presented in Section~\ref{sec:modelproblem}. Section~\ref{sec:vem} summarizes the theoretical framework of the VEM for the two-dimensional linear elastostatic problem. Also in this section, the VEM element stiffness matrix for the two-dimensional Poisson problem is given. The object-oriented implementation of \texttt{Veamy} is described and explained in Section~\ref{sec:implementation}. In Section~\ref{sec:meshgenerator}, some guidelines for the usage of \texttt{Veamy}'s built-in polygonal mesh generator are given. Several examples that demonstrate the usage of \texttt{Veamy} and a performance comparison between VEM and FEM are presented in Section~\ref{sec:sampleusage}. The paper ends with some concluding remarks in Section~\ref{sec:conclusions}.

\section{Model problem}
\label{sec:modelproblem}

The Galerkin weak formulation for the linear elastostatic problem is considered
for presenting the main ingredients of the VEM. Consider an elastic body that
occupies the open domain $\Omega \subset \Re^2$ and is
bounded by the one-dimensional surface $\Gamma$ whose
unit outward normal is $\vm{n}_\Gamma$. The boundary is assumed
to admit decompositions $\Gamma=\Gamma_g\cup\Gamma_f$ and
$\emptyset=\Gamma_g\cap\Gamma_f$, where $\Gamma_g$ is the essential (Dirichlet) boundary and $\Gamma_f$ is the natural (Neumann) boundary. The closure of the domain is $\overline{\Omega}\equiv\Omega\cup\Gamma$. Let
$\vm{u}(\vm{x}) : \Omega \rightarrow \Re^2$ be
the displacement field at a point $\vm{x}$ of the elastic
body when the body is subjected to external tractions
$\vm{f}(\vm{x}):\Gamma_f\rightarrow \Re^2$ and body forces $\vm{b}(\vm{x}):\Omega\rightarrow\Re^2$. The
imposed essential (Dirichlet) boundary conditions are
$\vm{g}(\vm{x}):\Gamma_g\rightarrow \Re^2$. The Galerkin weak formulation, with $\vm{v}$ being
the arbitrary test function, gives the following expression for the bilinear form:
\begin{equation}\label{eq:bilinearform1}
a(\vm{u},\vm{v})=\int_{\Omega}\bsym{\sigma}(\vm{u}):\bsym{\nabla}\vm{v}\,\diffx,
\end{equation}
where $\bsym{\sigma}$ is the Cauchy stress tensor and $\bsym{\nabla}$ is the gradient operator. The gradient of the displacement field can be decomposed into its symmetric ($\bsym{\nabla}_\mathrm{S}\vm{v}$) and skew-symmetric ($\bsym{\nabla}_\mathrm{AS}\vm{v}$) parts, as follows:
\begin{equation}\label{eq:gradv}
\bsym{\nabla}\vm{v}=\bsym{\nabla}_\mathrm{S}\vm{v}+\bsym{\nabla}_\mathrm{AS}\vm{v}=\bsym{\varepsilon}(\vm{v})+\bsym{\omega}(\vm{v}),
\end{equation}
where
\begin{equation}\label{eq:strain}
\bsym{\nabla}_\mathrm{S}\vm{v}=\bsym{\varepsilon}(\vm{v})=\frac{1}{2}\left(\bsym{\nabla}\vm{v}+\bsym{\nabla}^\transpose\vm{v}\right)
\end{equation}
is the strain tensor, and
\begin{equation}\label{eq:skew}
\bsym{\nabla}_\mathrm{AS}\vm{v}=\bsym{\omega}(\vm{v})=\frac{1}{2}\left(\bsym{\nabla}\vm{v}-\bsym{\nabla}^\transpose\vm{v}\right)
\end{equation}
is the skew-symmetric gradient tensor that represents rotations. The Cauchy stress tensor is related to the strain tensor by
\begin{equation}\label{eq:cauchystress}
\bsym{\sigma}=\mat{D}:\bsym{\varepsilon}(\vm{u}),
\end{equation}
where $\mat{D}$ is a fourth-order constant tensor that depends on the material of the elastic body.

Substituting~\eref{eq:gradv} into~\eref{eq:bilinearform1} and noting that
$\bsym{\sigma}(\vm{u}):\bsym{\omega}(\vm{v})=0$ because of the symmetry of the stress tensor,
results in the following simplification of the bilinear form:
\begin{equation}\label{eq:simpbilinearform}
a(\vm{u},\vm{v})=\int_{\Omega}\bsym{\sigma}(\vm{u}):\bsym{\varepsilon}(\vm{v})\,\diffx,
\end{equation}
which leads to the standard form of presenting the weak formulation:
find $\vm{u}(\vm{x})\in V$ such that
\begin{subequations}\label{eq:weakform}
\begin{align}
a(\vm{u},\vm{v}) = \ell_{b}(\vm{v}) &+ \ell_{f}(\vm{v})
\quad \forall \vm{v}(\vm{x})\in W,\label{eq:weakform_a}\\
a(\vm{u},\vm{v}) =& \int_{\Omega}\bsym{\sigma}(\vm{u}):\bsym{\varepsilon}(\vm{v})\,\diffx,\label{eq:weakform_b}\\
\ell_{b}(\vm{v}) = \int_{\Omega}\vm{b}\cdot\vm{v}\,\diffx &, \quad \ell_{f}(\vm{v})=\int_{\Gamma_f}\vm{f}\cdot\vm{v}\,\diffs,\label{eq:weakform_c}
\end{align}
\end{subequations}
where $V$ and $W$ are the displacement trial and test spaces defined as follows:
\begin{align*}
V &:=
\left\{\vm{u}(\vm{x}): \vm{u} \in
\mathcal{W}(\Omega) \subseteq [ H^{1}(\Omega)]^2, \ \vm{u} = \vm{g}
\ \textrm{on } \Gamma_g \right\},\\
W &:= \left\{\vm{v}(\vm{x}): \vm{v} \in \mathcal{W}(\Omega) \subseteq
[ H^{1}(\Omega) ]^2, \ \vm{v} = \vm{0} \ \textrm{on } \Gamma_g
\right\},
\end{align*}
where the space $\mathcal{W}(\Omega)$ includes linear displacement
fields.

In the Galerkin approximation, the domain $\Omega$ is partitioned into disjoint non
overlapping elements. This partition is known as a mesh. We denote by $E$ an element
having an area of $|E|$ and a boundary $\partial E$ that is formed by edges $e$ of length $|e|$.
The partition formed by these elements is denoted by $\mathcal{T}^h$, where $h$ is the maximum
diameter of any element in the partition. The set formed by the union of all the element edges
in this partition is denoted by $\mathcal{E}^h$, and the set formed by all the element edges
lying on $\Gamma_f$ is denoted by $\mathcal{E}_f^h$. On this partition, the trial and test displacement
fields are approximated using basis functions, and hence $\vm{u}$ and $\vm{v}$ are replaced by
the approximations $\vm{u}^h$ and $\vm{v}^h$, respectively. The bilinear and linear forms are then obtained
by summation of the contributions from the elements in the mesh, as follows:
\begin{equation*}
a(\vm{u}^h,\vm{v}^h)=\sum_{E\in\mathcal{T}^h}a_E(\vm{u}^h,\vm{v}^h),\quad \textrm{and} \quad \ell_b(\vm{v}^h)=\sum_{E\in\mathcal{T}^h}\ell_{b,E}(\vm{v}^h)\quad \textrm{and} \quad \ell_f(\vm{v}^h)=\sum_{e\in\mathcal{E}_f^h}\ell_{f,e}(\vm{v}^h).
\end{equation*}

In general, the weak form integrals are not available in closed-form expressions since
functions in $\mathcal{W}(E)$, and in particular its basis, are not necessarily polynomial
functions. Therefore, these integrals are evaluated using quadrature with the potential of
introducing quadrature errors making them mesh-dependent. If that is the case, the convergence
of the numerical solution will be affected. To reflect this,
a superscript $h$ is added to the symbols that represent the bilinear and linear forms.
Thus, the Galerkin solution  is sought as the solution of the global system
that results from the weak formulation described by the following discrete bilinear and linear forms:
\begin{equation*}
a^h(\vm{u}^h,\vm{v}^h)=\sum_{E\in\mathcal{T}^h}a_E^h(\vm{u}^h,\vm{v}^h), \quad \textrm{and} \quad \ell_b^h(\vm{v}^h)=\sum_{E\in\mathcal{T}^h}\ell_{b,E}^h(\vm{v}^h) \quad \textrm{and} \quad \ell_f^h(\vm{v}^h)=\sum_{e\in\mathcal{E}_f^h}\ell_{f,e}^h(\vm{v}^h),
\end{equation*}
respectively, with the corresponding discrete global trial and test spaces defined respectively as follows:
\begin{align*}
V^h &:=
\left\{\vm{u}^h(\vm{x})\in V: \vm{u}^h|_E \in
\mathcal{W}(E) \subseteq [ H^{1}(E)]^2 \ \forall E \in \mathcal{T}^h\right\},\\
W^h &:= \left\{\vm{v}^h(\vm{x})\in W: \vm{v}^h|_E \in \mathcal{W}(E) \subseteq
[ H^{1}(E)]^2 \ \forall E \in \mathcal{T}^h\right\}.
\end{align*}

In the preceding discussion, we have implied that $a_E^h$ is inexact
due to its evaluation using numerical quadrature --- in this case, $a_E^h$ is said to be \textit{uncomputable}.
The situation is completely different in the VEM approach: $a_E^h$ is not evaluated
using numerical quadrature. Instead, the displacement field is computed
through projection operators that are tailored to achieve an algebraic (exact)
evaluation of $a_E^h$ --- in this case, $a_E^h$ is said to be \textit{computable}.

\section{The virtual element method}
\label{sec:vem}

In standard two-dimensional finite element methods, the partition $\mathcal{T}^h$ is usually
formed by triangles and quadrilaterals. In the VEM, the partition is formed by elements with
arbitrary number of edges, where triangles and quadrilaterals are particular instances.
We refer to these more general elements as polygonal elements.

\subsection{The polygonal element}
\begin{figure}[!tbhp]
\centering
\epsfig{file = 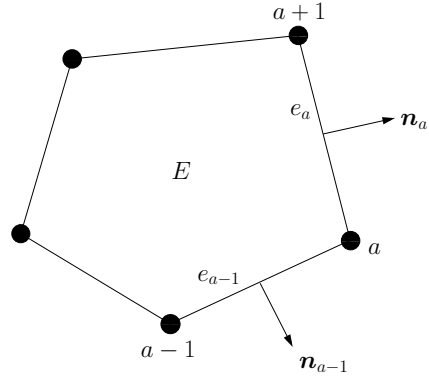, width = 0.37\textwidth}
\caption{Schematic representation of a polygonal element of $N=5$ edges}
\label{fig:1}
\end{figure}

Let the domain $\Omega$ be partitioned into disjoint non overlapping polygonal elements with straight edges. The number of edges and nodes of a polygonal element are denoted by $N$. The unit outward normal to the element boundary in the Cartesian coordinate system is denoted by $\vm{n}=[n_1 \quad n_2]^\transpose$. \fref{fig:1} presents a schematic representation of a polygonal element for $N=5$, where the edge $e_a$ of length $|e_a|$ and the edge $e_{a-1}$ of length $|e_{a-1}|$ are the element edges incident to node $a$, and $\vm{n}_a$ and $\vm{n}_{a-1}$ are the unit outward normals to these edges, respectively.

\subsection{Projection operators}
As in finite elements, for the numerical solution to converge monotonically it is required that the displacement approximation in the polygonal element can represent rigid body modes and constant strain states. This demands that the displacement approximation in the element is at least a linear polynomial~\cite{strang:2008:AAO}. In the VEM, projection operators are devised to extract the rigid body modes, the constant strain states and the linear polynomial part of the motion at the element level. The spaces where these components of the motion reside are given next.

The space of linear displacements over $E$ is defined as
\begin{equation}
\mathcal{P}(E):=\left\{\vm{a}+\mat{B}(\vm{x}-\overline{\vm{x}}):\vm{a}\in\Re^2,\,\mat{B}\in \Re^{2\times 2}\right\},
\label{eq:linearspace}
\end{equation}
where $\overline{\vm{x}}$ is defined through the mean value of a function $h$ over the element nodes given by
\begin{equation}
\overline{h}=\frac{1}{N}\sum_{j=1}^{N}h(\vm{x}_j),
\label{eq:h_bar}
\end{equation}
where $N$ is the number of nodes of coordinates $\vm{x}_j$ that define the polygonal element\footnote[1]{Eq.~\eref{eq:h_bar} in fact defines any `barred' term that appears in this paper.}; $\mat{B}$ is a second-order tensor
and thus can be uniquely expressed as the sum of a symmetric and a skew-symmetric tensor. Let the symmetric and skew-symmetric tensors be denoted by $\mat{B}_\mathrm{S}$ and $\mat{B}_\mathrm{AS}$, respectively. The spaces of rigid body modes and constant strain states over $E$ are defined, respectively, as follows:
\begin{equation}
\mathcal{R}(E):=\left\{\vm{a}+\mat{B}_\mathrm{AS}\cdot(\vm{x}-\overline{\vm{x}}):\vm{a}\in \Re^2,\,\mat{B}_\mathrm{AS}\in \Re^{2\times 2},\, \mat{B}_\mathrm{AS}^\transpose= -\mat{B}_\mathrm{AS}\right\},
\label{eq:rigidbodyspace}
\end{equation}
\begin{equation}
\mathcal{C}(E):=\left\{\mat{B}_\mathrm{S}\cdot(\vm{x}-\overline{\vm{x}}):\mat{B}_\mathrm{S}\in \Re^{2\times 2},\, \mat{B}_\mathrm{S}^\transpose= \mat{B}_\mathrm{S}\right\}.
\label{eq:constantstrainspace}
\end{equation}

Note that the space of linear displacements is the direct sum of the spaces given in~\eref{eq:rigidbodyspace} and~\eref{eq:constantstrainspace}, that is, $\mathcal{P}(E)=\mathcal{R}(E)+\mathcal{C}(E)$.

The extraction of the components of the displacement field in the three aforementioned spaces is achieved through
the following projection operators:
\begin{equation}\label{eq:pir}
\Pi_\mathcal{R}:
\mathcal{W}(E) \to \mathcal{R}(E), \quad \Pi_\mathcal{R}
\vm{r}=\vm{r}, \quad \forall \vm{r}\in \mathcal{R}(E)
\end{equation}
for extracting the rigid body modes,
\begin{equation}\label{eq:pic}
\Pi_\mathcal{C}: \mathcal{W}(E)
\to \mathcal{C}(E), \quad \Pi_\mathcal{C} \vm{c}=\vm{c}, \quad
\forall \vm{c}\in \mathcal{C}(E)
\end{equation} for extracting
the constant strain states, and
\begin{equation}\label{eq:pip}
\Pi_\mathcal{P}: \mathcal{W}(E)  \to \mathcal{P}(E), \quad
\Pi_\mathcal{P} \vm{p}=\vm{p}, \quad \forall \vm{p}\in \mathcal{P}(E)
\end{equation}
for extracting the linear polynomial part. And since $\mathcal{P}(E)=\mathcal{R}(E)+\mathcal{C}(E)$,
the projection operators satisfy the relation
\begin{equation}\label{eq:rcsum}
\Pi_\mathcal{P}=\Pi_\mathcal{R}+\Pi_\mathcal{C}.
\end{equation}

We know by definition that the space $\mathcal{W}(E)$ includes linear displacements.
This means that $\mathcal{W}(E) \supseteq \mathcal{P}(E)$. Thus,
any $\vm{u},\vm{v}\in \mathcal{W}(E)$ can be decomposed into three terms, as follows:
\begin{subequations}\label{eq:usplit}
\begin{align}
\vm{u}&=\Pi_\mathcal{R}\vm{u}+\Pi_\mathcal{C}\vm{u}+(\vm{u}-\Pi_\mathcal{P}\vm{u}),\\
\vm{v}&=\Pi_\mathcal{R}\vm{v}+\Pi_\mathcal{C}\vm{v}+(\vm{v}-\Pi_\mathcal{P}\vm{v}),
\end{align}
\end{subequations}
that is, into their rigid body modes, their constant strain states and their
additional non-polynomial or high-order functions, respectively.

The explicit forms of the projection operators that are defined through~\eref{eq:pir}-\eref{eq:pip}
are given in Ref.~\cite{Gain-Talischi-Paulino:2014} and are summarized as follows: let the cell-average
of the strain tensor be defined as
\begin{equation}\label{eq:volavg_strain}
\widehat{\bsym{\varepsilon}}(\vm{v})=\frac{1}{|E|}\int_E \bsym{\varepsilon}(\vm{v})\,\diffx = \frac{1}{2|E|}\int_{\partial
E}\left(\vm{v}\otimes\vm{n}+\vm{n}\otimes\vm{v}\right)\,\diffs,
\end{equation}
where the divergence theorem has been used to transform the volume integral into a surface integral.
Similarly, the cell-average of the skew-symmetric gradient tensor is defined as
\begin{equation}\label{eq:volavg_skewstrain}
\widehat{\bsym{\omega}}(\vm{v})=\frac{1}{|E|}\int_E \bsym{\omega}(\vm{v})\,\diffx=\frac{1}{2|E|}\int_{\partial
E}\left(\vm{v}\otimes\vm{n}-\vm{n}\otimes\vm{v}\right)\,\diffs.
\end{equation}
Note that $\widehat{\bsym{\varepsilon}}(\vm{v})$ and $\widehat{\bsym{\omega}}(\vm{v})$ are constant tensors in the element.

On using the preceding definitions, the projection of $\vm{v}$ onto the space of rigid body modes is written as
\begin{equation}\label{eq:pirv_final}
\Pi_\mathcal{R}\vm{v}=\widehat{\bsym{\omega}}(\vm{v})\cdot(\vm{x}-\overline{\vm{x}})+\overline{\vm{v}},
\end{equation}
where $\widehat{\bsym{\omega}}(\vm{v})\cdot(\vm{x}-\overline{\vm{x}})$ and $\overline{\vm{v}}$ are the rotation and translation modes of $\vm{v}$, respectively. And the projection of $\vm{v}$ onto the space of constant strain states is given by
\begin{equation}\label{eq:picv_final}
\Pi_\mathcal{C}\vm{v}=\widehat{\bsym{\varepsilon}}(\vm{v})\cdot(\vm{x}-\overline{\vm{x}}).
\end{equation}
Hence, by~\eref{eq:rcsum} the projection of $\vm{v}$ onto the space of linear displacements is written as
\begin{equation}\label{eq:pipv_final}
\Pi_\mathcal{P}\vm{v}=\Pi_\mathcal{R}\vm{v}+\Pi_\mathcal{C}\vm{v}=\widehat{\bsym{\varepsilon}}(\vm{v})\cdot(\vm{x}-\overline{\vm{x}})+\widehat{\bsym{\omega}}(\vm{v})\cdot(\vm{x}-\overline{\vm{x}})+\overline{\vm{v}}.
\end{equation}

The projection operator $\Pi_\mathcal{P}$ satisfies some important energy-orthogonality conditions that are
invoked when constructing the VEM bilinear form. The proofs can be found in Ref.~\cite{Gain-Talischi-Paulino:2014}.
The energy-orthogonality conditions are given next.

The projection $\Pi_\mathcal{P}$ satisfies:
\begin{subequations}
\begin{align}
a_E(\vm{p},\vm{v}-\Pi_\mathcal{P}\vm{v}) &= 0 \quad
\forall\vm{p}\in \mathcal{P}(E), \ \ \vm{v}\in \mathcal{W}(E),
\label{eq:ortoprop_pip} \\
a_E(\vm{c},\vm{v}-\Pi_\mathcal{P}\vm{v}) &= 0 \quad
\forall\vm{c}\in \mathcal{C}(E), \ \ \vm{v}\in \mathcal{W}(E).
\label{eq:ortoprop_pip_c}
\end{align}
\end{subequations}

The condition~\eref{eq:ortoprop_pip} means that $\vm{v}-\Pi_\mathcal{P}\vm{v}$
is energetically orthogonal to $\mathcal{P}$. The condition~\eref{eq:ortoprop_pip_c}
emanates from condition~\eref{eq:ortoprop_pip} after replacing $\vm{p}=\vm{r}+\vm{c}$
and using the fact that rigid body modes have zero strain, that is $a_E(\vm{r},\cdot)=0$.

\subsection{The VEM bilinear form}
Substituting the VEM decomposition~\eref{eq:usplit} into the bilinear form~\eref{eq:simpbilinearform} leads to the following splitting of the bilinear form at element level:
\begin{align}\label{eq:vem_strain_energy}
a_E(\vm{u},\vm{v}) &= a_E(\Pi_\mathcal{R}\vm{u}+\Pi_\mathcal{C}\vm{u}+(\vm{u}-\Pi_\mathcal{P}\vm{u}),\Pi_\mathcal{R}\vm{v}+\Pi_\mathcal{C}\vm{v}+(\vm{v}-\Pi_\mathcal{P}\vm{v}))\nonumber\\
&=  a_E(\Pi_\mathcal{C}\vm{u},\Pi_\mathcal{C}\vm{v})
                     + a_E(\vm{u}-\Pi_\mathcal{P}\vm{u},\vm{v}-\Pi_\mathcal{P}\vm{v}),
\end{align}
where the symmetry of the bilinear form, the fact that $\Pi_\mathcal{R}\vm{u}$ and $\Pi_\mathcal{R}\vm{v}$ do not contribute in the bilinear form (both have zero strain as they belong to the space of rigid body modes), and the energy-orthogonality condition~\eref{eq:ortoprop_pip_c} have been used.

The first term on the right-hand side of~\eref{eq:vem_strain_energy} is the
bilinear form associated with the constant strain states that provides
consistency (it leads to the \textit{consistency} stiffness) and the second term
is the bilinear form associated with the additional non-polynomial or high-order functions
that provides stability (it leads to the \textit{stability} stiffness). We come back to
these concepts later in this section.

\subsection{Projection matrices}
\label{sec:projection_matrices}

The projection matrices are constructed by discretizing the projection operators. We begin by writing the projections $\Pi_\mathcal{R}\vm{v}$ and $\Pi_\mathcal{C}\vm{v}$ in terms of their space basis. To this end, consider the two-dimensional Cartesian space and the skew-symmetry of $\widehat{\bsym{\omega}}\equiv\widehat{\bsym{\omega}}(\vm{v})$\footnote[3]{Note that $\widehat{\omega}_{11}=\widehat{\omega}_{22}=0$ and $\widehat{\omega}_{21}=-\widehat{\omega}_{12}$.}. The projection~\eref{eq:pirv_final} can be written as follows:
\begin{equation}\label{eq:pirv_final_alt}
\Pi_\mathcal{R}\vm{v}=\vm{r}_1\overline{v}_1+\vm{r}_2\overline{v}_2+\vm{r}_3\widehat{\omega}_{12},
\end{equation}
where the basis for the space of rigid body modes is:
\begin{equation}\label{eq:basis_rigid_body}
\vm{r}_1 = \smat{1 & 0}^\transpose,\,\, \vm{r}_2 = \smat{0 & 1}^\transpose,\,\, \vm{r}_3 = \smat{(x_2-\overline{x}_2) & -(x_1-\overline{x}_1)}^\transpose.
\end{equation}

Similarly, on considering  the symmetry of $\widehat{\bsym{\varepsilon}}\equiv\widehat{\bsym{\varepsilon}}(\vm{v})$, the projection~\eref{eq:picv_final} can be written as
\begin{equation}\label{eq:picv_final_alt}
\Pi_\mathcal{C}\vm{v}=\vm{c}_1\widehat{\varepsilon}_{11}+\vm{c}_2\widehat{\varepsilon}_{22}+\vm{c}_3\widehat{\varepsilon}_{12},
\end{equation}
where the basis for the space of constant strain states is:
\begin{equation}\label{eq:basis_const_strain}
\vm{c}_1 = \smat{(x_1-\overline{x}_1) & 0}^\transpose,\,\, \vm{c}_2 = \smat{0 & (x_2-\overline{x}_2)}^\transpose,\,\, \vm{c}_3 = \smat{(x_2-\overline{x}_2) & (x_1-\overline{x}_1)}^\transpose.
\end{equation}

On each polygonal element of $N$ edges with nodal coordinates denoted by $\vm{x}_a=[x_{1a} \quad x_{2a}]^\transpose$,
the trial and test displacements are locally approximated as
\begin{equation}\label{eq:disc_displacement}
\vm{u}^h(\vm{x})=\sum_{a=1}^N\phi_a(\vm{x}) \vm{u}_a,\quad\vm{v}^h(\vm{x})=\sum_{b=1}^N\phi_b(\vm{x}) \vm{v}_b,
\end{equation}
where $\phi_a(\vm{x})$ and $\phi_b(\vm{x})$ are assumed to be the canonical basis functions having the
Kronecker delta property (i.e., Lagrange-type functions), and $\vm{u}_a = [u_{1a} \quad u_{2a}]^\transpose$
and $\vm{v}_b = [v_{1b} \quad v_{2b}]^\transpose$ are nodal displacements.
The canonical basis functions are also used to locally approximate the
components of the basis for the space of rigid body modes:
\begin{equation}\label{eq:disc_r_basis} \vm{r}_\alpha^h(\vm{x})
=\sum_{a=1}^N\phi_a(\vm{x})\vm{r}_\alpha(\vm{x}_a), \quad
\alpha=1,\ldots,3
\end{equation}
and the components of the basis for the space of constant strain states:
\begin{equation}\label{eq:disc_c_basis}
\vm{c}_\beta^h(\vm{x})
= \sum_{a=1}^N \phi_a(\vm{x})\vm{c}_\beta(\vm{x}_a),\quad
\beta=1,\ldots,3.
\end{equation}

The discrete version of the projection to extract the rigid body modes
is obtained by substituting~\eref{eq:disc_displacement}
and \eref{eq:disc_r_basis} into~\eref{eq:pirv_final_alt},
which yields
\begin{equation}\label{eq:disc_proj_rigid_body}
\Pi_\mathcal{R}\vm{v}^h = \mat{N}\mat{P}_\mathcal{R}\mat{q},
\end{equation}
where
\begin{equation}\label{eq:matrix_N}
\mat{N} =
\left[(\mat{N})_1 \quad \cdots \quad (\mat{N})_a \quad \cdots \quad
(\mat{N})_N\right] \, , \,\,\, (\mat{N})_a=\smat{\phi_a & 0 \\ 0 &
\phi_a},
\end{equation}
\begin{equation} \mat{q}
= \left[\vm{v}_1^\transpose \quad \cdots \quad \vm{v}_a^\transpose
\quad \cdots \quad \vm{v}_N^\transpose\right]^\transpose \, ,
\,\,\, \vm{v}_a=[v_{1a} \quad v_{2a}]^\transpose
\end{equation}
and
\begin{equation}\label{eq:matrix_pr}
\mat{P}_\mathcal{R}=\mat{H}_\mathcal{R}\mat{W}_\mathcal{R}^\transpose
\end{equation}
with
\begin{equation}\label{eq:matrix_hr}
\mat{H}_\mathcal{R}=\smat{ (\mat{H}_\mathcal{R})_1 & \cdots &
(\mat{H}_\mathcal{R})_a & \cdots & (\mat{H}_\mathcal{R})_N }^\transpose,
\quad (\mat{H}_\mathcal{R})_a = \smat{ 1 & 0 \\ 0 & 1\\ (x_{2a}-\overline{x}_2) & -(x_{1a}-\overline{x}_1)}^\transpose
\end{equation}
and
\begin{equation}\label{eq:matrix_wr}
\mat{W}_\mathcal{R}=\smat{
(\mat{W}_\mathcal{R})_1 & \cdots & (\mat{W}_\mathcal{R})_a &
\cdots & (\mat{W}_\mathcal{R})_N}^\transpose, \quad (\mat{W}_\mathcal{R})_a =
\smat{ \overline{\phi}_a & 0 \\ 0 & \overline{\phi}_a \\ q_{2a} & -q_{1a}}^\transpose.
\end{equation}

In~\eref{eq:matrix_wr}, $q_{ia}$ appeared because of the discretization of
$\widehat{\omega}_{12}$ (see~\eref{eq:volavg_skewstrain}) and is given by
\begin{equation}\label{eq:qia}
q_{ia}=\frac{1}{2|E|}\int_{\partial E}\phi_a n_i\,\diffs,\quad i=1,2.
\end{equation}

Similarly, substituting~\eref{eq:disc_displacement} and
\eref{eq:disc_c_basis} into~\eref{eq:picv_final_alt}
leads to the following discrete version of the
projection to extract the constant strain states:
\begin{equation}\label{eq:disc_proj_const_strains}
\Pi_\mathcal{C}\vm{v}^h = \mat{N}\mat{P}_\mathcal{C}\mat{q},
\end{equation}
where
\begin{equation}\label{eq:matrix_pc}
\mat{P}_\mathcal{C}=\mat{H}_\mathcal{C}\mat{W}_\mathcal{C}^\transpose
\end{equation}
with
\begin{equation}\label{eq:matrix_hc}
\mat{H}_\mathcal{C}=\smat{ (\mat{H}_\mathcal{C})_1 & \cdots &
(\mat{H}_\mathcal{C})_a & \cdots & (\mat{H}_\mathcal{C})_N}^\transpose,
\quad (\mat{H}_\mathcal{C})_a = \smat{ (x_{1a}-\overline{x}_1) & 0  \\ 0 & (x_{2a}-\overline{x}_2) \\ (x_{2a}-\overline{x}_2) & (x_{1a}-\overline{x}_1) }^\transpose
\end{equation}
and
\begin{equation}\label{eq:matrix_wc}
\mat{W}_\mathcal{C}=\smat{
(\mat{W}_\mathcal{C})_1 & \cdots & (\mat{W}_\mathcal{C})_a &
\cdots & (\mat{W}_\mathcal{C})_N}^\transpose, \quad (\mat{W}_\mathcal{C})_a =
\smat{ 2q_{1a} & 0\\ 0 & 2q_{2a}\\ q_{2a} &
q_{1a}}^\transpose.
\end{equation}

In~\eref{eq:matrix_wc}, $q_{ia}$ is also given by~\eref{eq:qia} but in this case
it stems from the discretization of $\widehat{\varepsilon}_{ij}$ (see~\eref{eq:volavg_strain}).

The matrix form of the projection to extract the polynomial part of the displacement field is then $\mat{P}_\mathcal{P} = \mat{P}_\mathcal{R} + \mat{P}_\mathcal{C}$.

For the development of the element \textit{consistency} stiffness matrix, it will be useful to have the following alternative expression for the discrete projection to extract the constant strain states:
\begin{align}\label{eq:disc_proj_const_strains_alt_b}
\Pi_\mathcal{C}\vm{v}^h &= \vm{c}_1\widehat{\varepsilon}_{11}+\vm{c}_2\widehat{\varepsilon}_{22}+\vm{c}_3\widehat{\varepsilon}_{12}\nonumber\\
&=\smat{\vm{c}_1 & \vm{c}_2 & \vm{c}_3}\sum_{b=1}^N\smat{2q_{1b} & 0\\ 0 & 2q_{2b}\\ q_{2b} & q_{1b}}\smat{v_{1b} \\ v_{2b}}\nonumber\\
&=\mat{c}\,\mat{W}_\mathcal{C}^\transpose\,\mat{q}.
\end{align}

\subsection{VEM element stiffness matrix}
\label{sec:vem_element_stiffness}

The decomposition given in~\eref{eq:vem_strain_energy} is used to construct the approximate mesh-dependent bilinear form $a_E^h(\vm{u},\vm{v})$ in a way that is computable at the element level. To this end, we approximate the quantity $a_E(\vm{u}-\Pi_\mathcal{P}\vm{u},\vm{v}-\Pi_\mathcal{P}\vm{v})$, which is uncomputable, with a computable one given by $s_E(\vm{u}-\Pi_\mathcal{P}\vm{u},\vm{v}-\Pi_\mathcal{P}\vm{v})$ and define
\begin{align}\label{eq:vem_element_decomposition}
a_E^h(\vm{u},\vm{v}) := a_E(\Pi_\mathcal{C}\vm{u},\Pi_\mathcal{C}\vm{v})
+ s_E(\vm{u}-\Pi_\mathcal{P}\vm{u},\vm{v}-\Pi_\mathcal{P}\vm{v}),
\end{align}
where its right-hand side, as it will be revealed in the sequel, is computed algebraically. The decomposition~\eref{eq:vem_element_decomposition} has been proved to be endowed with the following crucial properties for establishing convergence~\cite{BeiraoDaVeiga-Brezzi-Cangiani-Manzini-Marini-Russo:2013,BeiraodaVeiga-Brezzi-Marini:2013}:

For all $h$ and for all $E$ in $\mathcal{T}^h$
\begin{itemize}
\item \textit{Consistency}: $\forall \vm{p} \in \mathcal{P}(E)$ and $\forall \vm{v}^h\in V^h|_E$
\begin{equation}\label{eq:consistency_cond}
a_E^h(\vm{p},\vm{v}^h)=a_E(\vm{p},\vm{v}^h).
\end{equation}
\item \textit{Stability}: $\exists$ two constants $\alpha_*>0$ and $\alpha^*>0$, independent of $h$ and of $E$, such that
\begin{equation}\label{eq:stability_cond}
\forall\vm{v}^h\in V^h|_E, \quad \alpha_*a_E(\vm{v}^h,\vm{v}^h)\leq a_E^h(\vm{v}^h,\vm{v}^h)\leq \alpha^*a_E(\vm{v}^h,\vm{v}^h).
\end{equation}
\end{itemize}

The discrete version of the VEM element bilinear form~\eref{eq:vem_element_decomposition} is constructed as follows. Substitute~\eref{eq:disc_proj_const_strains_alt_b} into the first term of the right-hand side of~\eref{eq:vem_element_decomposition}
(note that when $\vm{u}^h$ is used instead of $\vm{v}^h$, $\mat{q}$ is replaced by  the column vector
of nodal displacements $\mat{d}$, which has the same structure of $\mat{q}$); use~\eref{eq:disc_proj_rigid_body} and \eref{eq:disc_proj_const_strains} to obtain $\Pi_\mathcal{P}\vm{v}^h=\Pi_\mathcal{R}\vm{v}^h+\Pi_\mathcal{C}\vm{v}^h=\mat{N}\mat{P}_\mathcal{P}\mat{q}$, where $\mat{P}_\mathcal{P} = \mat{H}_\mathcal{R}\mat{W}_\mathcal{R}^\transpose+\mat{H}_\mathcal{C}\mat{W}_\mathcal{C}^\transpose$. Also, note that $\vm{v}^h=\mat{N}\mat{q}$. Then, substitute the expressions for $\Pi_\mathcal{P}\vm{v}^h$ and $\vm{v}^h$ into the second term of the right-hand side of~\eref{eq:vem_element_decomposition}. This yields
\begin{align}\label{eq:disc_vem_strain_energy}
a_E^h(\vm{u}^h,\vm{v}^h) &= a_E(\mat{c}\,\mat{W}_\mathcal{C}^\transpose\,\mat{d},\mat{c}\,\mat{W}_\mathcal{C}^\transpose\,\mat{q})
+ s_E(\mat{N}\mat{d}-\mat{N}\mat{P}_\mathcal{P}\mat{d},\mat{N}\mat{q}-\mat{N}\mat{P}_\mathcal{P}\mat{q})  \nonumber\\
&=\mat{q}^\transpose\mat{W}_\mathcal{C}\,a_E(\mat{c}^\transpose,\mat{c})\,\mat{W}_\mathcal{C}^\transpose\mat{d}
+ \mat{q}^\transpose(\mat{I}_{2N}-\mat{P}_\mathcal{P})^\transpose\,
s_E(\mat{N}^\transpose,\mat{N})\,(\mat{I}_{2N}-\mat{P}_\mathcal{P})\,\mat{d}\nonumber\\
 &=
 \mat{q}^\transpose|E|\,\mat{W}_\mathcal{C}\,\mat{D}\,\mat{W}_\mathcal{C}^\transpose\mat{d} +
 \mat{q}^\transpose(\mat{I}_{2N}-\mat{P}_\mathcal{P})^\transpose\,\mat{S}_E\,(\mat{I}_{2N}-\mat{P}_\mathcal{P})\,\mat{d},
\end{align}
where $\mat{I}_{2N}$ is the identity ($2N \times 2N$) matrix and $\mat{S}_E=s_E(\mat{N}^\transpose,\mat{N})$. Using Voigt notation and observing that $\upvarepsilon(\mat{c})=\smat{\varepsilon_{11}(\mat{c}) & \varepsilon_{22}(\mat{c}) & \varepsilon_{12}(\mat{c})}^\transpose=\mat{I}_{3}$ (the identity (3$\times$3) matrix), in~\eref{eq:disc_vem_strain_energy} we have used that $a_E(\mat{c}^\transpose,\mat{c})=\int_E\upvarepsilon^\transpose(\mat{c})
\mat{D}\upvarepsilon(\mat{c})\,\diffx=\mat{D}\int_E\,\diffx=|E|\mat{D}$, where $\mat{D}$ is the constitutive matrix for an isotropic linear elastic material given by
\begin{equation}
\mat{D} =
\frac{E_Y}{(1+\nu)(1-2\nu)}\smat{1-\nu & \nu & 0\\ \nu & 1-\nu & 0\\ 0 & 0 & 2(1-2\nu)}
\end{equation}
for plane strain condition, and
\begin{equation}
\mat{D} =
\frac{E_Y}{(1-\nu^2)}\smat{1 & \nu & 0\\ \nu & 1 & 0\\ 0 & 0 & 2(1-\nu)}
\end{equation}
for plane stress condition, where $E_Y$ is the Young's modulus and $\nu$ is the Poisson's ratio.

The first term on the right-hand side of~\eref{eq:disc_vem_strain_energy} is the \textit{consistency}
part of the discrete VEM element bilinear form that provides patch test satisfaction when the solution
is a linear displacement field (condition~\eref{eq:consistency_cond} is satisfied). The second term on
the right-hand side of~\eref{eq:disc_vem_strain_energy} is the \textit{stability} part of the discrete
VEM element bilinear form and is dependent on the matrix $\mat{S}_E=s_E(\mat{N}^\transpose,\mat{N})$. This matrix
must be chosen such that condition~\eref{eq:stability_cond} holds without putting at risk
condition~\eref{eq:consistency_cond} already taken care of by the consistency part. There are
quite a few possibilities for this matrix (see for instance~\cite{BeiraoDaVeiga-Brezzi-Cangiani-Manzini-Marini-Russo:2013,BeiraodaVeiga-Brezzi-Marini:2013,Gain-Talischi-Paulino:2014}). Herein, we adopt $\mat{S}_E$ given by~\cite{Gain-Talischi-Paulino:2014}
\begin{equation}\label{eq:alt_stability}
\mat{S}_E=\alpha_E\,\mat{I}_{2N},\quad \alpha_E=\gamma\frac{|E|\textrm{trace}(\mat{D})}{\textrm{trace}(\mat{H}_\mathcal{C}^\transpose\mat{H}_\mathcal{C})},
\end{equation}
where $\alpha_E$ is the scaling parameter and $\gamma$ is typically set to 1.

From~\eref{eq:disc_vem_strain_energy}, the final expression for the VEM element stiffness matrix
is obtained respectively as the summation of the element \textit{consistency} and
\textit{stability} stiffness matrices, as follows:
\begin{equation}\label{eq:disc_stiffness}
\mat{K}_E=
|E|\,\mat{W}_\mathcal{C}\,\mat{D}\,\mat{W}_\mathcal{C}^\transpose+(\mat{I}_{2N}-\mat{P}_\mathcal{P})^\transpose\,\mat{S}_E\,(\mat{I}_{2N}-\mat{P}_\mathcal{P}),
\end{equation}
where we recall that $\mat{P}_\mathcal{P} = \mat{H}_\mathcal{R}\mat{W}_\mathcal{R}^\transpose+\mat{H}_\mathcal{C}\mat{W}_\mathcal{C}^\transpose$. Note that $\mat{H}_\mathcal{R}$ and $\mat{H}_\mathcal{C}$, which are given in~\eref{eq:matrix_hr} and~\eref{eq:matrix_hc}, respectively, are easily computed using the nodal coordinates of the element. However, in order to compute $\mat{W}_\mathcal{R}$ and $\mat{W}_\mathcal{C}$ (see their expressions in~\eref{eq:matrix_wr} and~\eref{eq:matrix_wc}, respectively), we need some knowledge of the basis functions so that $\overline{\phi}_a$ and $q_{ia}$ can be determined. Observe that $\overline{\phi}_a$ is computed using~\eref{eq:h_bar}, which requires the knowledge of the basis functions at the element nodes. And $q_{ia}$ is computed using~\eref{eq:qia}, which requires the knowledge of the basis functions on the element edges. Hence, everything we need to know about the basis functions is their behavior on the element boundary.

We have already mentioned that the basis functions in the VEM are assumed to be Lagrange-type functions. This provides everything we need to know about them on the boundary of an element: basis functions are piecewise linear (edge by edge) and continuous on the element edges, and have the Kronecker delta property. Therefore, $\overline{\phi}_a$ can be computed simply as
\begin{equation}\label{eq:known_basisfunctions_1}
\overline{\phi}_a=\frac{1}{N}\sum_{j=1}^N\phi_a(\vm{x}_j)=\frac{1}{N},
\end{equation}
and $q_{ia}$ can be computed exactly using a trapezoidal rule, which gives
\begin{equation}\label{eq:known_basisfunctions_2}
q_{ia}=\frac{1}{2|E|}\int_{\partial E}\phi_a n_i\,\diffs= \frac{1}{4|E|}\left(|e_{a-1}|\{n_i\}_{a-1}+|e_a|\{n_i\}_a\right),\quad i=1,2,
\end{equation}
where $\{n_i\}_a$ are the components of $\vm{n}_a$ and $|e_a|$ is the length of the edge incident to node $a$ as defined in~\fref{fig:1}.

The adoption of~\eref{eq:known_basisfunctions_1} and \eref{eq:known_basisfunctions_2} in the VEM, results in an algebraic evaluation
of the element stiffness matrix. This also means that the basis functions are not evaluated explicitly --- in fact, they are never computed. Thus, basis functions are said to be \textit{virtual}. In addition, the knowledge of the basis functions in the interior of the element is not required, although the linear approximation of the displacement field everywhere in the element is computable through the projection~\eref{eq:pipv_final}. Therefore, a more specific discrete global trial space than the one already given in Section~\ref{sec:modelproblem} can be built by assembling element by element the local space defined as~\cite{BeiraoDaVeiga-Brezzi-Cangiani-Manzini-Marini-Russo:2013,BeiraodaVeiga-Lovadina-Mora:2015}
\begin{equation}
V^h|_E := \left\{\vm{v}^h \in [H^1(E)\cap C^0(E)]^2:\Delta \vm{v}^h=\vm{0}\,\,\textrm{in}\,\, E, \, \vm{v}^h|_e = \mathcal{P}(e)\,\,\,\forall e\in\partial E\right\}.
\end{equation}

\subsection{VEM element body and traction force vectors}
\label{sec:vembodyforcevector}

For linear displacements, the body force can be approximated by a piecewise constant. Typically, this piecewise constant approximation is defined as the cell-average $\vm{b}^h=\frac{1}{|E|}\int_E\vm{b}\,\diffx=\widehat{\vm{b}}$. Thus, the body force part of the discrete VEM element linear form can be computed as follows~\cite{BeiraoDaVeiga-Brezzi-Cangiani-Manzini-Marini-Russo:2013,BeiraodaVeiga-Brezzi-Marini:2013,artioli:AO2DVEM:2017}:
\begin{equation}\label{eq:discrete_element_body_force_linear_form}
\ell_{b,E}^h(\vm{v}^h)=\int_E\vm{b}^h\cdot\overline{\vm{v}}^h\,\diffx=|E|\widehat{\vm{b}}\cdot\overline{\vm{v}}^h=
\mat{q}^\transpose|E|\,\overline{\mat{N}}^\transpose\widehat{\vm{b}},
\end{equation}
where
\begin{equation}\label{eq:matrix_N_bar}
\overline{\mat{N}} =
\left[(\overline{\mat{N}})_1 \quad \cdots \quad (\overline{\mat{N}})_a \quad \cdots \quad
(\overline{\mat{N}})_N\right] \, , \,\,\, (\overline{\mat{N}})_a=\smat{\overline{\phi}_a & 0 \\ 0 &
\overline{\phi}_a}.
\end{equation}
Hence, the VEM element body force vector is given by
\begin{equation}
\mat{f}_{b,E}=|E|\,\overline{\mat{N}}^\transpose\widehat{\vm{b}}.
\end{equation}

The traction force part of the VEM element linear form is similar to the integral expression given in~\eref{eq:discrete_element_body_force_linear_form}, but the integral is one dimension lower. Therefore,
on considering the element edge as a two-node one-dimensional element, the VEM element traction
force vector can be computed on an element edge lying on the natural (Neumann) boundary, as follows:
\begin{equation}
\mat{f}_{f,e}=|e|\,\overline{\mat{N}}_\Gamma^\transpose\,\widehat{\vm{f}},
\end{equation}
where
\begin{equation}
\overline{\mat{N}}_\Gamma=\smat{\overline{\phi}_1 & 0 & \overline{\phi}_2 & 0\\ 0 &
\overline{\phi}_1 & 0 & \overline{\phi}_2}=\smat{\frac{1}{N} & 0 & \frac{1}{N} & 0\\ 0 &
\frac{1}{N} & 0 & \frac{1}{N}}=\smat{\frac{1}{2} & 0 & \frac{1}{2} & 0\\ 0 &
\frac{1}{2} & 0 & \frac{1}{2}}
\end{equation}
and $\widehat{\vm{f}}=\frac{1}{|e|}\int_e\vm{f}\,\diffs$.

\subsection{$L^2$-norm and $H^1$-seminorm of the error}
\label{sec:norms}

To assess the accuracy and convergence of the VEM, two global error measures are used. The relative $L^2$-norm of the displacement error defined as
\begin{equation}
\frac{||\vm{u}-\Pi_\mathcal{P}\vm{u}^h||_{L^2(\Omega)}}{||\vm{u}||_{L^2(\Omega)}}
=\sqrt{\frac{\sum_E\int_E\left(\vm{u}-\Pi_\mathcal{P}\vm{u}^h\right)^\transpose
       \left(\vm{u}-\Pi_\mathcal{P}\vm{u}^h\right)\,\diffx}
       {\sum_E\int_E\vm{u}^\transpose\vm{u}\,\diffx}},
\end{equation}
and the relative $H^1$-seminorm of the displacement error given by
\begin{equation}
\frac{||\vm{u}-\Pi_\mathcal{P}\vm{u}^h||_{H^1(\Omega)}}{||\vm{u}||_{H^1(\Omega)}}
=\sqrt{\frac{\sum_E\int_E\left(\upvarepsilon(\vm{u})-\upvarepsilon(\Pi_\mathcal{C}\vm{u}^h)\right)^\transpose
       \mat{D}\left(\upvarepsilon(\vm{u})-\upvarepsilon(\Pi_\mathcal{C}\vm{u}^h)\right)\,\diffx}
       {\sum_E\int_E\upvarepsilon(\vm{u})^\transpose\mat{D}\upvarepsilon(\vm{u})\,\diffx}},
\end{equation}
where the strain appears in Voigt notation and
$\upvarepsilon(\Pi_\mathcal{C}\vm{u}^h)=\bsym{\nabla}_\mathrm{S}(\Pi_\mathcal{C}\vm{u}^h)=\widehat{\upvarepsilon}(\vm{u}^h)$ (see~\eref{eq:strain} and~\eref{eq:picv_final}).

\subsection{VEM element stiffness matrix for the Poisson problem}
\label{sec:vempoisson}

The VEM formulation for the Poisson problem is derived similarly to the VEM formulation for the linear elastostatic problem. However, herein we give the VEM stiffness matrix for the Poisson problem by reducing the solution dimension in the two-dimensional linear elastostatic VEM formulation. The following reductions are used: the displacement field reduces to the scalar field $v(\vm{x})$, the strain is simplified to $\bsym{\varepsilon}(v)=\bsym{\nabla} v$, the rotations become $\bsym{\omega}(v)=\vm{0}$, and the constitutive matrix is replaced by the identity ($2\times 2$) matrix. Hence, the VEM projections for the Poisson problem become $\Pi_\mathcal{R}v=\overline{v}$ and $\Pi_\mathcal{C}u=\widehat{\bsym{\nabla}}(v)\cdot(\vm{x}-\overline{\vm{x}})$. The matrices that result from the discretization of the projection operators are simplified to
\begin{equation}\label{eq:matrix_hr_poisson}
\mat{H}_\mathcal{R}=\smat{ (\mat{H}_\mathcal{R})_1 & \cdots &
(\mat{H}_\mathcal{R})_a & \cdots & (\mat{H}_\mathcal{R})_N}^\transpose,
\quad (\mat{H}_\mathcal{R})_a = 1,
\end{equation}
\begin{equation}\label{eq:matrix_wr_poisson}
\mat{W}_\mathcal{R}=\smat{
(\mat{W}_\mathcal{R})_1 & \cdots & (\mat{W}_\mathcal{R})_a &
\cdots & (\mat{W}_\mathcal{R})_N}^\transpose, \quad (\mat{W}_\mathcal{R})_a =
\frac{1}{N},
\end{equation}
\begin{equation}\label{eq:matrix_hc_poisson}
\mat{H}_\mathcal{C}=\smat{ (\mat{H}_\mathcal{C})_1 & \cdots &
(\mat{H}_\mathcal{C})_a & \cdots & (\mat{H}_\mathcal{C})_N}^\transpose,
\quad (\mat{H}_\mathcal{C})_a = \smat{ (x_{1a}-\overline{x}_1) & (x_{2a}-\overline{x}_2) },
\end{equation}
\begin{equation}\label{eq:matrix_wc_poisson}
\mat{W}_\mathcal{C}=\smat{
(\mat{W}_\mathcal{C})_1 & \cdots & (\mat{W}_\mathcal{C})_a &
\cdots & (\mat{W}_\mathcal{C})_N}^\transpose, \quad (\mat{W}_\mathcal{C})_a =
\smat{ 2q_{1a} & 2q_{2a} }.
\end{equation}
On using the preceding matrices, the projection matrix is $\mat{P}_\mathcal{P} = \mat{H}_\mathcal{R}\mat{W}_\mathcal{R}^\transpose+\mat{H}_\mathcal{C}\mat{W}_\mathcal{C}^\transpose$ and the final expression for the VEM element stiffness matrix is written as
\begin{equation}\label{eq:disc_stiffness_poisson}
\mat{K}_E=
|E|\,\mat{W}_\mathcal{C}\mat{W}_\mathcal{C}^\transpose+(\mat{I}_{N}-\mat{P}_\mathcal{P})^\transpose(\mat{I}_{N}-\mat{P}_\mathcal{P}),
\end{equation}
where $\mat{I}_{N}$ is the identity ($N\times N$) matrix and $\mat{S}_E=\mat{I}_{N}$ has been used in the stability stiffness as this represents a suitable choice for $\mat{S}_E$ in the Poisson problem~\cite{BeiraoDaVeiga-Brezzi-Cangiani-Manzini-Marini-Russo:2013}.

\section{Object-oriented implementation of VEM in C++}
\label{sec:implementation}

In this section, we introduce \texttt{Veamy}, a library that implements the VEM for the linear elastostatic and Poisson problems in two dimensions using object-oriented programming in C++. For the purpose of comparison with the VEM, a module implementing the standard FEM is available within \texttt{Veamy} for the solution of the two-dimensional linear elastostatic problem using three-node triangular finite elements. In \texttt{Veamy}, entities such as element, degree of freedom, constraint, among others, are represented by C++ classes.

\texttt{Veamy} uses the following external libraries:
\begin{itemize}
\item Triangle~\cite{shewchuk96b}, a two-dimensional quality mesh generator and Delaunay triangulator.
\item Clipper~\cite{clipperweb}, an open source freeware library for clipping and offsetting lines and polygons.
\item Eigen~\cite{eigenweb}, a C++ template library for linear algebra.
\end{itemize}

Triangle and Clipper are used in the implementation of \texttt{Delynoi}~\cite{delynoiweb}, a polygonal mesh generator
that is based on the constrained Voronoi diagram. The usage of our polygonal mesh generator is
covered in Section~\ref{sec:meshgenerator}.

\texttt{Veamy} is free and open source software and is available from Netlib (\url{http://www.netlib.org/numeralgo/}) as
the na51 package. In addition, a website (\url{http://camlab.cl/software/veamy/}) is available, where the software is maintained. After downloading and uncompressing the software, the main directory ``Veamy-2.1/'' is created. This directory is organized as follows. The source code that implements the VEM is provided in the folder ``veamy/'' and the subfolders therein. External libraries that are used by \texttt{Veamy} are provided in the folder ``lib/.'' The folder ``matplots/'' contains MATLAB functions that are useful for plotting meshes and the VEM solution, and for writing a \texttt{PolyMesher}~\cite{Talischi:POLYM:2012} mesh and boundary conditions to a text file that is readable by \texttt{Veamy}. A detailed software documentation with graphical content can be found in the tutorial manual that is provided in the folder ``docs/.'' Several tests are located in the folder ``test/.'' Some of these tests are covered in the tutorial manual and in Section~\ref{sec:sampleusage} of this paper. Veamy supports Linux and Mac OS machines only and compiles with g++, the GNU C++ compiler (GCC 7.3 or newer versions should be used). The installation procedure and the content that comprises the software are described in detail in the README.txt file (and also in the tutorial manual), which can be found in the main directory.

The core design of \texttt{Veamy} is presented in three UML diagrams that are intended to explain the numerical methods implemented (\fref{fig:2}), the problem conditions inherent to the linear elastostatic and Poisson problems (\fref{fig:3}), and the computation of the $L^2$-norm and $H^1$-seminorm of the errors (\fref{fig:4}).

\subsection{Numerical methods}
\begin{figure}[!tbhp]
\centering
\epsfig{file = 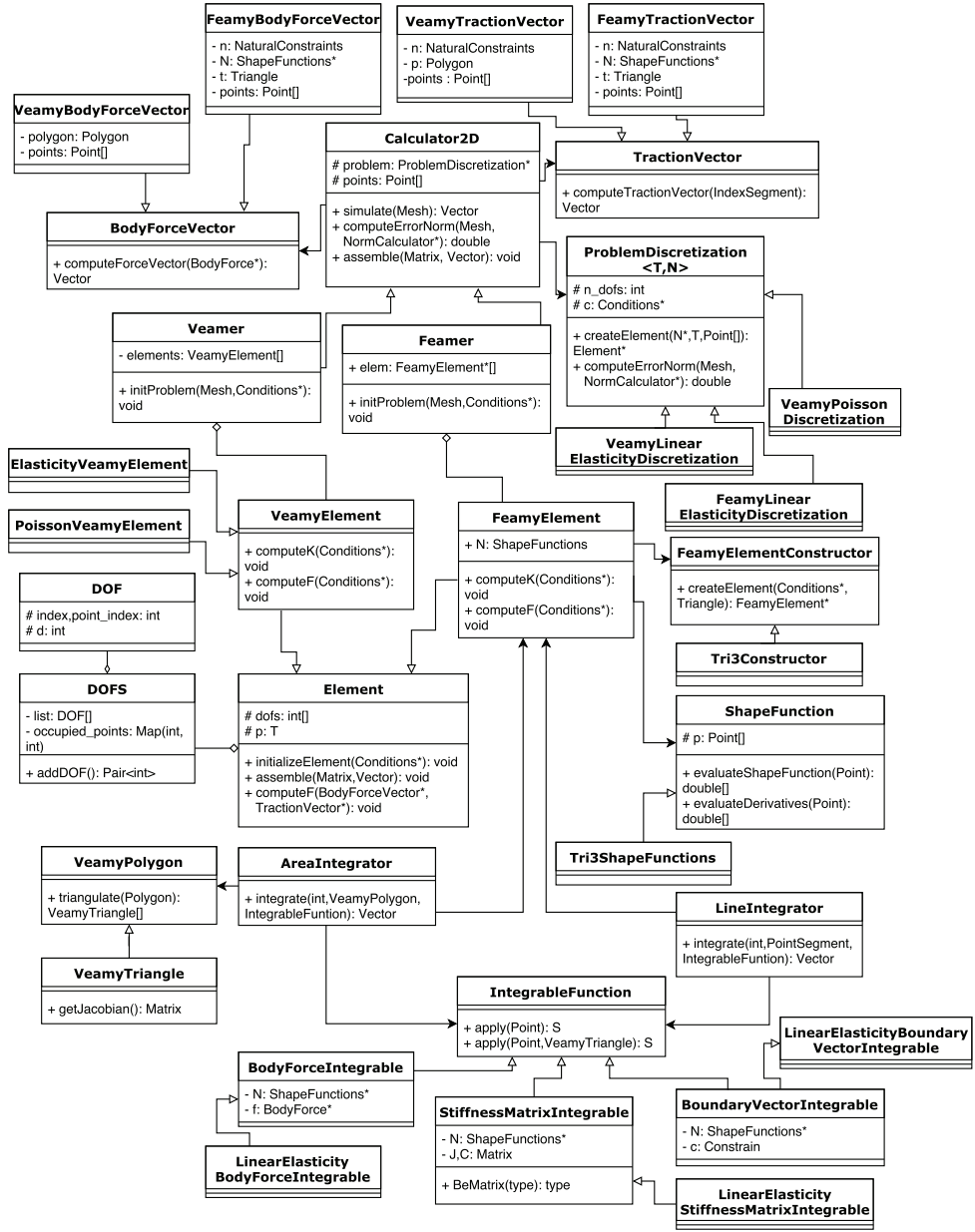, width = 0.85\textwidth}
\caption{UML diagram for the \texttt{Veamy} library. VEM and FEM modules}
\label{fig:2}
\end{figure}

The \texttt{Veamy} library is divided into two modules, one that implements the VEM and another one that implements the FEM. \fref{fig:2} summarizes the implementation of these methods. Two abstract classes are central to the \texttt{Veamy} library, \texttt{Calculator2D} and \texttt{Element}. \texttt{Calculator2D} is designed in the spirit of the controller design pattern. It receives the \texttt{ProblemDiscretization} subclasses with all their associated problem conditions, creates the required structures, applies the boundary conditions and runs the simulation. \texttt{Calculator2D}, as an abstract class, has a number of methods that all inherited classes must implement. The two most important are the one in charge of creating the elements, and the one in charge of computing the element stiffness matrix and the element (body and traction) force vector. We implement two concrete \texttt{Calculator2D} classes, called \texttt{Veamer} and \texttt{Feamer}, with the former representing the controller for the VEM and the latter for the FEM.

On the other hand, \texttt{Element} is the class that encapsulates the behavior of each element in the domain. It is in charge of keeping the degrees of freedom of the element and its associated stiffness matrix and force vector. \texttt{Element} contains methods to create and assign degrees of freedom, assemble the element stiffness matrix and the element force vector into the global ones. An \texttt{Element} has the information of its defining polygon (the three-node triangle is the lowest-order polygon) along with its degrees of freedom. \texttt{Element} has two inherited classes, \texttt{VeamyElement} and \texttt{FeamyElement}, which represent elements of the VEM and FEM, respectively. They are in charge of the computation of the element stiffness matrix and the element force vector. Algorithm~\ref{algo:1} summarizes the implementation of the linear elastostatic VEM element stiffness matrix in the \texttt{VeamyElement} class using the notation presented in Sections~\ref{sec:projection_matrices} and \ref{sec:vem_element_stiffness}.

\begin{algorithm}[H]
\SetAlgoCaptionSeparator{\quad}
\DontPrintSemicolon
\SetArgSty{textrm}
\SetAlgoLined
 $\mat{H}_\mathcal{R}=\mat{0}$, $\mat{W}_\mathcal{R}=\mat{0}$, $\mat{H}_\mathcal{C}=\mat{0}$, $\mat{W}_\mathcal{C}=\mat{0}$\;
 \For{each node in the polygonal element}{Get incident edges\;
 Compute the unit outward normal vector to each incident edge\;
 Compute $(\mat{H}_\mathcal{R})_a$ and $(\mat{H}_\mathcal{C})_a$, and insert them into $\mat{H}_\mathcal{R}$ and $\mat{H}_\mathcal{C}$, respectively\;
 Compute $(\mat{W}_\mathcal{R})_a$ and $(\mat{W}_\mathcal{C})_a$, and insert them into $\mat{W}_\mathcal{R}$ and $\mat{W}_\mathcal{C}$, respectively}
 Compute $\mat{I}_{2N}$, $\mat{P}_\mathcal{R}$, $\mat{P}_\mathcal{C}$, $\mat{P}_\mathcal{P}$, $\mat{D}$\;
 Compute $\mat{S}_E$\;
 \KwOut{$\mat{K}_E=|E|\,\mat{W}_\mathcal{C}\,\mat{D}\,\mat{W}_\mathcal{C}^\transpose+
 (\mat{I}_{2N}-\mat{P}_\mathcal{P})^\transpose\,\mat{S}_E\,(\mat{I}_{2N}-\mat{P}_\mathcal{P})$}
 \label{algo:1}
 \caption{Implementation of the VEM element stiffness matrix for the linear elastostatic problem in the \texttt{VeamyElement} class}
\end{algorithm}

The element force vector is computed with the aid of the abstract classes \texttt{BodyForceVector} and \texttt{TractionVector}. Each of them has two concrete subclasses named \texttt{VeamyBodyForceVector} and \texttt{FeamyBodyForceVector}, and \texttt{VeamyTractionVector} and \texttt{FeamyTractionVector}, respectively.

Even though we have implemented the three-node triangular finite element only as a means to comparison with the VEM, we decided to define \texttt{FeamyElement} as an abstract class so that more advanced elements can be implemented if desired. Finally, each \texttt{FeamyElement} concrete implementation has a \texttt{ShapeFunction} concrete subclass, representing the shape functions that are used to interpolate the solution inside the element. For the three-node triangular finite element, we include the \texttt{Tri3ShapeFunctions} class.

One of the structures related to all \texttt{Element} classes is called \texttt{DOF}. It describes a single degree of freedom. The degree of freedom is associated with the nodal points of the mesh according to the \texttt{ProblemDiscretization} subclasses. So, in the linear elastostatic problem each nodal point has two associated \texttt{DOF} instances and in the Poisson problem just one \texttt{DOF} instance. The \texttt{DOF} instances are kept in a list inside a container class called \texttt{DOFS}.

Although the VEM matrices are computed algebraically, the FEM matrices in general require numerical integration both inside the element (area integration) and on the edges that lie on the natural boundary (line integration). Thus, we have implemented two classes, \texttt{AreaIntegrator} and \texttt{LineIntegrator}, which contain methods that integrate a given function inside the element and on its boundary. There are several classes related to the numerical integration. \texttt{IntegrableFunction} is a template interface that has a method called \texttt{apply} that must be implemented. This method receives a sample point and must be implemented so that it returns the evaluation of a function at the sample point. We include three concrete \texttt{IntegrableFunction} implementations, one for the body force, another one for the stiffness matrix and the last one for the boundary vector.

\subsection{Problem conditions}
\begin{figure}[!tbhp]
\centering
\epsfig{file = 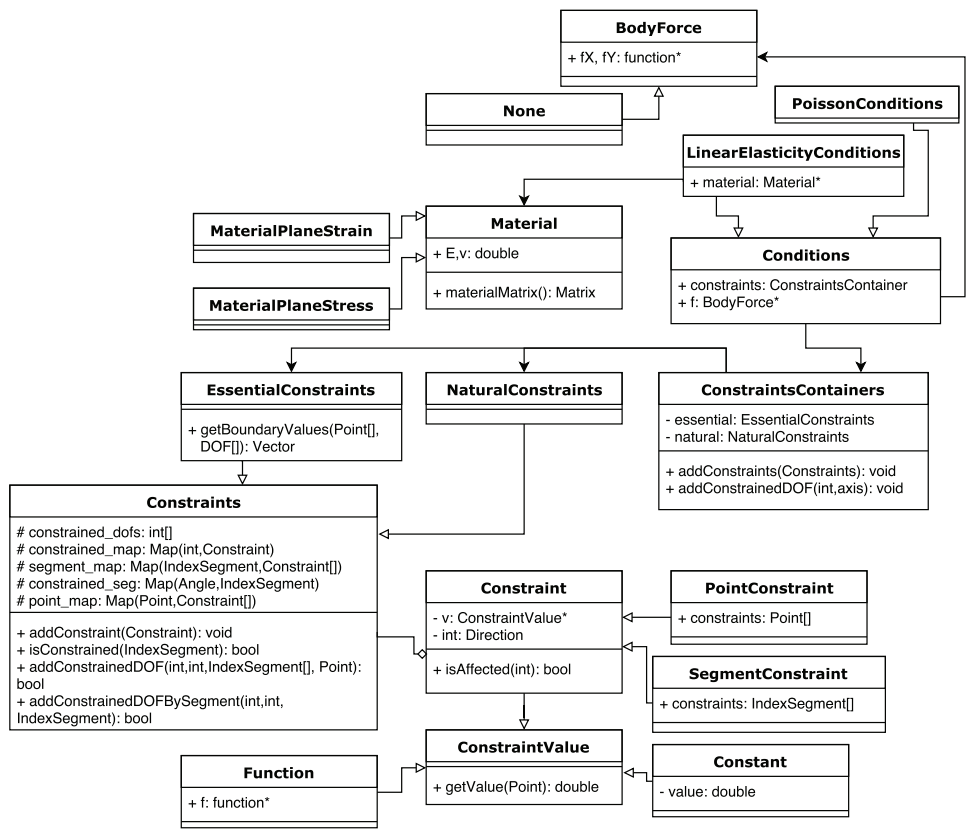, width = 0.85\textwidth}
\caption{UML diagram for the \texttt{Veamy} library. Problem conditions}
\label{fig:3}
\end{figure}

\fref{fig:3} presents the classes for the problem conditions used in the linear elastostatic and Poisson problems. The problem conditions are kept in a structure called \texttt{Conditions} that contains the physical properties of the material (\texttt{Material} class), the boundary conditions and the body force. \texttt{BodyForce} is a class that contains two functions pointers that represent the body force in each of the two axes of the Cartesian coordinate system. These two functions must be instantiated by the user to include a body force in the problem. By default, \texttt{Conditions} creates an instance of the \texttt{None} class, which is a subclass of \texttt{BodyForce} that represents the nonexistence of body forces. \texttt{Material} is an abstract class that keeps the elastic constants associated with the material properties (Young's modulus and Poisson's ratio) and has an abstract function that computes the material matrix; \texttt{Material} has two subclasses, \texttt{MaterialPlaneStress} and \texttt{MaterialPlaneStrain}, which return the material matrix for the plane stress and plane strain states, respectively.

To model the boundary conditions, we have created a number of classes: \texttt{Constraint} is an abstract class that represents a single constraint --- a constraint can be an essential (Dirichlet) boundary condition or a natural (Neumann) boundary condition. \texttt{PointConstraint} and \texttt{SegmentConstraint} are concrete classes implementing \texttt{Constraint} and representing a constraint at a point and on a segment of the domain, respectively. \texttt{Constraints} is the class that manages all the constraints in the system, and the relationship between them and the degrees of freedom; \texttt{EssentialConstraints} and \texttt{NaturalConstraints} inherit from \texttt{Constraints}. Finally, \texttt{ConstraintsContainers} is a utility class that contains \texttt{EssentialConstraints} and \texttt{NaturalConstraints} instances. \texttt{Constraint} keeps a list of domain segments subjected to a given condition, the value of this condition, and a certain direction (vertical, horizontal or both). The interface called \texttt{ConstraintValue} is the method to control the way the user inputs the constraints: to add any constraint, the user must choose between a constant value (\texttt{Constant} class) and a function (\texttt{Function} class), or implement a new class inheriting from \texttt{ConstraintValue}.

\subsection{Norms of the error}
\begin{figure}[!tbhp]
\centering
\epsfig{file = 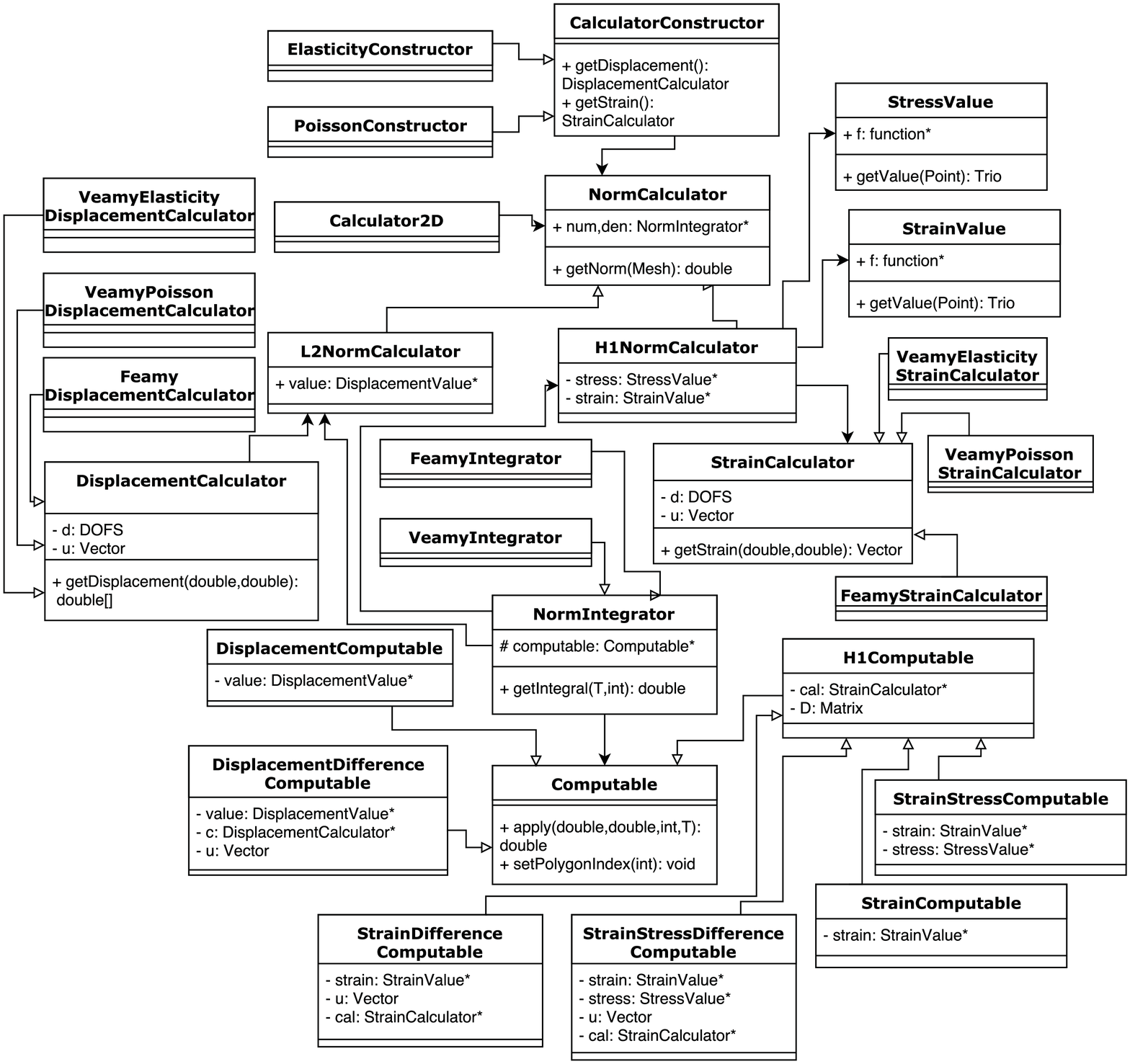, width = 0.85\textwidth}
\caption{UML diagram for the \texttt{Veamy} library. Computation of the $L^2$-norm and $H^1$-seminorm of the error}
\label{fig:4}
\end{figure}

As shown in \fref{fig:4}, \texttt{Veamy} provides functionalities for computing the relative $L^2$-norm and $H^1$-seminorm of the error through the classes \texttt{L2NormCalculator} and \texttt{H1NormCalculator}, respectively, which inherit from the abstract class \texttt{NormCalculator}. Each \texttt{NormCalculator} instance has two instances of what we call the \texttt{NormIntegrator} classes: \texttt{VeamyIntegrator} and \texttt{FeamyIntegrator}. These are in charge of integrating the norms integrals in the VEM and FEM approaches, respectively. In these \texttt{NormIntegrator} classes, the integrands of the norms integrals are represented by the \texttt{Computable} class. Depending on the integrand, we define various \texttt{Computable} subclasses: \texttt{DisplacementComputable}, \texttt{DisplacementDifferenceComputable}, \texttt{H1Computable} and its subclasses, \texttt{StrainDifferenceComputable}, \texttt{StrainStressDifferenceComputable}, \texttt{StrainComputable} and \texttt{StrainStressComputable}. Finally, \texttt{DisplacementCalculator} and \texttt{StrainCalculator} (and their subclasses) permit to obtain the numerical displacement and the numerical strain, respectively; and \texttt{StrainValue} and \texttt{StressValue} classes represent the exact value of the strains and stresses at the quadrature points, respectively.

\subsection{Computation of nodal displacements}
\label{subsec:displacementcomputation}
Each simulation is represented by a single \texttt{Calculator2D} instance, which is in charge of conducting the simulation through its \texttt{simulate} method until the displacement solution is obtained. The procedure is similar to a finite element simulation. The implementation of the \texttt{simulate} method is summarized in Algorithm~\ref{algo:2}.

\begin{algorithm}[H]
\SetAlgoCaptionSeparator{\quad}
\DontPrintSemicolon
\SetArgSty{textrm}
\SetAlgoLined
\KwIn{Mesh}
 Initialization of the global stiffness matrix and the global force vector\;
 \For{each element in the mesh}{Compute the element stiffness matrix\;
 Compute the element force vector\;
 Assemble the element stiffness matrix and the element force vector into global ones}
 Apply natural boundary conditions to the global force vector\;
 Impose the essential boundary conditions into the global matrix system\;
 Solve the resulting global matrix system of linear equations\;
 \KwOut{Column vector containing the nodal displacements solution}
 \label{algo:2}
 \caption{Implementation of the \texttt{simulate} method in the \texttt{Calculator2D} class}
\end{algorithm}

The resulting matrix system of linear equations is solved using appropriate solvers available in the Eigen library~\cite{eigenweb} for linear algebra.

\section{Polygonal mesh generator}
\label{sec:meshgenerator}

In this section, we provide some guidelines for the usage of our polygonal mesh generator \texttt{Delynoi}~\cite{delynoiweb}.

\subsection{Domain definition}
\begin{figure}[!tbhp]
\centering
\mbox{
\subfigure[]{\label{fig:5a}
\epsfig{file = 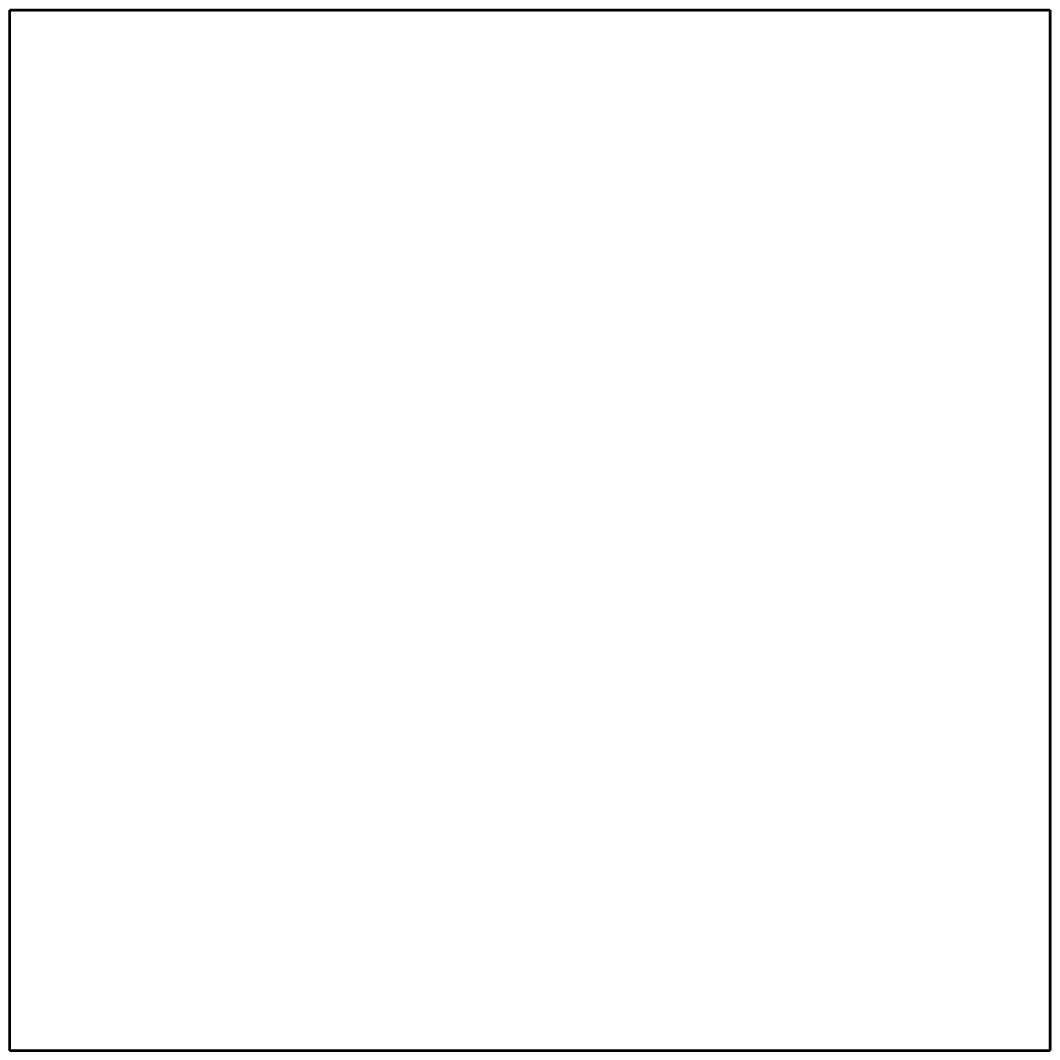, width = 0.16\textwidth}}
\subfigure[]{\label{fig:5b}
\epsfig{file = 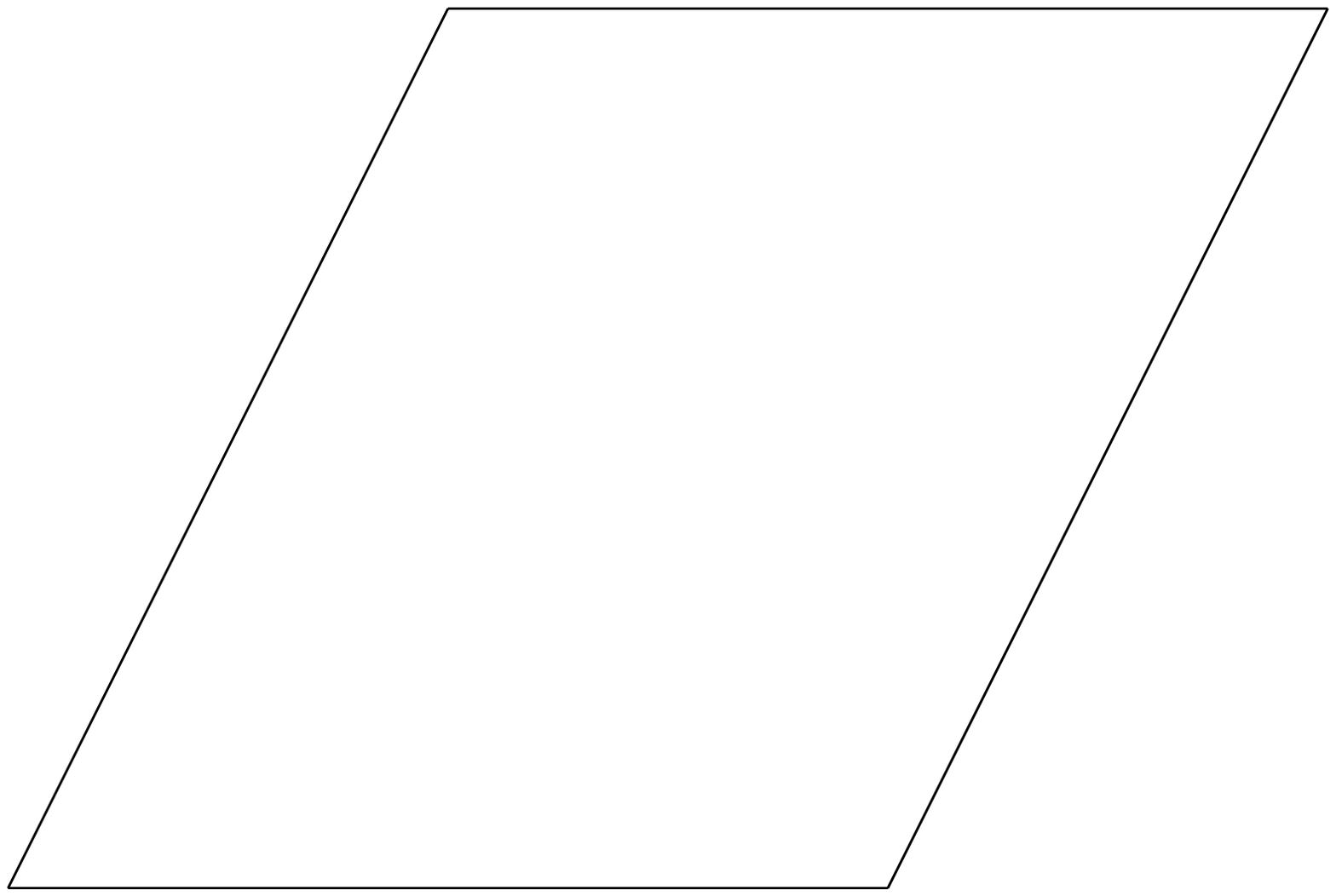, width = 0.2\textwidth}}
\subfigure[]{\label{fig:5c}
\epsfig{file = 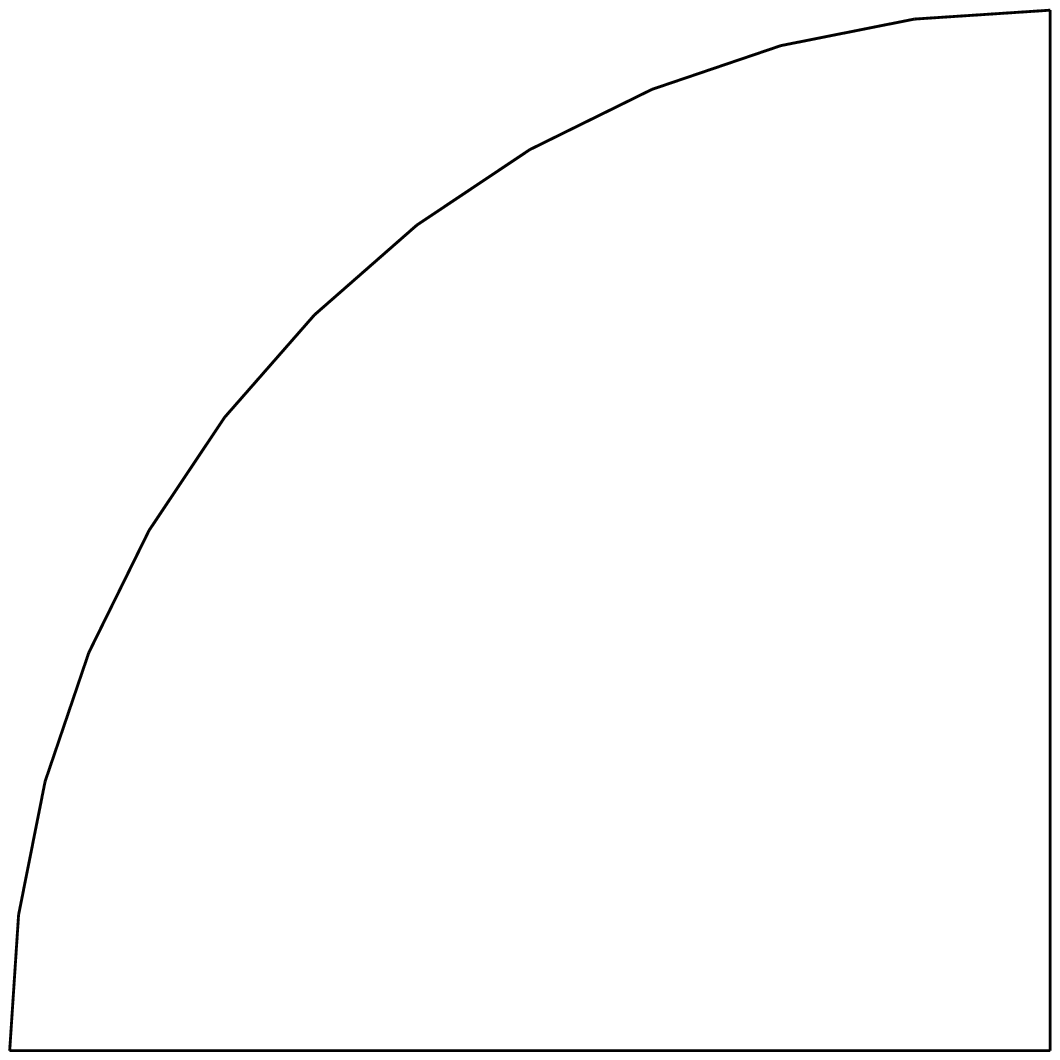, width = 0.16\textwidth}}
\subfigure[]{\label{fig:5d}
\epsfig{file = 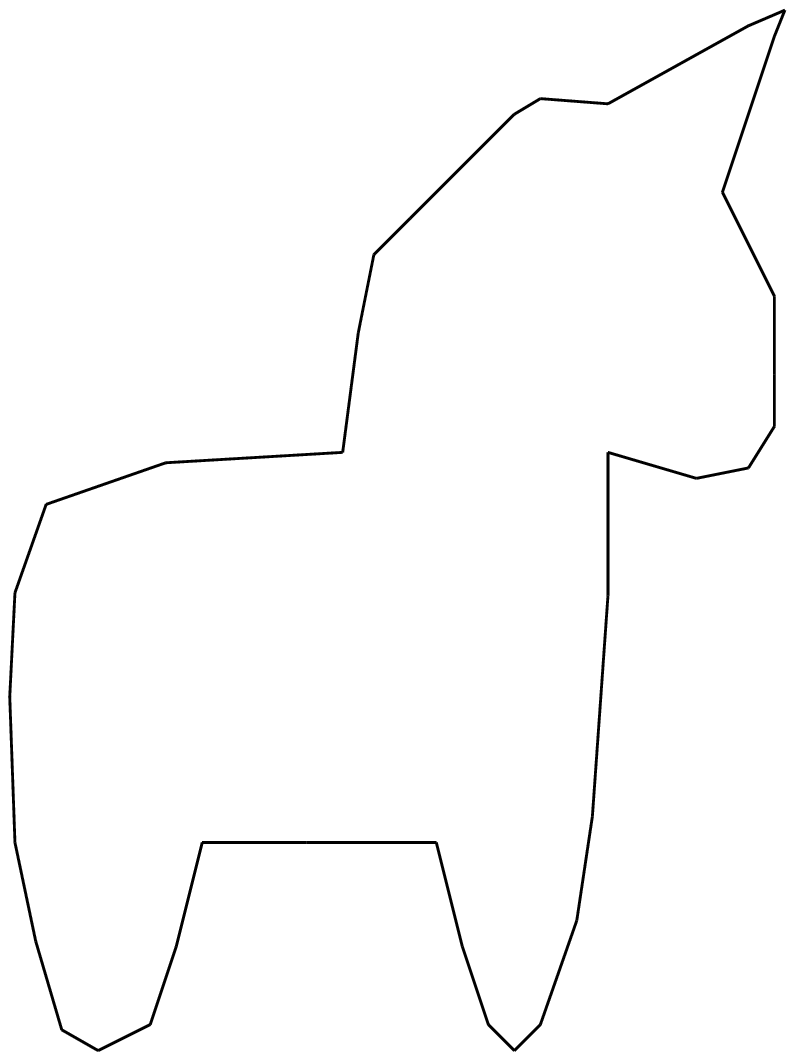, width = 0.16\textwidth}}
}
\caption{Domain examples. (a) Square domain, (b) rhomboid domain, (c) quarter circle domain, (d) unicorn-shaped domain}
\label{fig:5}
\end{figure}

\begin{figure}[!tbhp]
\centering
\mbox{
\subfigure[]{\label{fig:6a}
\epsfig{file = 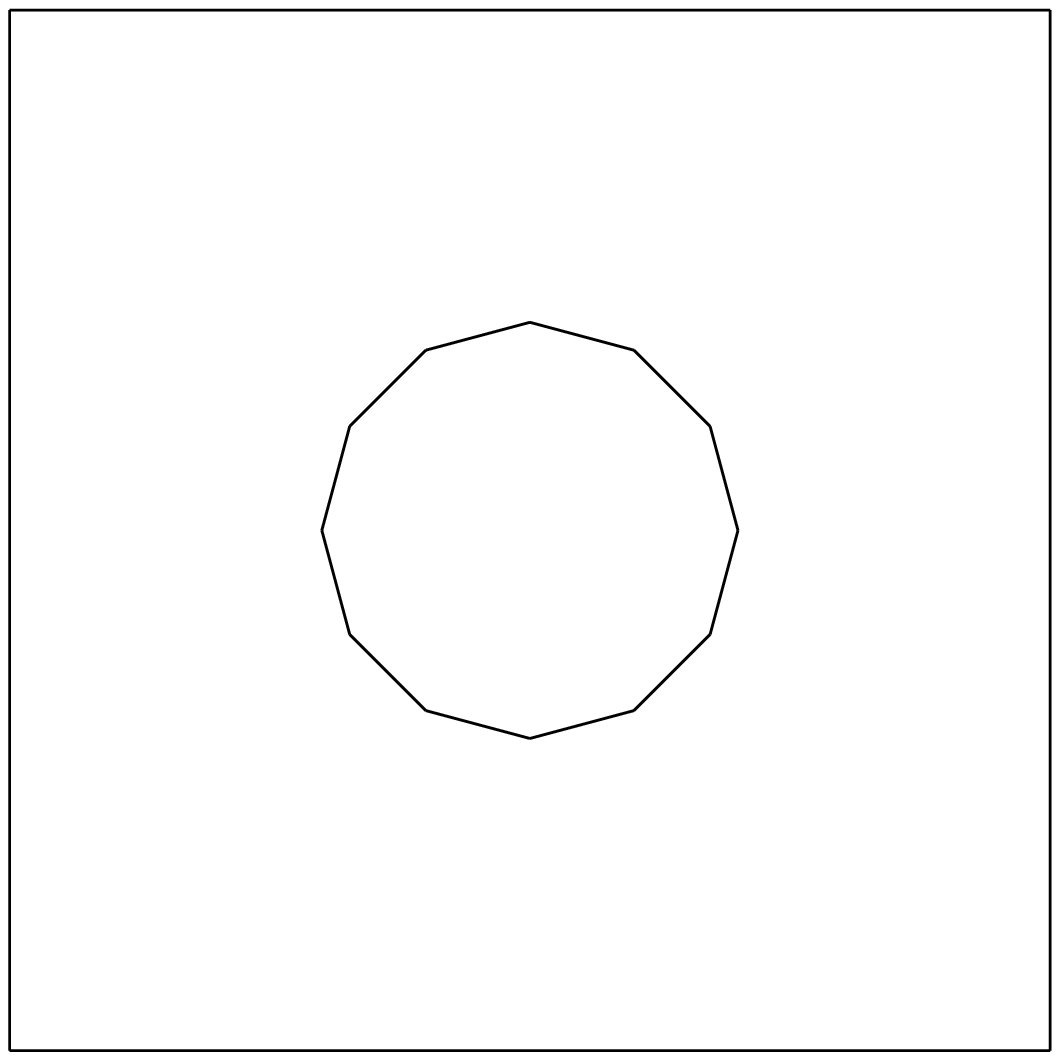, width = 0.18\textwidth}}
\subfigure[]{\label{fig:6b}
\epsfig{file = 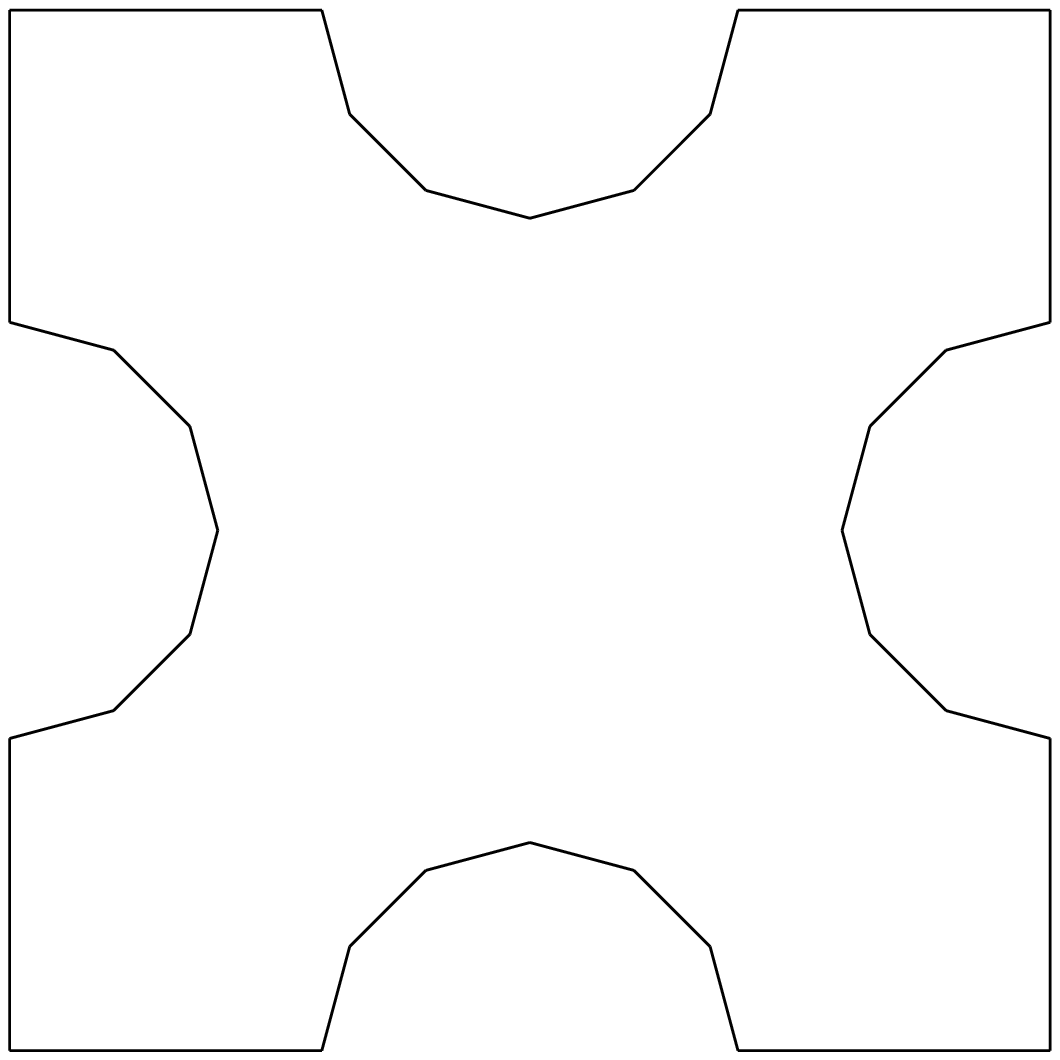, width = 0.18\textwidth}}
\subfigure[]{\label{fig:6c}
\epsfig{file = 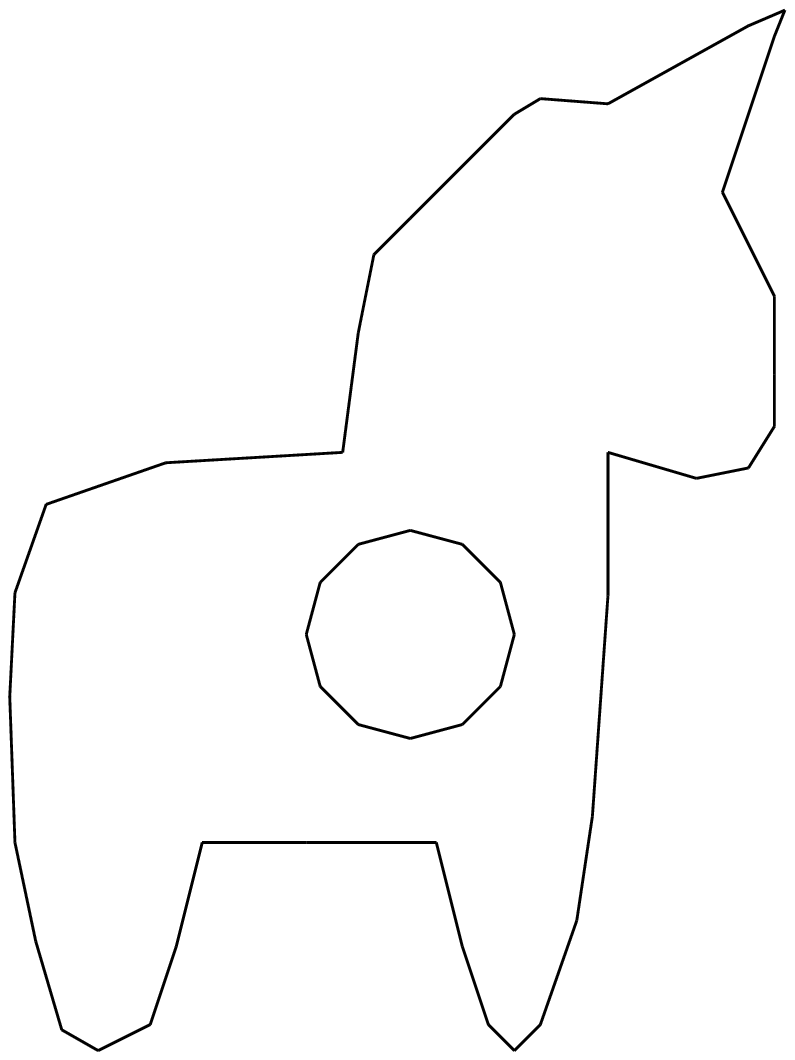, width = 0.16\textwidth}}
\subfigure[]{\label{fig:6d}
\epsfig{file = 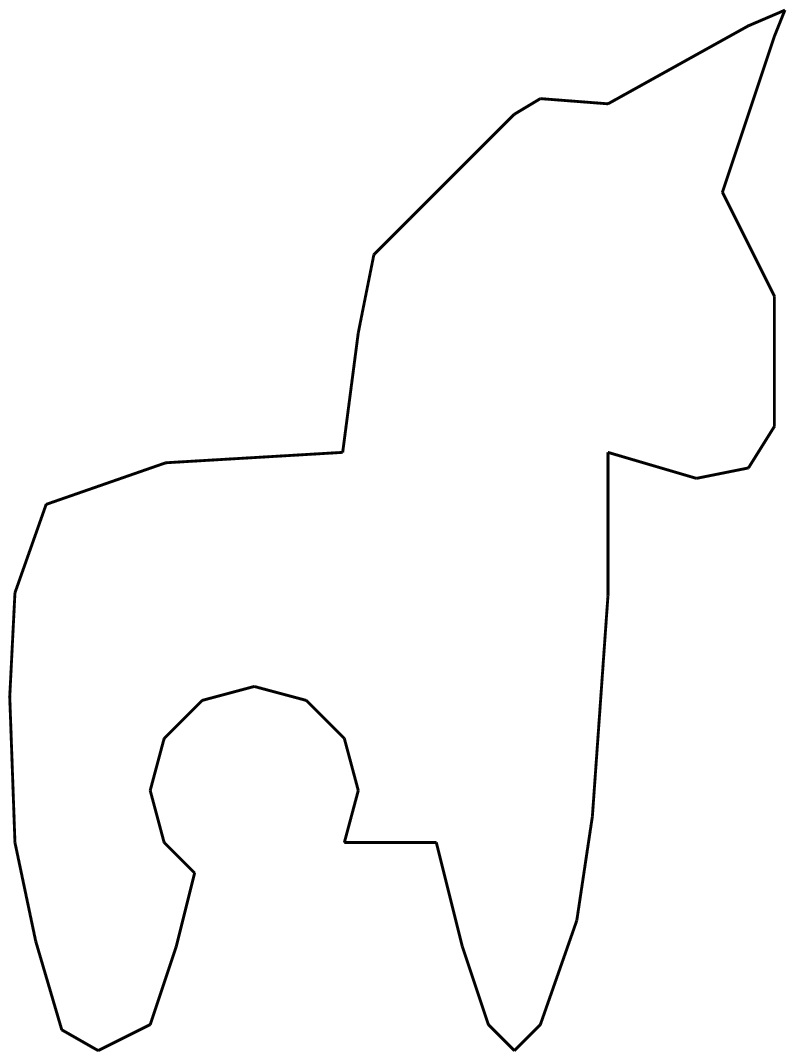, width = 0.16\textwidth}}
}
\caption{Examples of domains with holes. (a) Square with an inner hole, (b) square with four intersecting holes, (c) unicorn-shaped domain with an inner hole, (d) unicorn-shaped domain with an intersecting hole}
\label{fig:6}
\end{figure}

The domain is defined by creating its boundary from a counterclockwise list of points. Some examples of domains created in \texttt{Delynoi} are shown in \fref{fig:5}. We include the possibility of adding internal or intersecting holes to the domain as additional objects that are independent of the domain boundary. Some examples of domains created in \texttt{Delynoi} with one and several intersecting holes are shown in \fref{fig:6}.

Listing~\ref{lst:square_domain} shows the code to generate a square domain and a quarter circle domain. More domain definitions
are given in Section~\ref{sec:sampleusage} as part of \texttt{Veamy}'s sample usage problems.

\begin{lstlisting}[language=C++, caption={Definition of square and quarter circle domains}, label=lst:square_domain]
std::vector<Point> square_points = {Point(0,0), Point(10,0), Point(10,10), Point(0,10)};
Region square(square_points);
std::vector<Point> qc_points = {Point(0,0), Point(10,0), Point(10,10)};
std::vector<Point> quarter = delynoi_utilities::generateArcPoints(Point(10,0), 10, 90.0, 180.0);
qc_points.insert(quarter_circle_points.end(), quarter.begin(), quarter.end());
Region quarter_circle(qc_points);
\end{lstlisting}

To add a circular hole to the center of the square domain already defined, first the required hole is created and then added to
the domain as shown in Listing~\ref{lst:square_domain_hole}.

\begin{lstlisting}[language=C++, caption={Adding a circular hole to the center of the square domain}, label=lst:square_domain_hole]
Hole circular = CircularHole(Point(5,5), 2);
square.addHole(circular);
\end{lstlisting}

\subsection{Mesh generation rules}
\begin{figure}[!tbhp]
\centering
\mbox{
\subfigure[]{\label{fig:7a}
\epsfig{file = 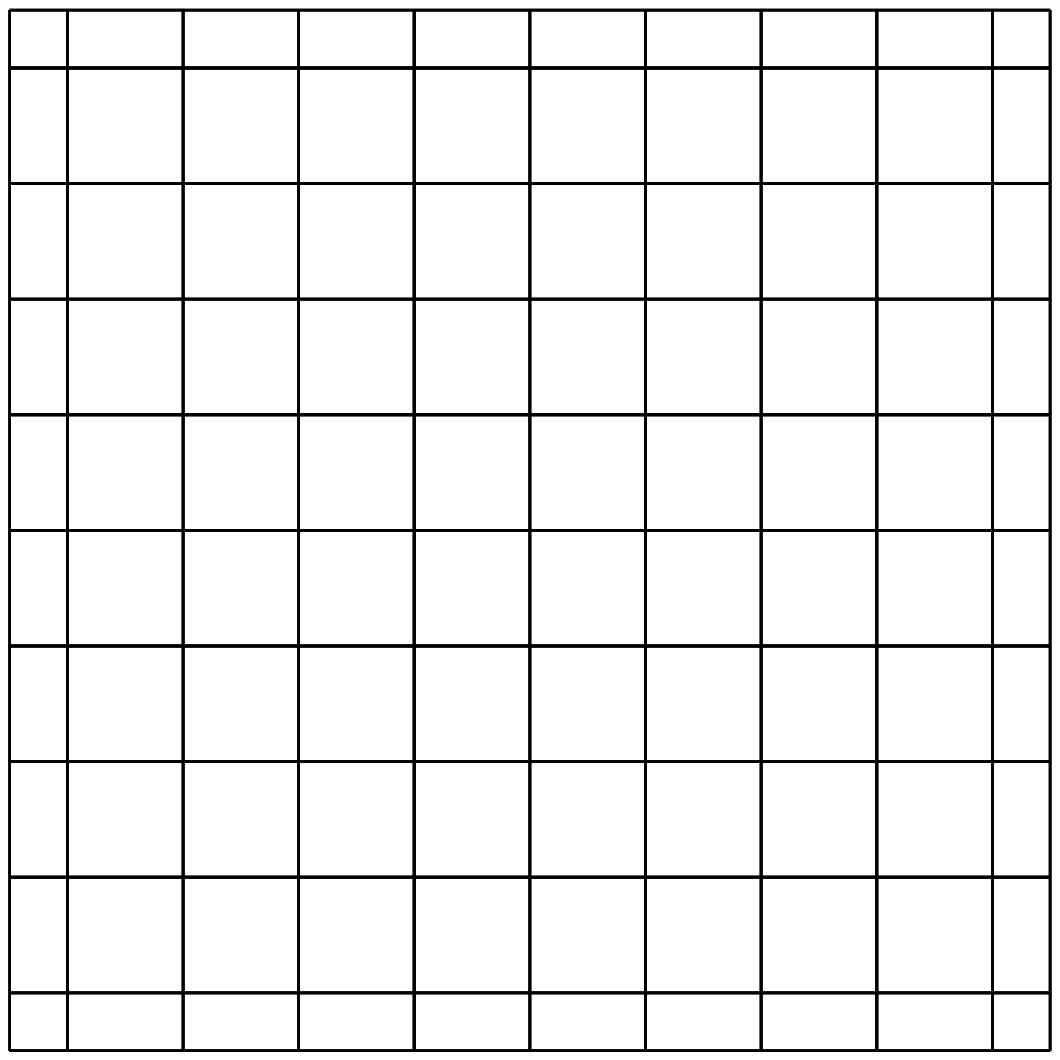, width = 0.18\textwidth}}
\subfigure[]{\label{fig:7b}
\epsfig{file = 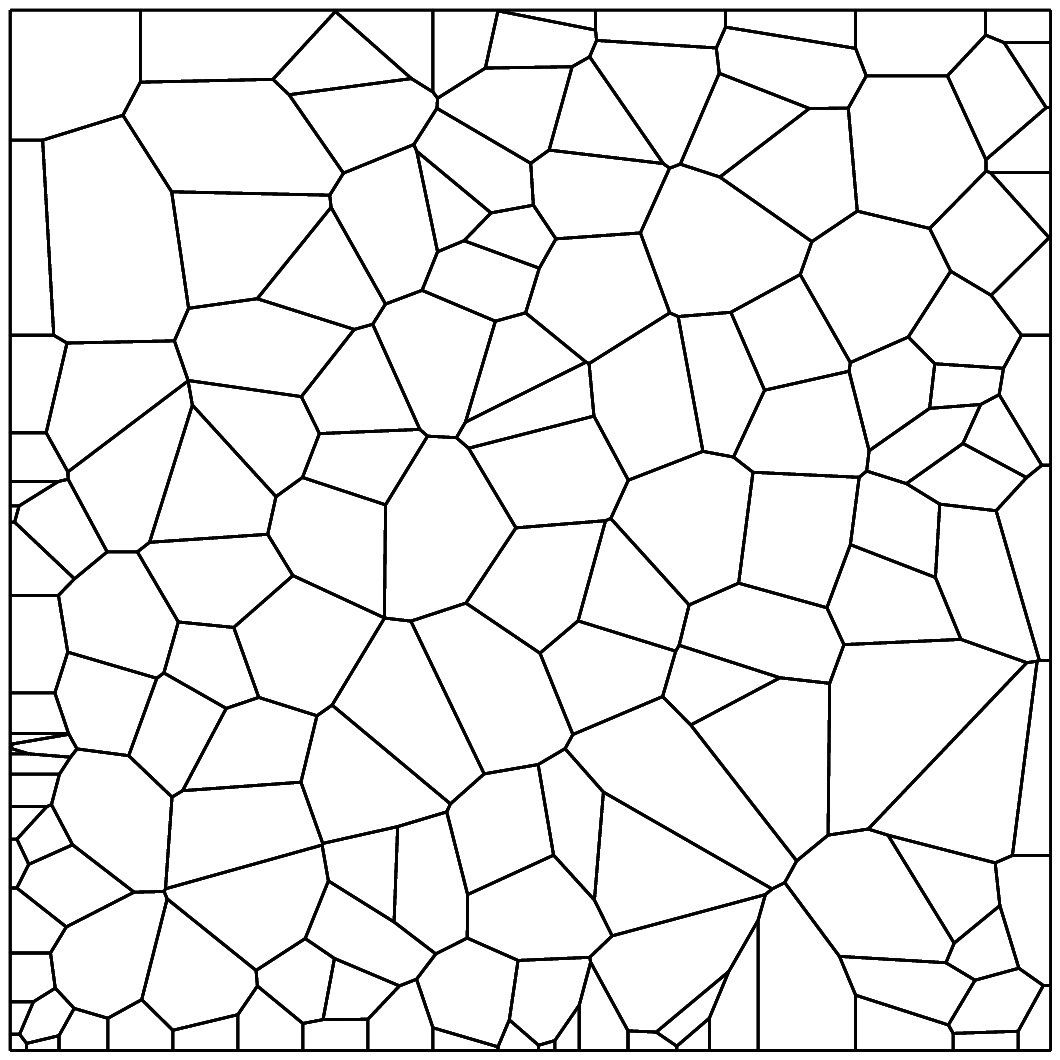, width = 0.18\textwidth}}
\subfigure[]{\label{fig:7c}
\epsfig{file = 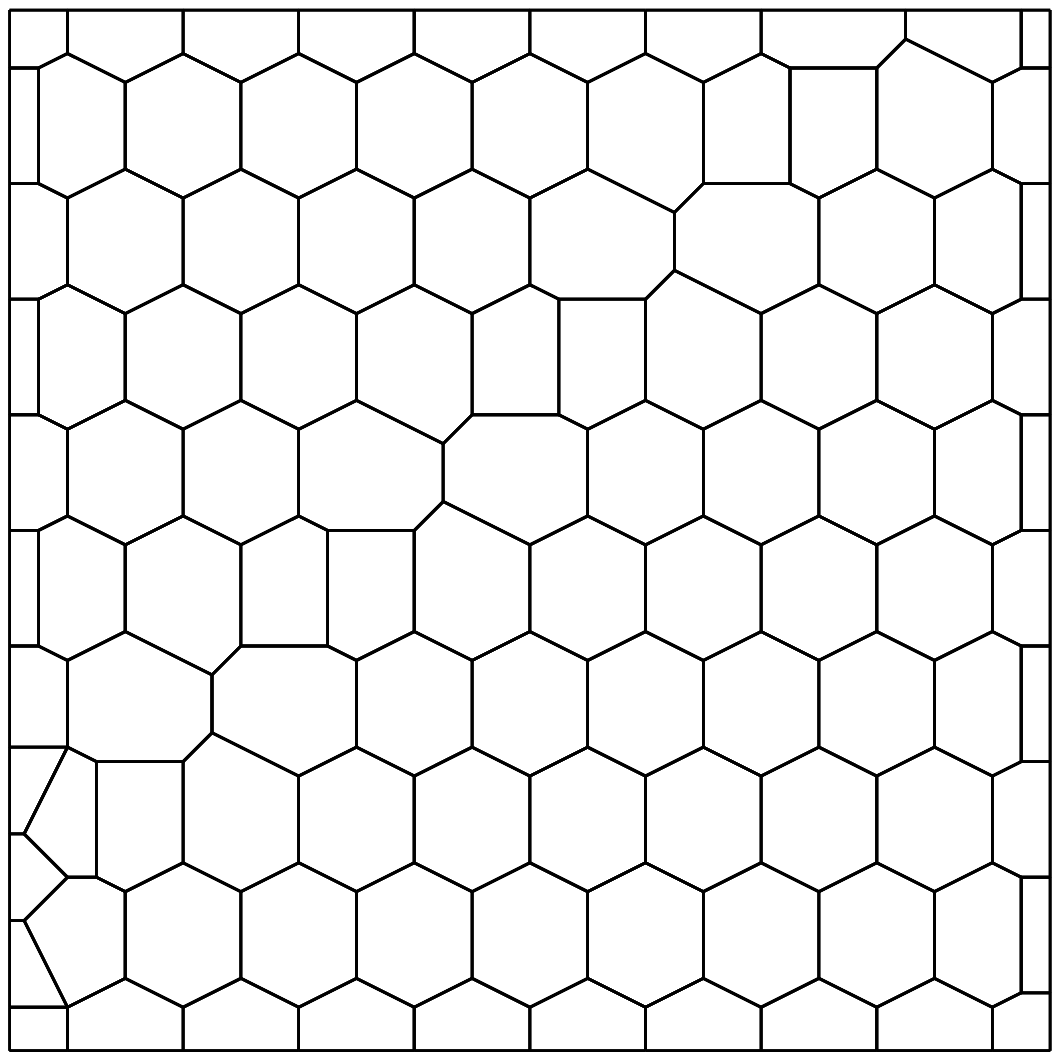, width = 0.18\textwidth}}
\subfigure[]{\label{fig:7d}
\epsfig{file = 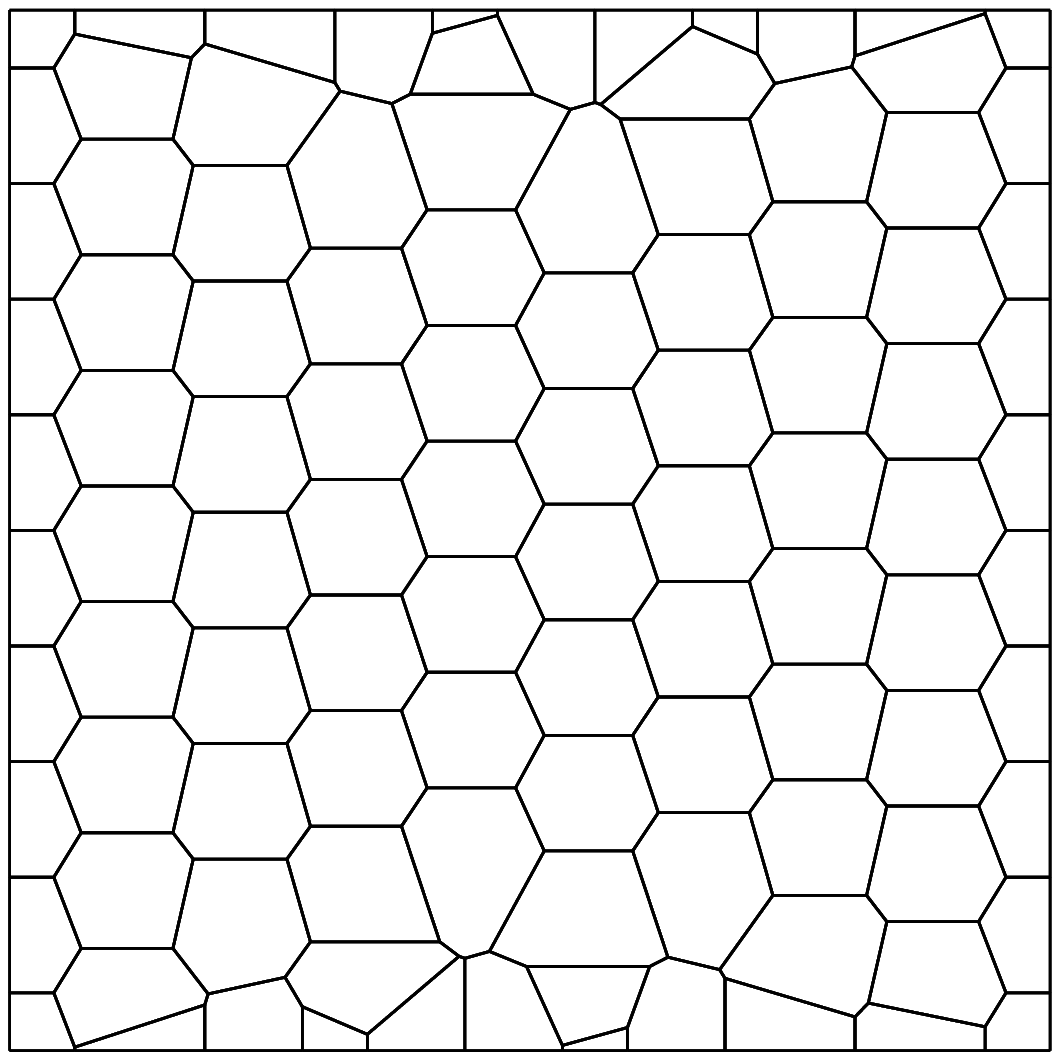, width = 0.18\textwidth}}
}
\caption{Polygonal mesh generation on a square domain using different rules. (a) \texttt{constant}, (b) \texttt{random\_double}, (c) \texttt{ConstantAlternating}, (d) \texttt{sine}}
\label{fig:7}
\end{figure}

We include a number of different rules for the generation of the seeds points for the Voronoi diagram. These rules are \texttt{constant},
\texttt{random\_double}, \texttt{ConstantAlternating} and \texttt{sine}. The \texttt{constant} method generates uniformly distributed seeds points;
the \texttt{random\_double} method generates random seeds points; the \texttt{ConstantAlternating} method generates seeds points by displacing
alternating the points along one Cartesian axis. \fref{fig:7} presents some examples of meshes generated on a square domain using different
rules. We show how to generate constant (uniform) and random points for a given domain in Listing~\ref{lst:generation_tests_points}.

\begin{lstlisting}[language=C++, caption={Generation of constant (uniform) and random points}, label=lst:generation_tests_points]
dom1.generateSeedPoints(PointGenerator(functions::constant(), functions::constant()), nX, nY);
dom2.generateSeedPoints(PointGenerator(functions::random_double(0,maxX), functions::random_double(0,maxY)), nX, nY);
// nX, nY: horizontal and vertical divisions along sides of the bounding box
\end{lstlisting}

\begin{figure}[!tbhp]
\centering
\mbox{
\subfigure[]{\label{fig:8a}
\epsfig{file = 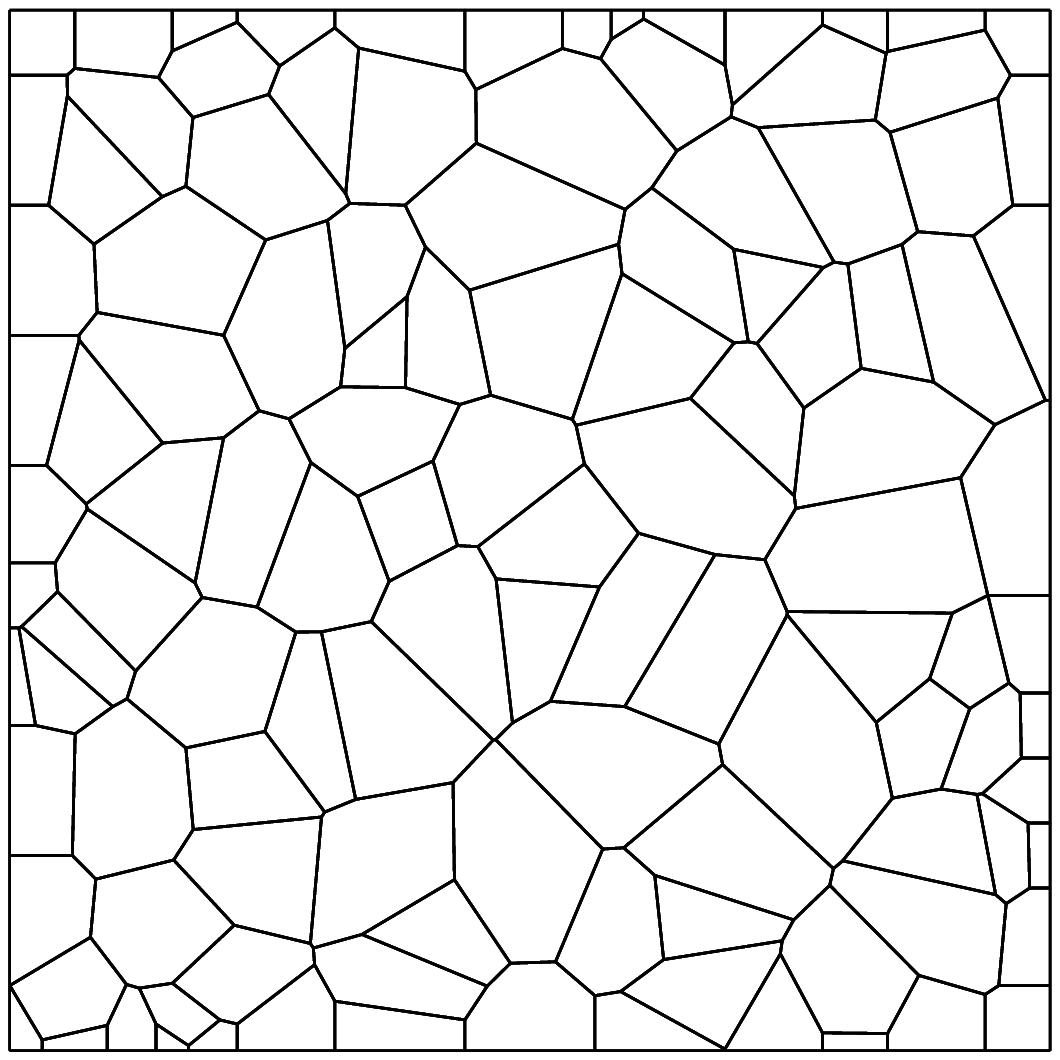, width = 0.18\textwidth}}
\subfigure[]{\label{fig:8b}
\epsfig{file = 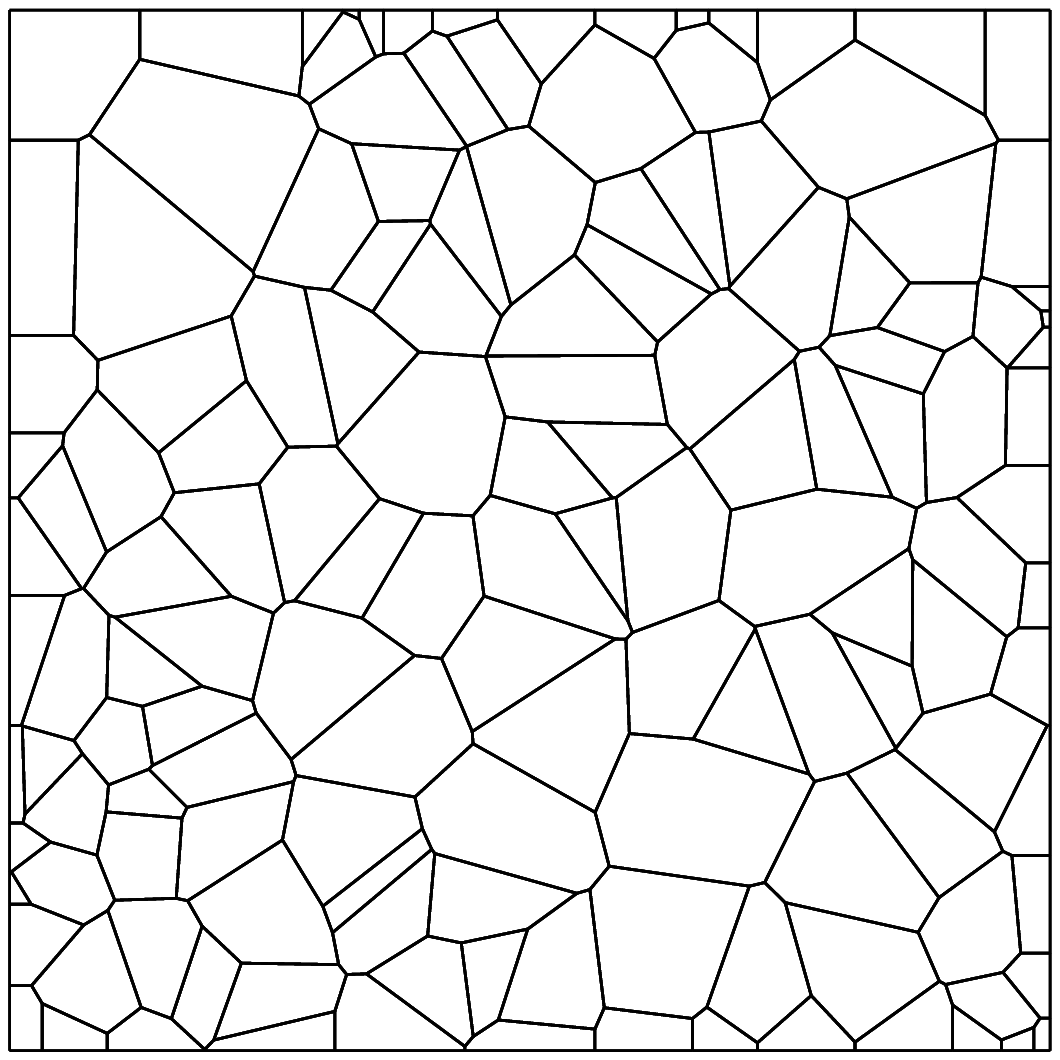, width = 0.18\textwidth}}
\subfigure[]{\label{fig:8c}
\epsfig{file = 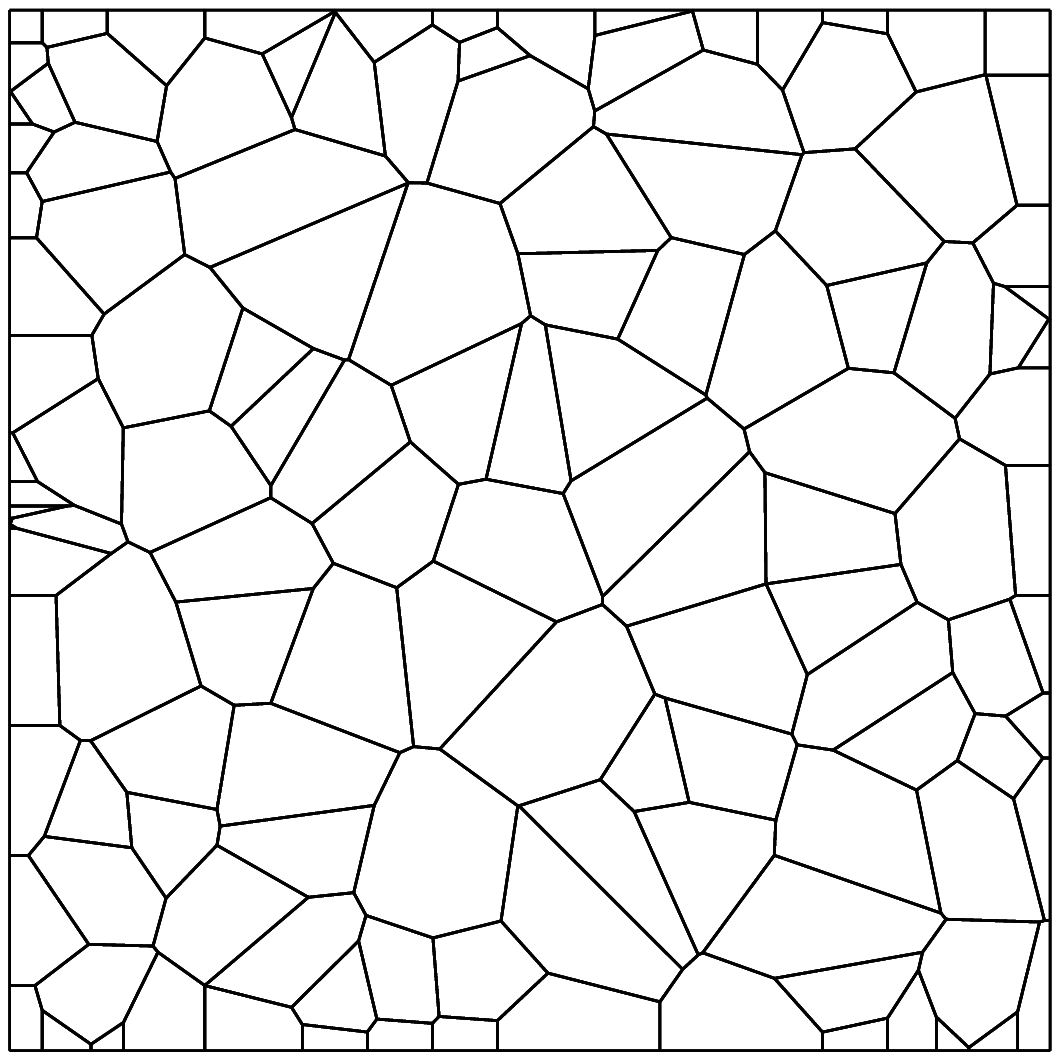, width = 0.18\textwidth}}
\subfigure[]{\label{fig:8d}
\epsfig{file = 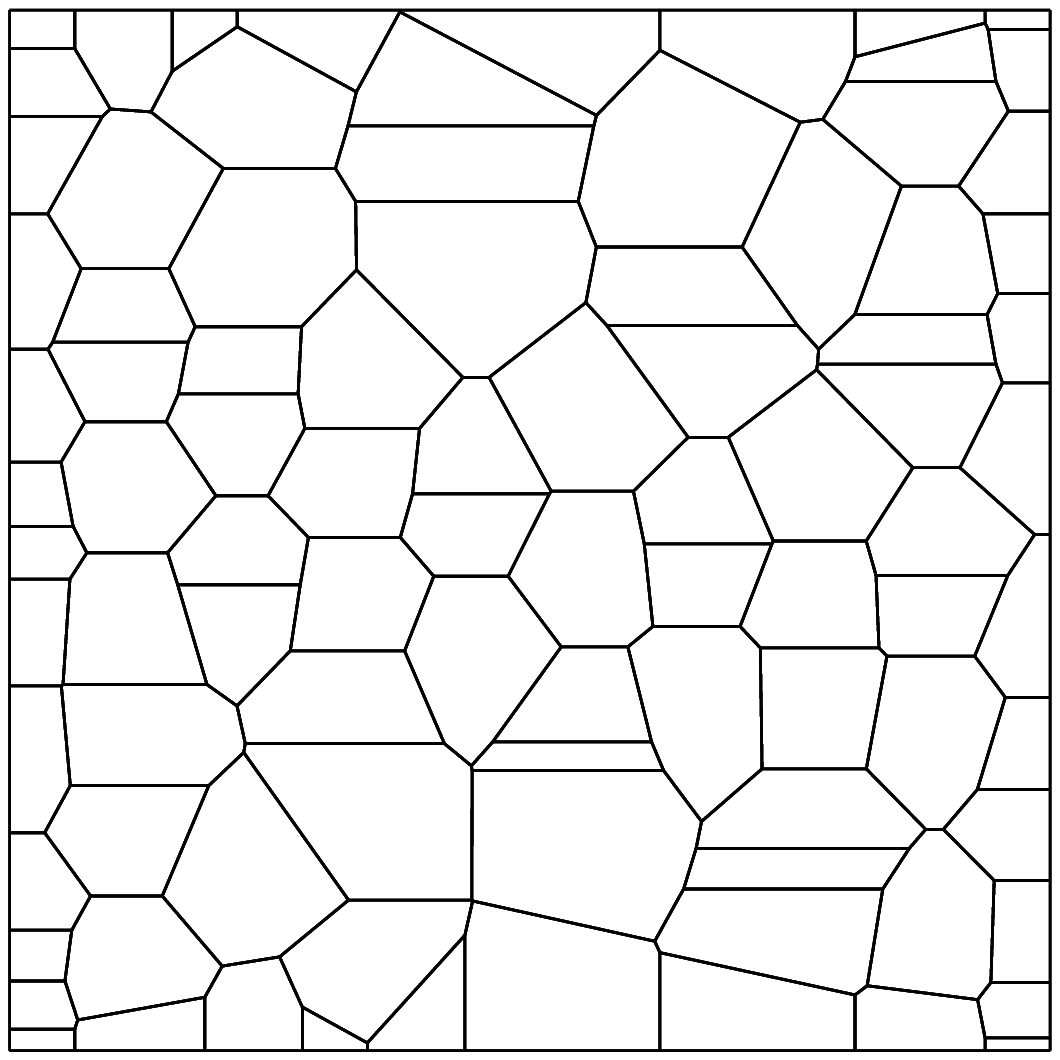, width = 0.18\textwidth}}
}
\caption{Polygonal mesh generation on a square domain using different rules with random noise. (a) \texttt{constant} with noise, (b) \texttt{random\_double} with noise, (c) \texttt{ConstantAlternating} with noise, (d) \texttt{sine} with noise}
\label{fig:8}
\end{figure}

We also include the possibility of adding noise to the generation rules. For this, we implement a random noise function that adds a random displacement to each seed point. \fref{fig:8} depicts some examples of generation rules with random noise. Listing~\ref{lst:generation_with_noise} presents the code to add random noise to the \texttt{constant} generation rule on a square domain.

\begin{lstlisting}[language=C++, caption={Generation of constant (uniform) points with random noise}, label=lst:generation_with_noise]
Functor* n = noise::random_double_noise(functions::constant(), minNoise, maxNoise);
square.generateSeedPoints(PointGenerator(n,n,nX, nY));
// nX, nY: horizontal and vertical divisions along sides of the bounding box
\end{lstlisting}

\subsection{Mesh generation on complicated domains}
\begin{figure}[!tbhp]
\centering
\mbox{
\subfigure[]{\label{fig:9a}
\epsfig{file = 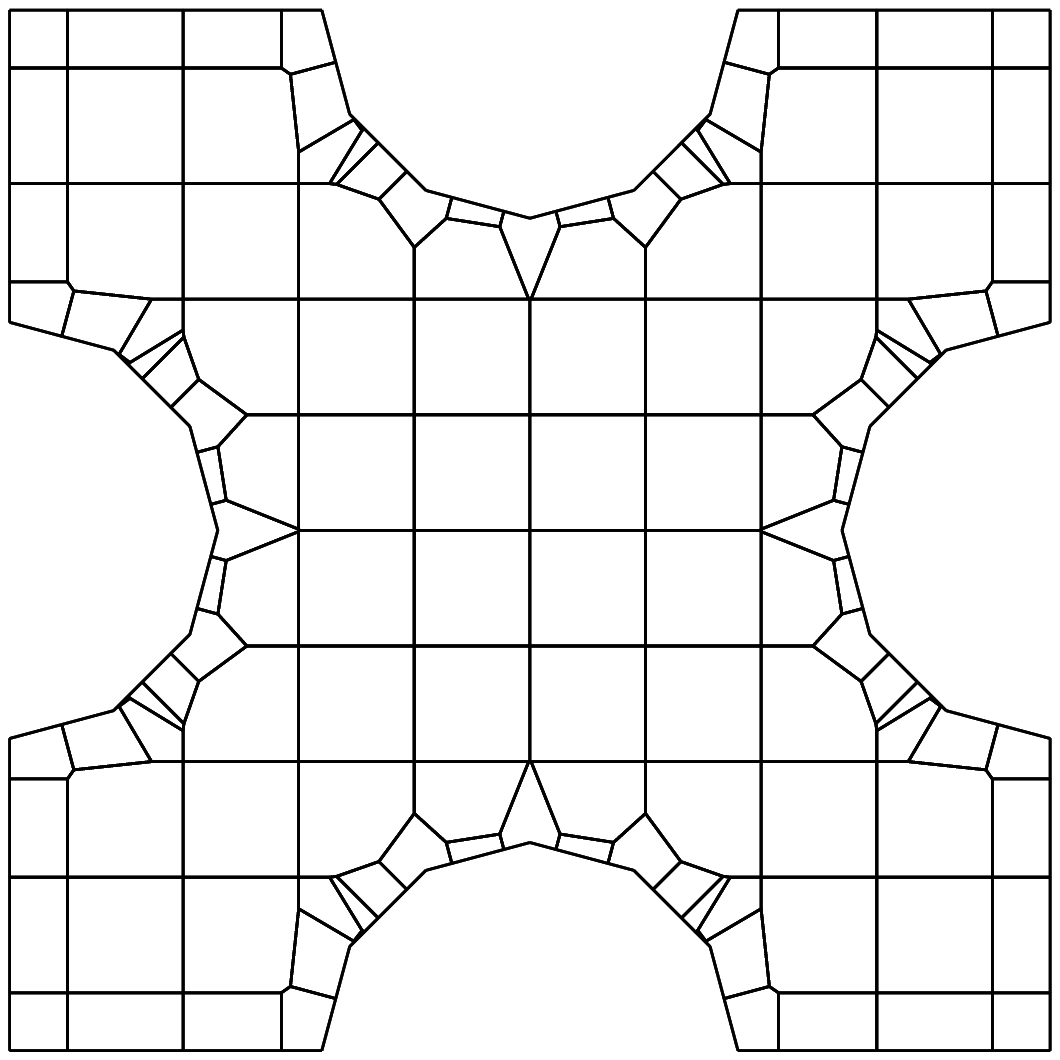, width = 0.2\textwidth}}
\subfigure[]{\label{fig:9b}
\epsfig{file = 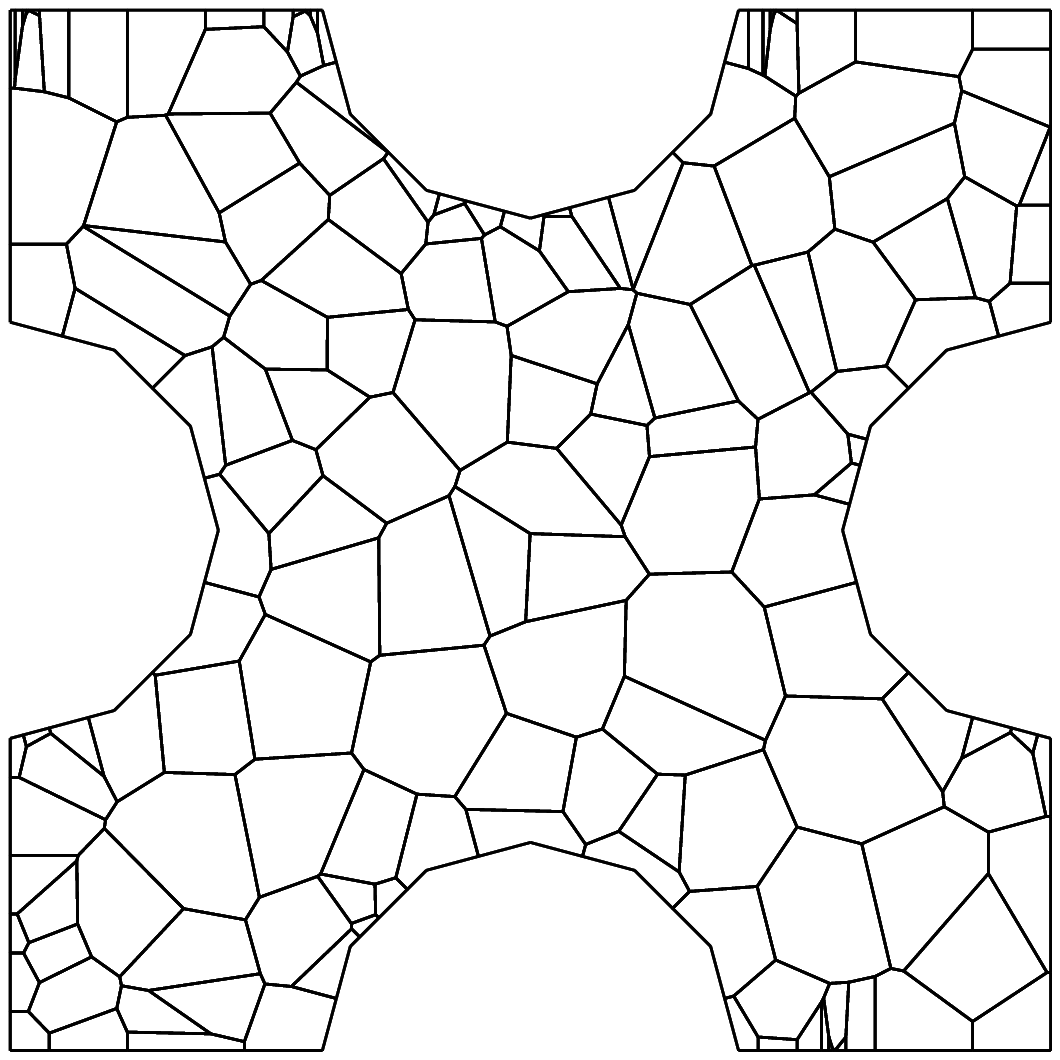, width = 0.2\textwidth}}
}
\caption{Examples of polygonal meshes in complicated domains. (a) Square with four intersecting holes and \texttt{constant} generation rule, and (b) square with four intersecting holes and \texttt{random\_double} generation rule}
\label{fig:9}
\end{figure}

\begin{figure}[!tbhp]
\centering
\mbox{
\subfigure[]{\label{fig:10a}
\epsfig{file = 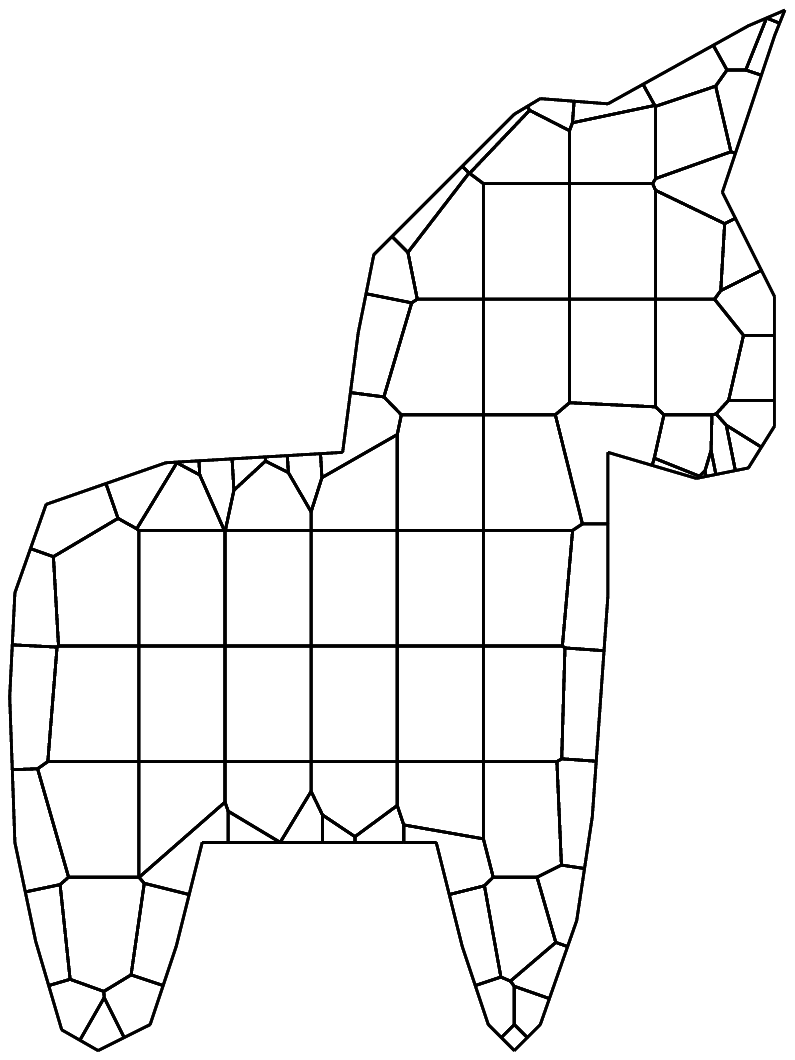, width = 0.25\textwidth}}
\subfigure[]{\label{fig:10b}
\epsfig{file = 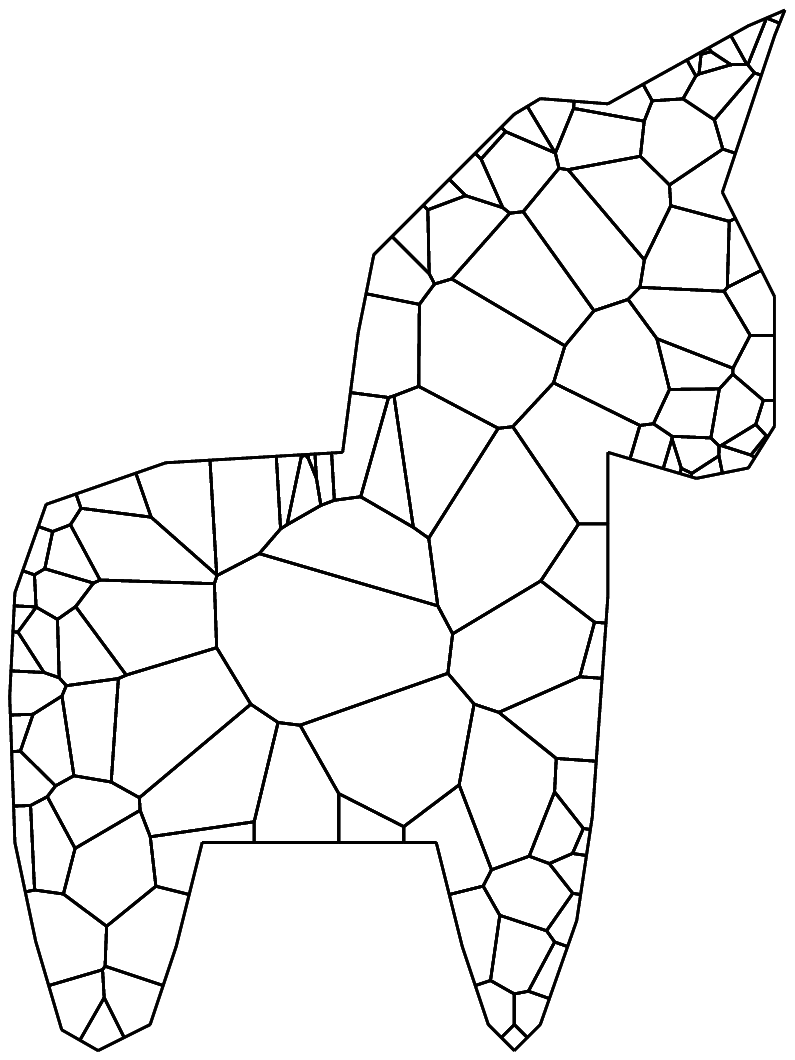, width = 0.25\textwidth}}
\subfigure[]{\label{fig:10c}
\epsfig{file = 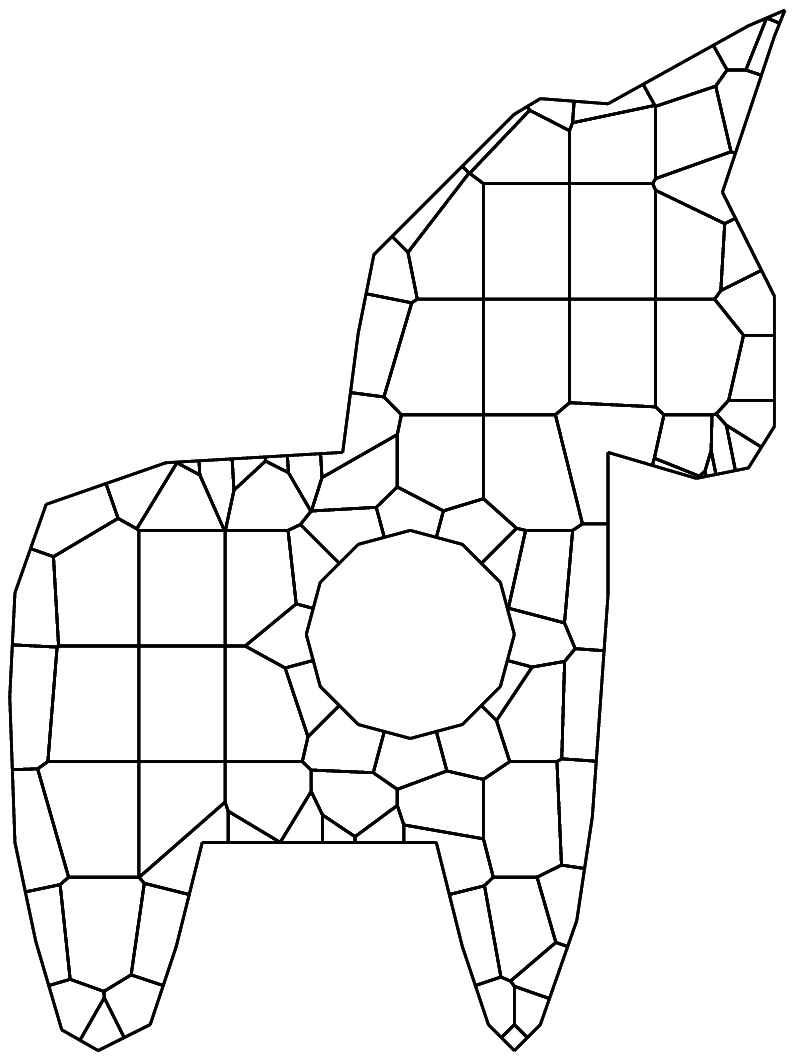, width = 0.18\textwidth}}
}
\mbox{
\subfigure[]{\label{fig:10d}
\epsfig{file = 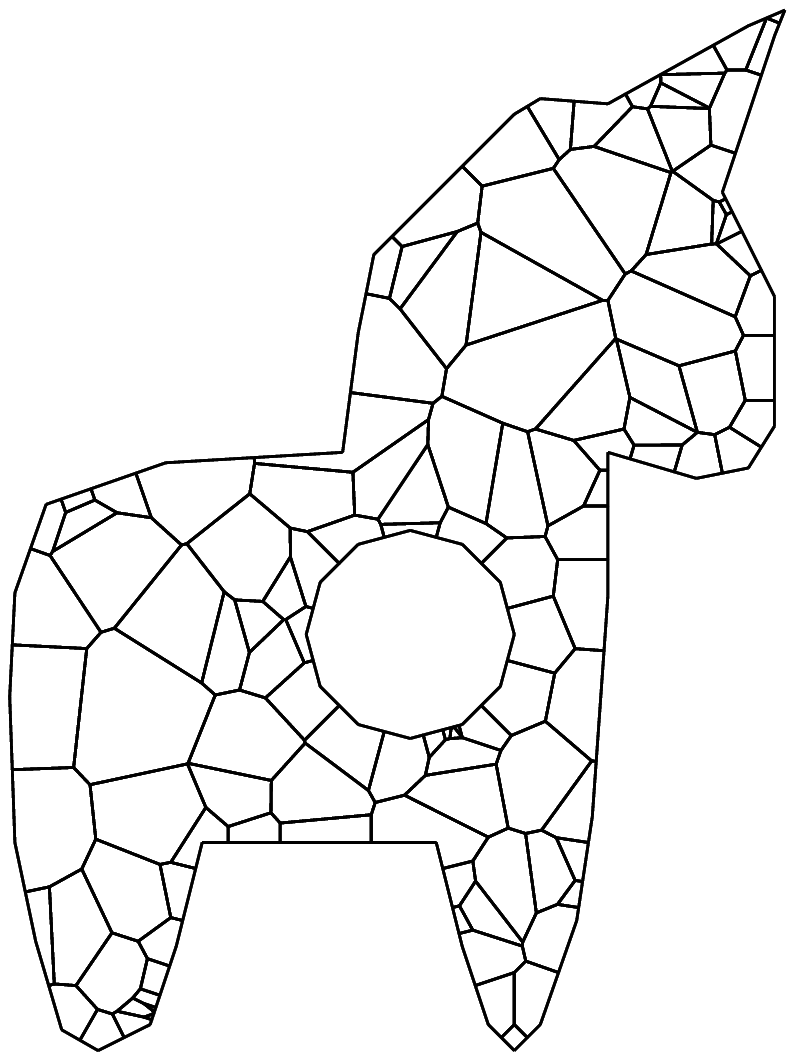, width = 0.25\textwidth}}
\subfigure[]{\label{fig:10e}
\epsfig{file = 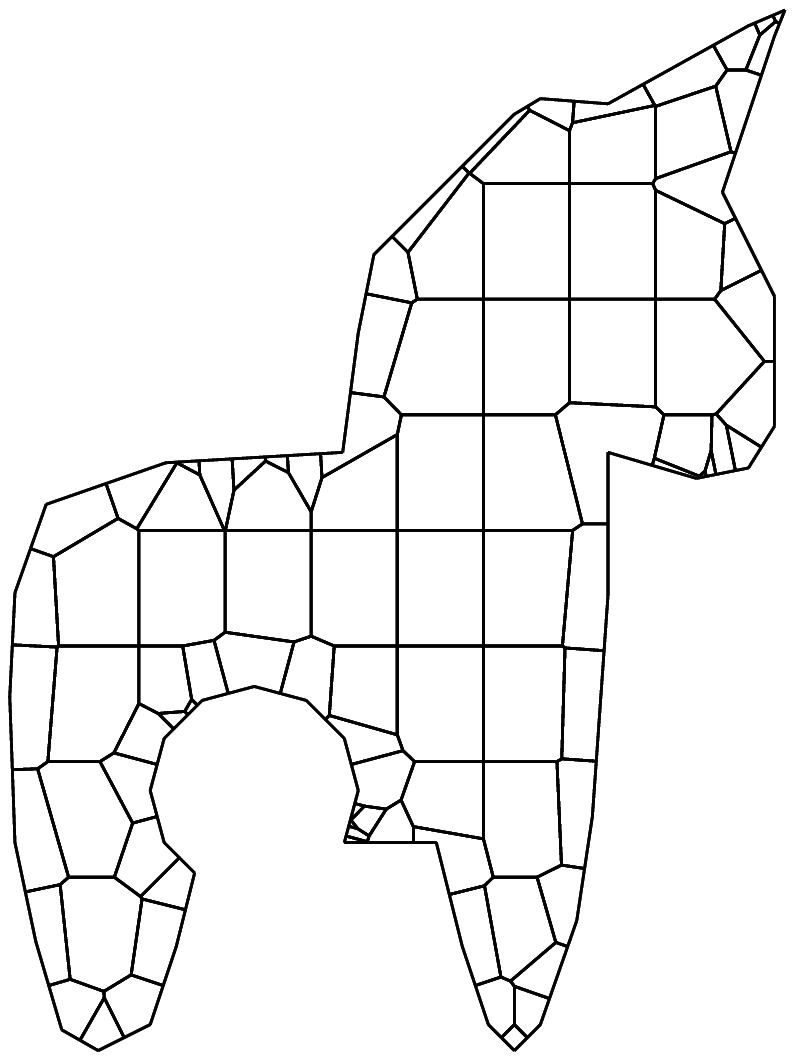, width = 0.19\textwidth}}
\subfigure[]{\label{fig:10f}
\epsfig{file = 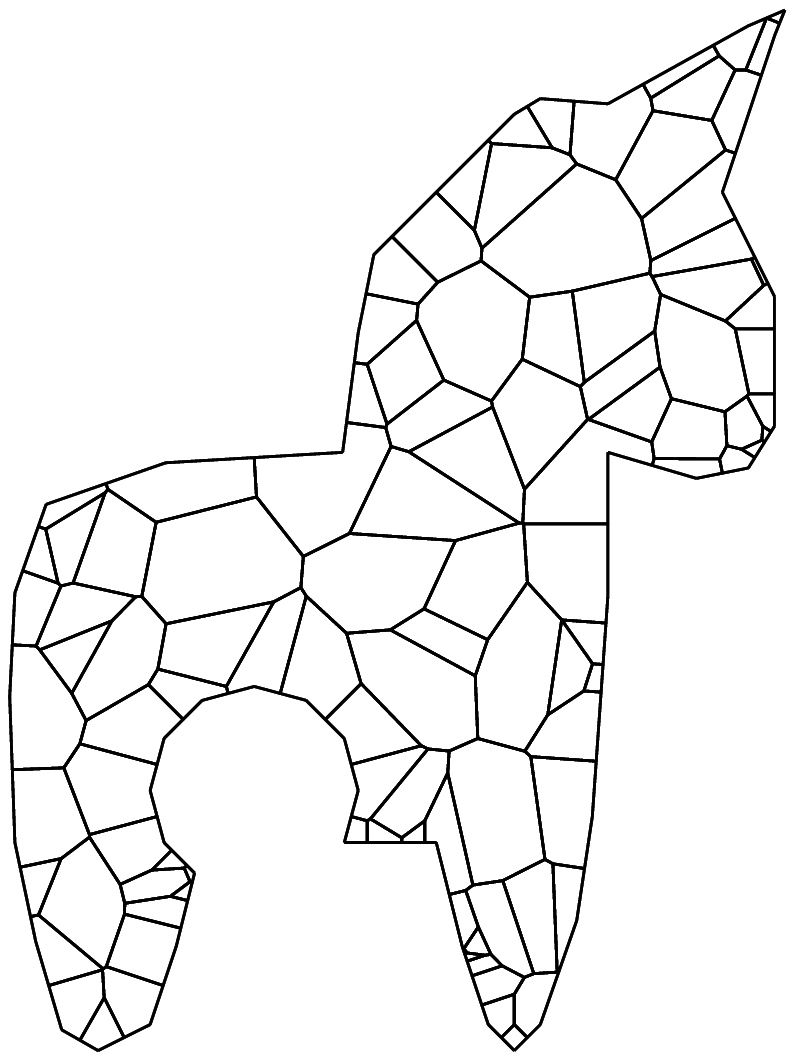, width = 0.24\textwidth}}
}
\caption{Examples of polygonal meshes in complicated domains. Unicorn-shaped domain with (a) \texttt{constant} generation rule, (b) \texttt{random\_double} generation rule, (c) inner hole and \texttt{constant} generation rule, (d) inner hole and \texttt{random\_double} generation rule, (e) intersecting hole and \texttt{constant} generation rule, and (f) intersecting hole and \texttt{random\_double} generation rule}
\label{fig:10}
\end{figure}

Finally, we present some examples of meshes generated on some complicated domains using \texttt{constant} and \texttt{random\_double} rules. \fref{fig:9} shows polygonal meshes for a square domain with four intersecting holes and \fref{fig:10} depicts polygonal meshes for the unicorn-shaped domain without
holes and with different configuration of holes.

\section{Sample usage}
\label{sec:sampleusage}
This section illustrates the usage of \texttt{Veamy} through several examples. For each example, a main C++ file is written to setup the problem. This is the only file that needs to be written by the user in order to run a simulation in \texttt{Veamy}. The setup file for each of the examples that are considered in this section is included in the folder ``test/.'' To be able to run these examples, it is necessary to compile the source code. A tutorial manual that provides complete instructions on how to prepare, compile and run the examples is included in the folder ``docs/.''

\subsection{Cantilever beam subjected to a parabolic end load}
\label{sec:beamexample}

\begin{figure}[!tbhp]
\centering
\epsfig{file = 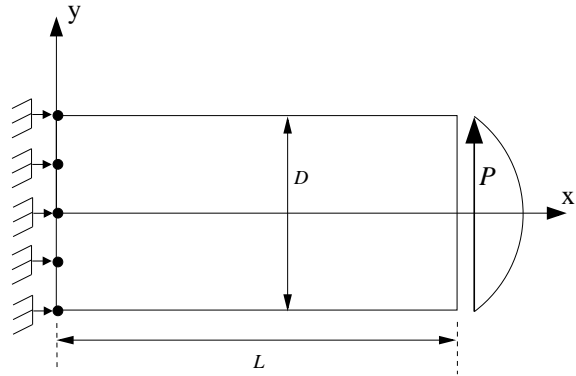, width = 0.5\textwidth}
\caption{Model geometry and boundary conditions for the cantilever beam problem}
\label{fig:11}
\end{figure}

The VEM solution for the displacement field on a cantilever beam of unit thickness subjected to a parabolic end load $P$ is computed using \texttt{Veamy}. \fref{fig:11} illustrates the geometry and boundary conditions. Plane strain state is assumed. The essential boundary conditions on the clamped edge are applied according to the analytical solution given by
Timoshenko and Goodier~\cite{timoshenko:1970:TOE}:
\begin{subequations}\label{beam_exact_sol}
\begin{align*}
u_{x} &=  -\frac{Py}{6\overline{E}_Y I}\left((6L-3x)x + (2+\overline{\nu})y^2 - \frac{3D^2}{2}(1+\overline{\nu})\right),\\
u_{y} &= \frac{P}{6\overline{E}_Y I}\left(3\overline{\nu}y^{2}(L-x)+(3L-x)x^{2}\right),
\end{align*}
\end{subequations}
where $\overline{E}_Y=E_Y/\left(1-\nu^{2}\right)$ with the Young's modulus
set to $E_Y=1\times 10^7$ psi, and $\overline{\nu}=\nu/\left(1-\nu\right)$ with
the Poisson's ratio set to $\nu=0.3$; $L=8$ in. is the length of
the beam, $D=4$ in. is the height of the beam, and $I$ is the
second-area moment of the beam section. The total load on the
traction boundary is $P=-1000$ lbf.

\subsubsection{Setup file}
The setup instructions for this problem are provided in the main C++ file ``ParabolicMain.cpp'' that is located in the folder ``test/.'' Additionally, the structure of the setup file is explained in detail in the tutorial manual that is located in the folder ``docs/.'' The interested reader is referred therein to learn more about this setup file.

\subsubsection{Post processing}

\begin{figure}[!tbhp]
\centering
\mbox{
\subfigure[]{\label{fig:12a}
\epsfig{file = 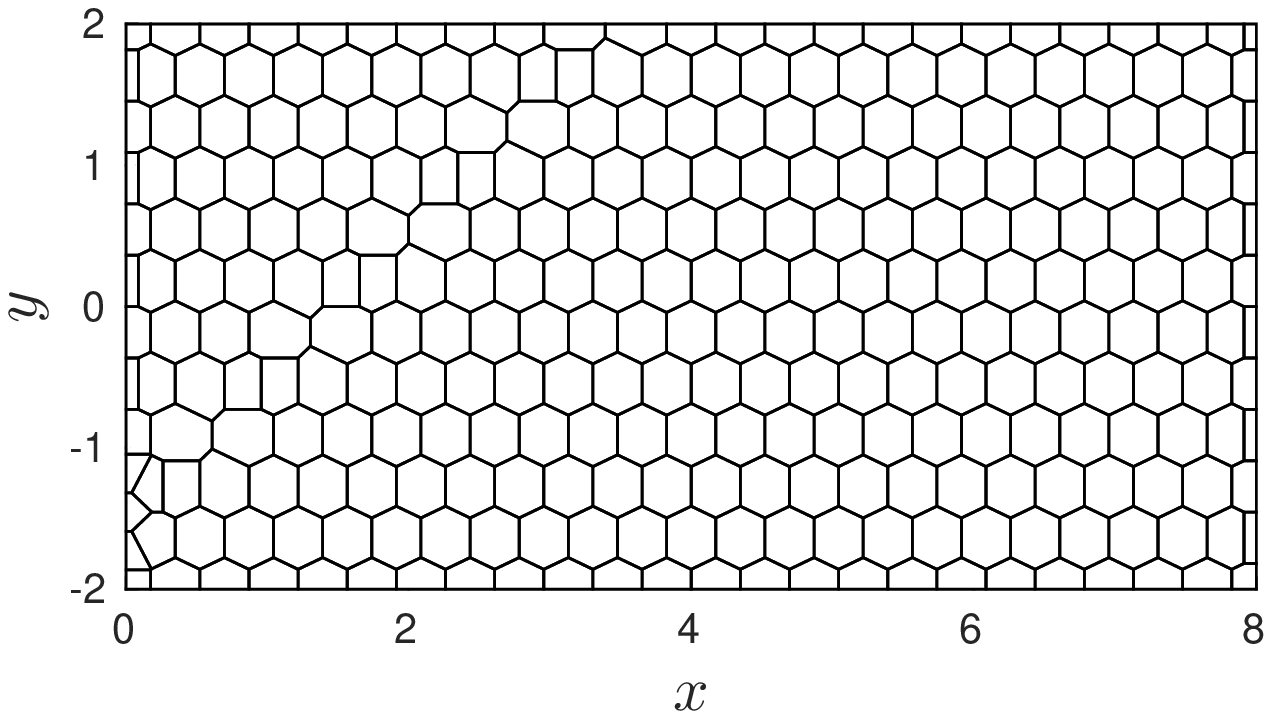, width = 0.42\textwidth}
}
\subfigure[]{\label{fig:12b}
\epsfig{file = 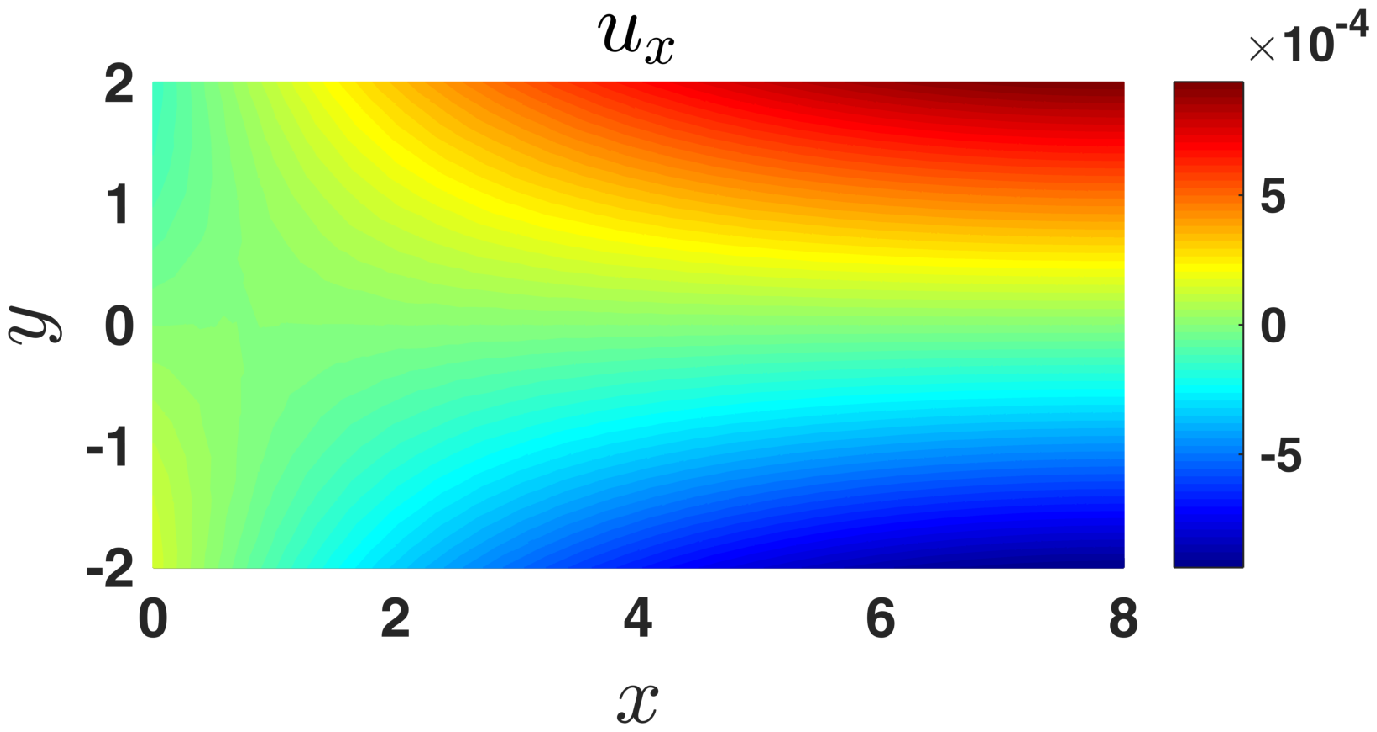, width = 0.44\textwidth}
}
}
\mbox{
\subfigure[]{\label{fig:12c}
\epsfig{file = 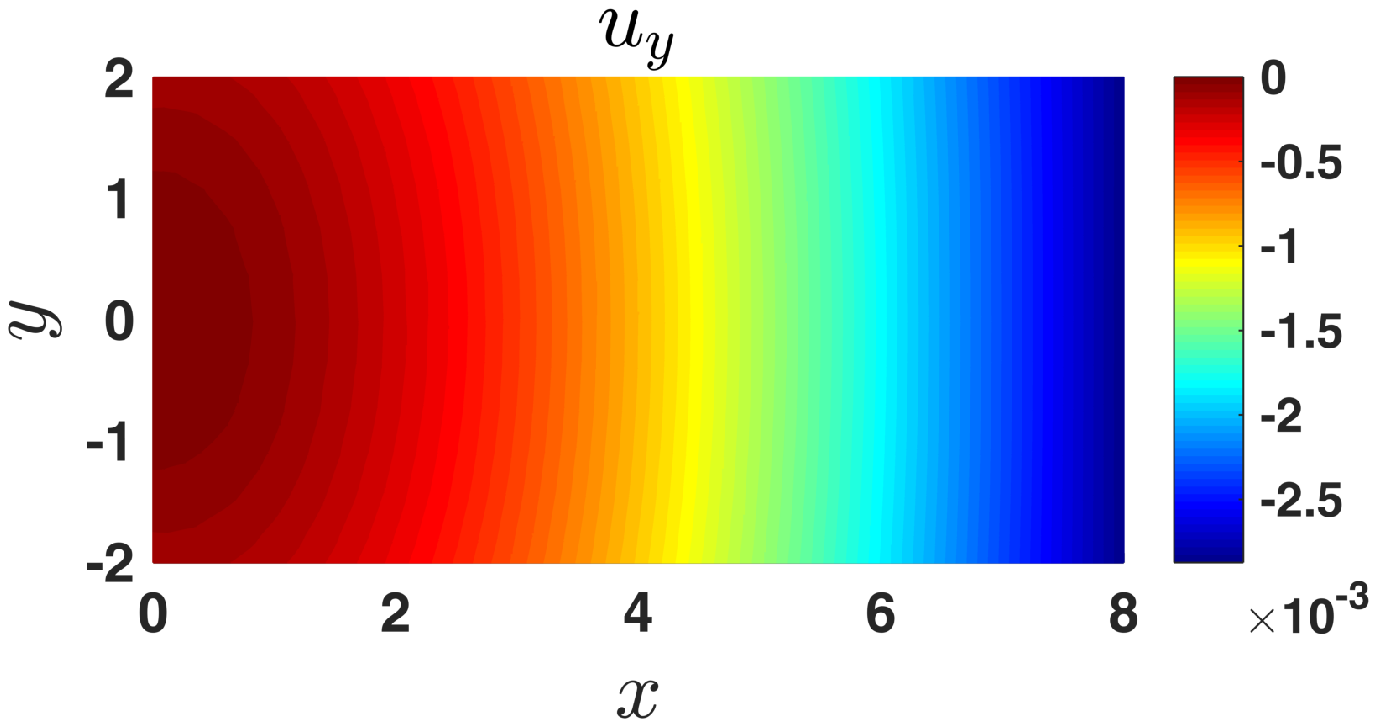, width = 0.44\textwidth}
}
\subfigure[]{\label{fig:12d}
\epsfig{file = 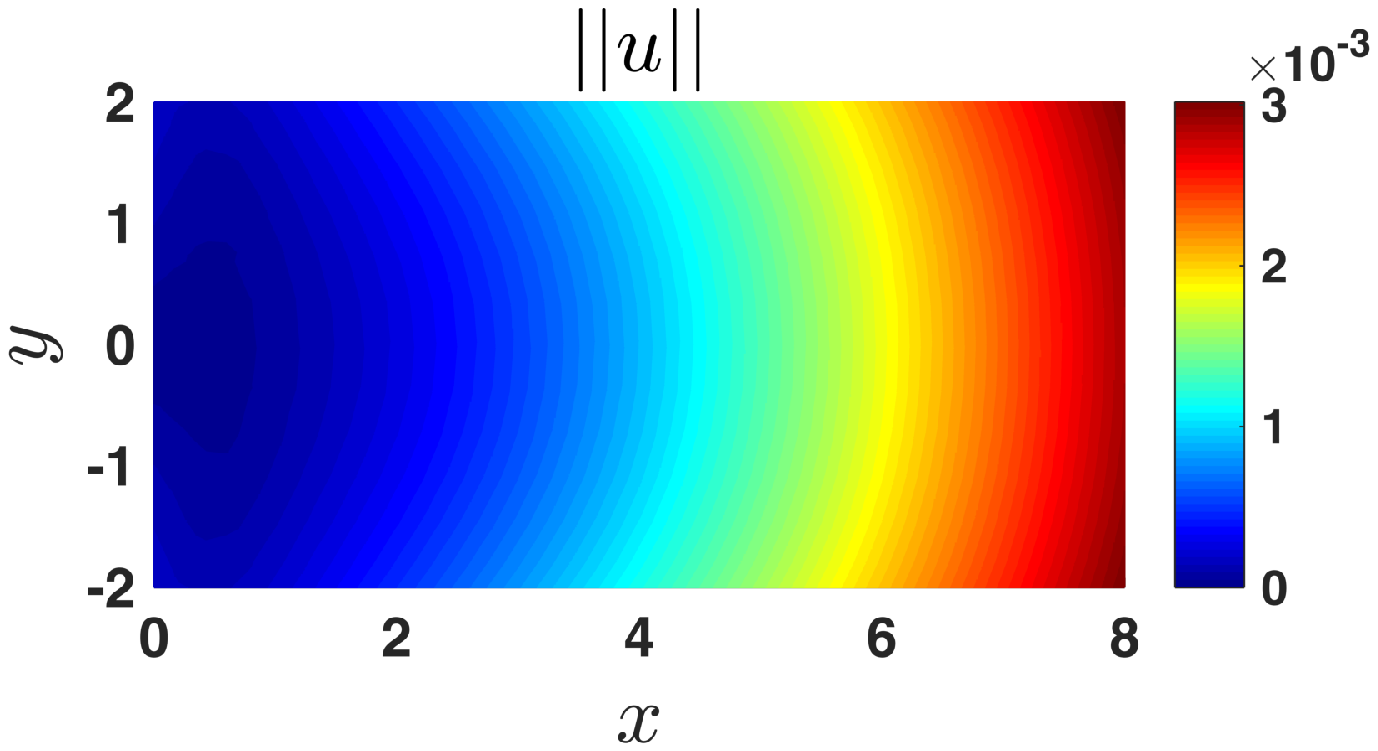, width = 0.44\textwidth}
}
}
\caption{Solution for the cantilever beam subjected to a parabolic end load using \texttt{Veamy}. (a) Polygonal mesh, (b) VEM horizontal displacements, (c) VEM vertical displacements, (d) norm of the VEM displacements}
\label{fig:12}
\end{figure}

\texttt{Veamy} does not provide a post processing interface. The user may opt for a post processing interface of their choice. Here we visualize the displacements using a MATLAB function written for this purpose. This MATLAB function is provided in the folder ``matplots/'' as the file ``plotPolyMeshDisplacements.m.'' In addition, a file named ``plotPolyMesh.m'' that serves for plotting the mesh is provided in the same folder. \fref{fig:12} presents the polygonal mesh used and the VEM solutions.

\subsubsection{VEM performance}
\begin{figure}[!tbhp]
\centering
\mbox{
\subfigure[]{\label{fig:13a}
\epsfig{file = 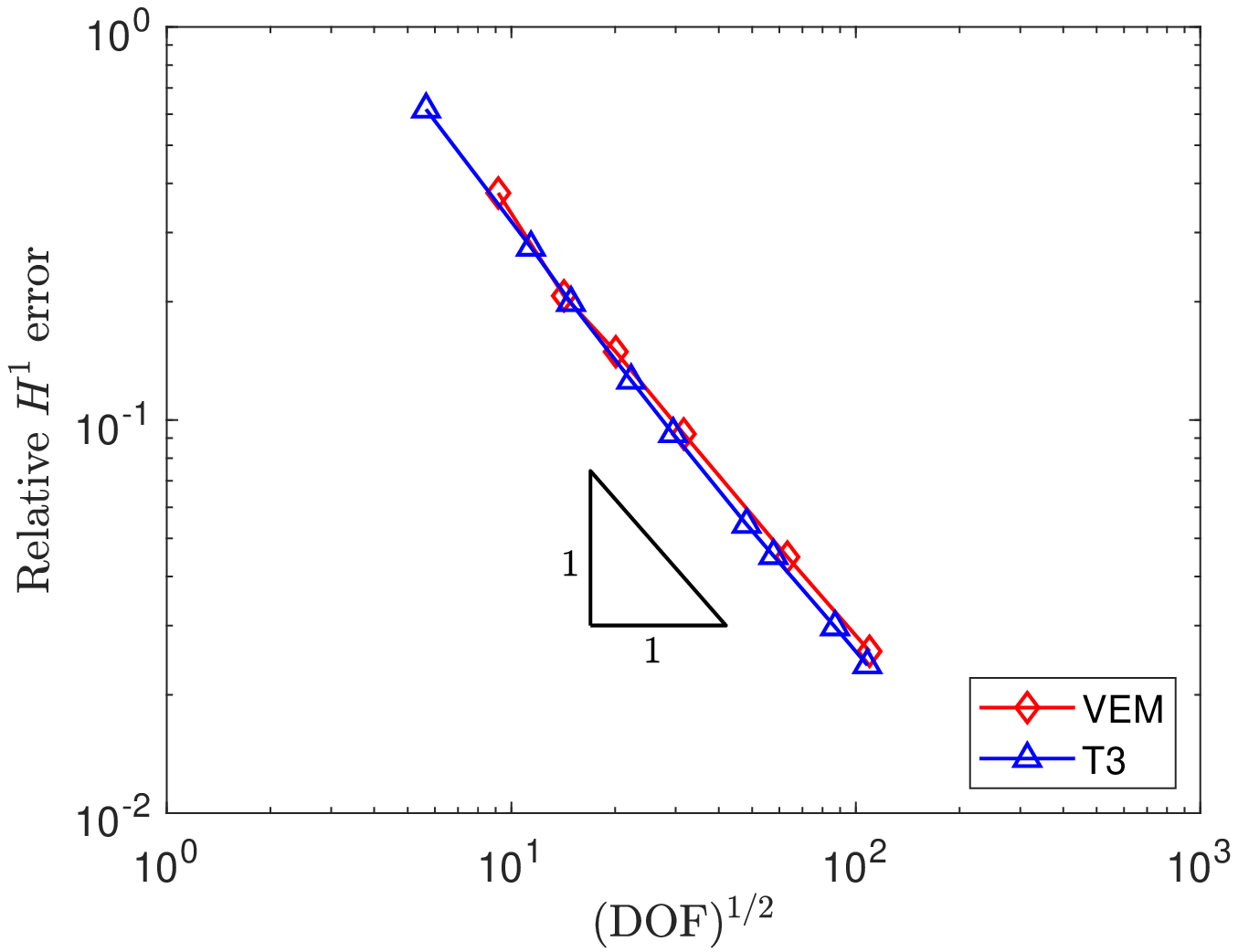, width = 0.48\textwidth}
}
\subfigure[]{\label{fig:13b}
\epsfig{file = 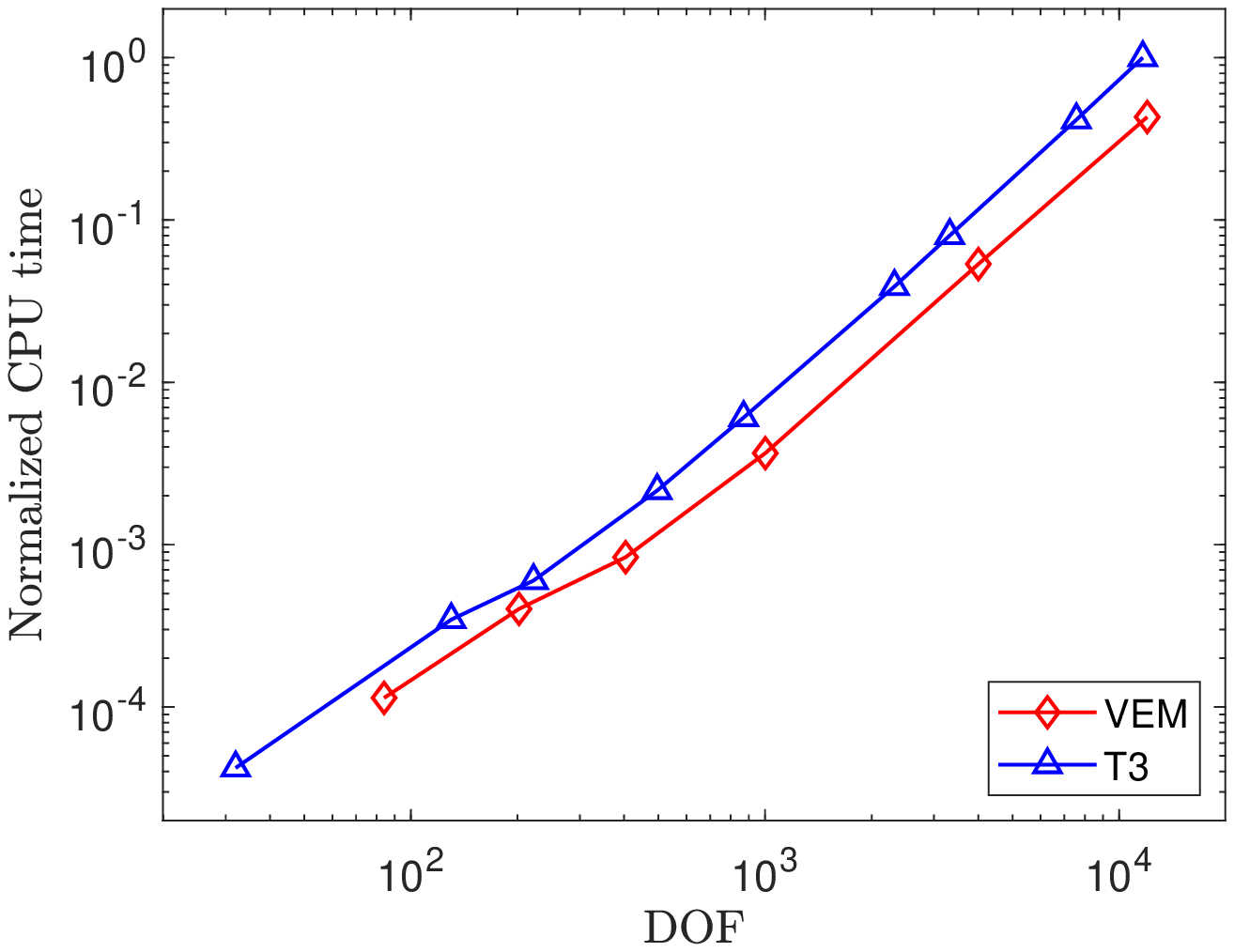, width = 0.48\textwidth}
}
}
\caption{Cantilever beam subjected to a parabolic end load. Performance comparison between the VEM (polygonal elements) and the FEM (three-node triangles ($T3$)). (a) $H^1$-seminorn of the error as a function of the number of degrees of freedom and (b) normalized CPU time as a function of the number of degrees of freedom}
\label{fig:13}
\end{figure}

A performance comparison between VEM and FEM is conducted. For the FEM simulations, the \texttt{Feamy} module is used. The main C++ setup files for these tests are located in the folder ``test/'' and named as ``ParabolicMainVEMnorms.cpp'' for the VEM and ``FeamyParabolicMainNorms.cpp'' for the FEM using three-node triangles ($T3$). The meshes used for these tests were written to text files, which are located in the folder ``test/test\_files/.'' \texttt{Veamy} implements a function named \texttt{createFromFile} that is used to read these mesh files. The performance of the two methods are compared in~\fref{fig:13}, where the $H^1$-seminorm of the error and the normalized CPU time are each plotted as a function of the number of degrees of freedom (DOF). The normalized CPU time is defined as the ratio of the CPU time of a particular model analyzed to the maximum CPU time found for any of the models analyzed. From \fref{fig:13} it is observed that for equal number of degrees of freedom both methods deliver similar accuracy and the computational costs are about the same.

\subsection{Using a \texttt{PolyMesher} mesh}
\label{sec:mbbbeamexample}

In order to conduct a simulation in \texttt{Veamy} using a \texttt{PolyMesher} mesh, a MATLAB
function named \texttt{PolyMesher2Veamy}, which needs to be called within \texttt{PolyMesher},
was especially devised to read this mesh and write it to a text file that is readable by \texttt{Veamy}. This MATLAB function
is provided in the folder ``matplots/.'' Function \texttt{PolyMesher2Veamy} receives five \texttt{PolyMesher}'s
data structures (\texttt{Node}, \texttt{Element}, \texttt{NElem}, \texttt{Supp}, \texttt{Load}) and writes
a text file containing the mesh and boundary conditions. \texttt{Veamy} implements a function
named \texttt{initProblemFromFile} that is able to read this text file and solve the problem straightforwardly.

\begin{figure}[!tbhp]
\centering
\epsfig{file = 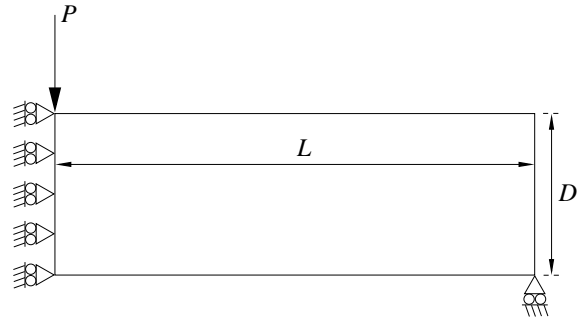, width = 0.5\textwidth}
\caption{MBB beam problem definition as per Section 6.1 in Ref.~\cite{Talischi:POLYM:2012}}
\label{fig:14}
\end{figure}

\begin{figure}[!tbhp]
\centering
\mbox{
\subfigure[]{\label{fig:15a}
\epsfig{file = 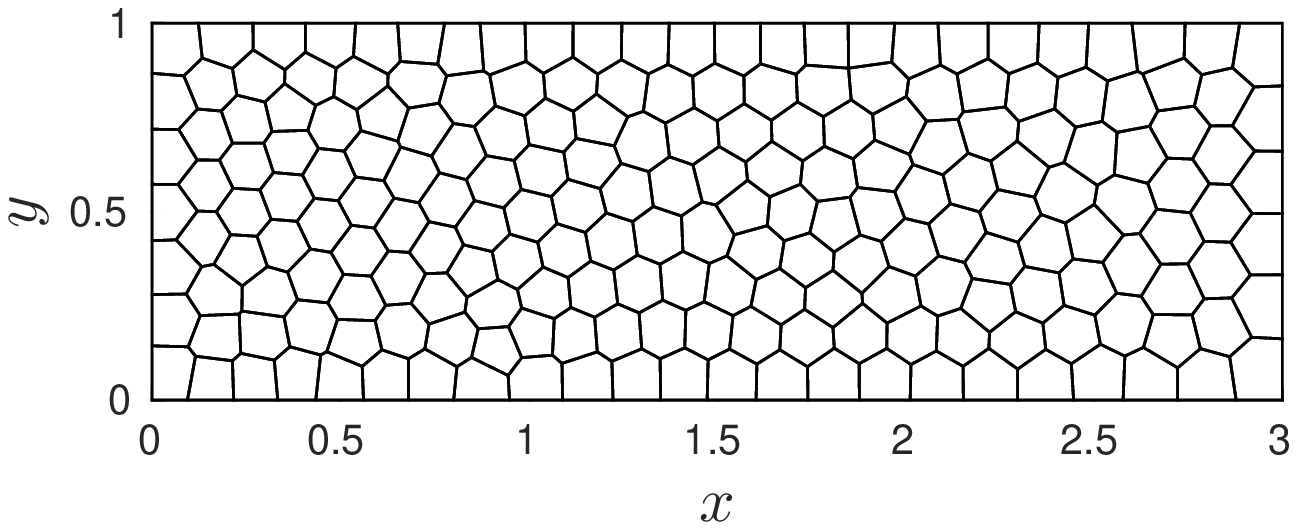, width = 0.43\textwidth}
}
\subfigure[]{\label{fig:15b}
\epsfig{file = 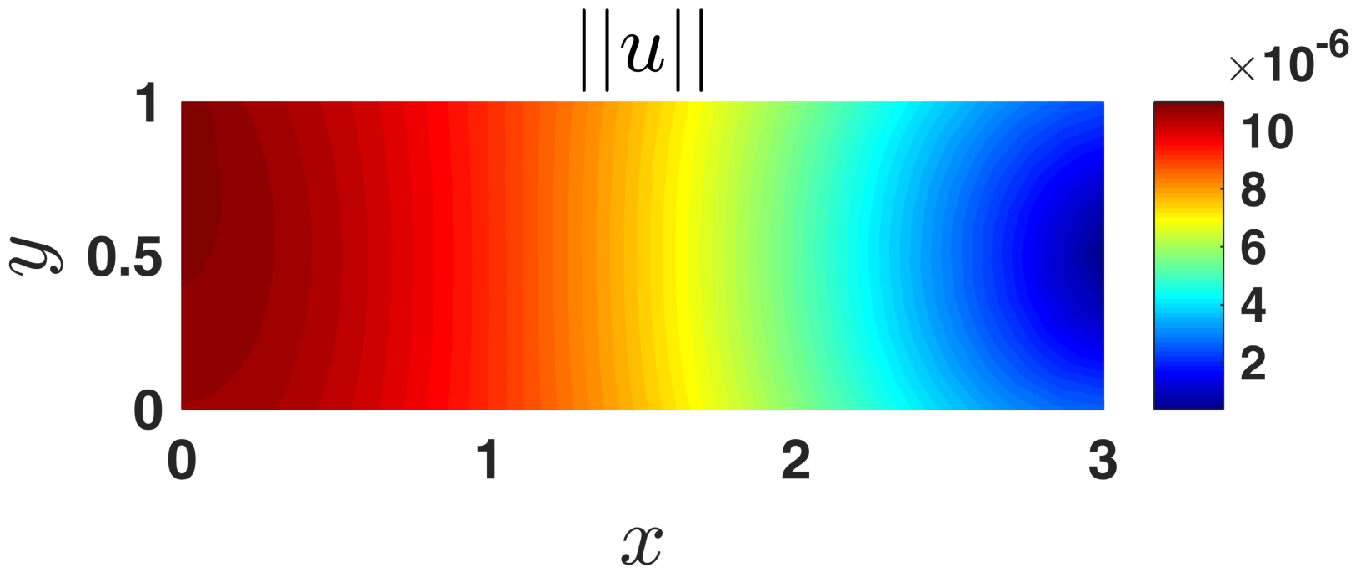, width = 0.48\textwidth}
}
}
\caption{Solution for the MBB beam problem using \texttt{Veamy}. (a) Polygonal mesh and (b) norm of the VEM displacements}
\label{fig:15}
\end{figure}

As a demonstration of the potential that is offered to the simulation when \texttt{Veamy} interacts with \texttt{PolyMesher}, the MBB beam problem of Section 6.1 in Ref.~\cite{Talischi:POLYM:2012} is considered. The MBB problem is shown in~\fref{fig:14}, where $L=3$ in., $D=1$ in. and $P=0.5$ lbf. The following material parameters are considered: $E_Y = 1\times 10^7$ psi, $\nu = 0.3$ and plane strain condition is assumed. The file containing the translated mesh with boundary conditions is provided in the folder ``test/test\_files/'' under the name ``\texttt{polymesher2veamy.txt}.'' The complete setup instructions for this problem are provided in the file ``PolyMesherMain.cpp'' that is located in the folder ``test/.'' The polygonal mesh and the VEM solution are presented in \fref{fig:15}.

\subsection{Perforated Cook's membrane}
\label{sec:cook}
\begin{figure}[!tbhp]
\centering
\mbox{
\subfigure[]{\label{fig:16a}
\epsfig{file = 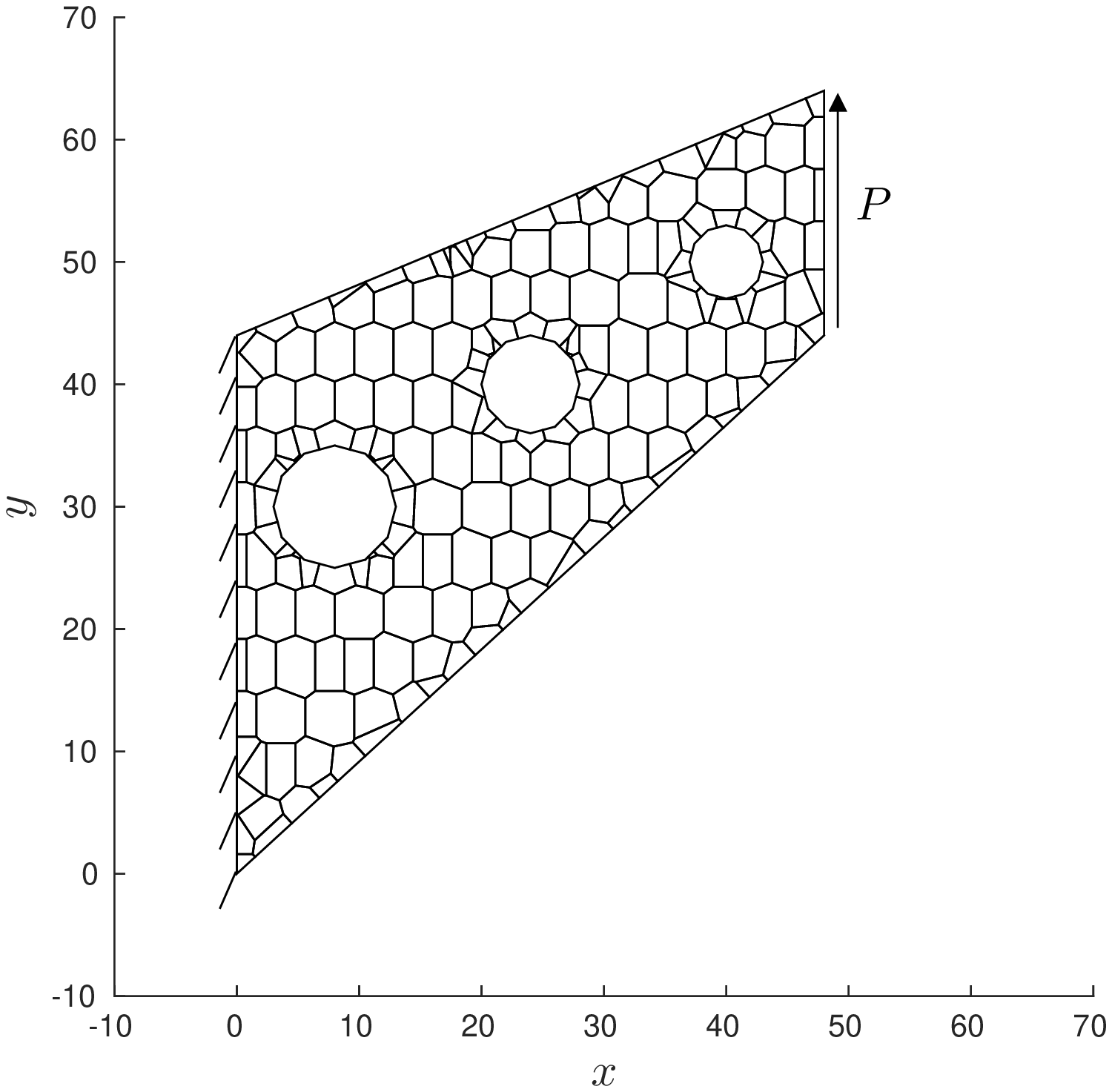, width = 0.35\textwidth}
}
\subfigure[]{\label{fig:16b}
\epsfig{file = 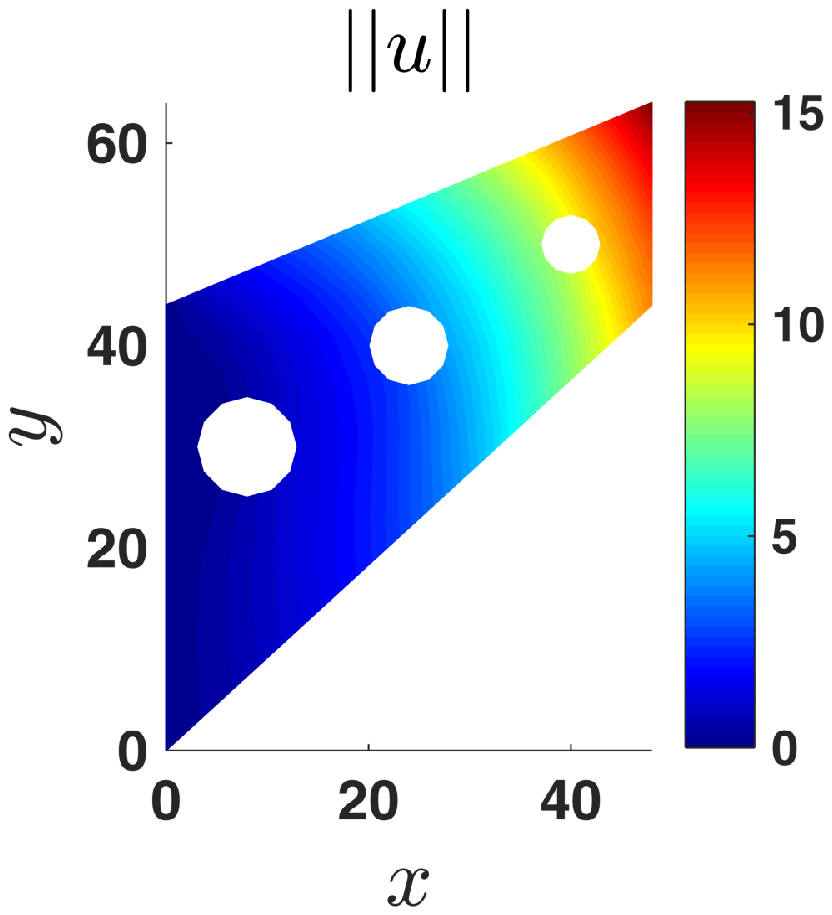, width = 0.48\textwidth}
}
}
\caption{Solution for the perforated Cook's membrane problem using \texttt{Veamy}. (a) Model geometry, polygonal mesh and boundary conditions, and (b) norm of the VEM displacements}
\label{fig:16}
\end{figure}

In this example, a perforated Cook's membrane is considered. The objective of this problem is to show more advanced domain definitions and mesh generation capabilities offered by \texttt{Veamy}. The complete setup instructions for this problem are provided in the file ``CookTestMain.cpp'' that is located in the folder ``test/.'' The following material parameters are considered: $E_Y = 250$ MPa, $\nu = 0.3$ and plane strain condition is assumed. The model geometry, the polygonal mesh and boundary conditions, and the VEM solution are presented in \fref{fig:16}.

\subsection{A toy problem}
\label{sec:toy}
\begin{figure}[!tbhp]
\centering
\mbox{
\subfigure[]{\label{fig:17a}
\epsfig{file = 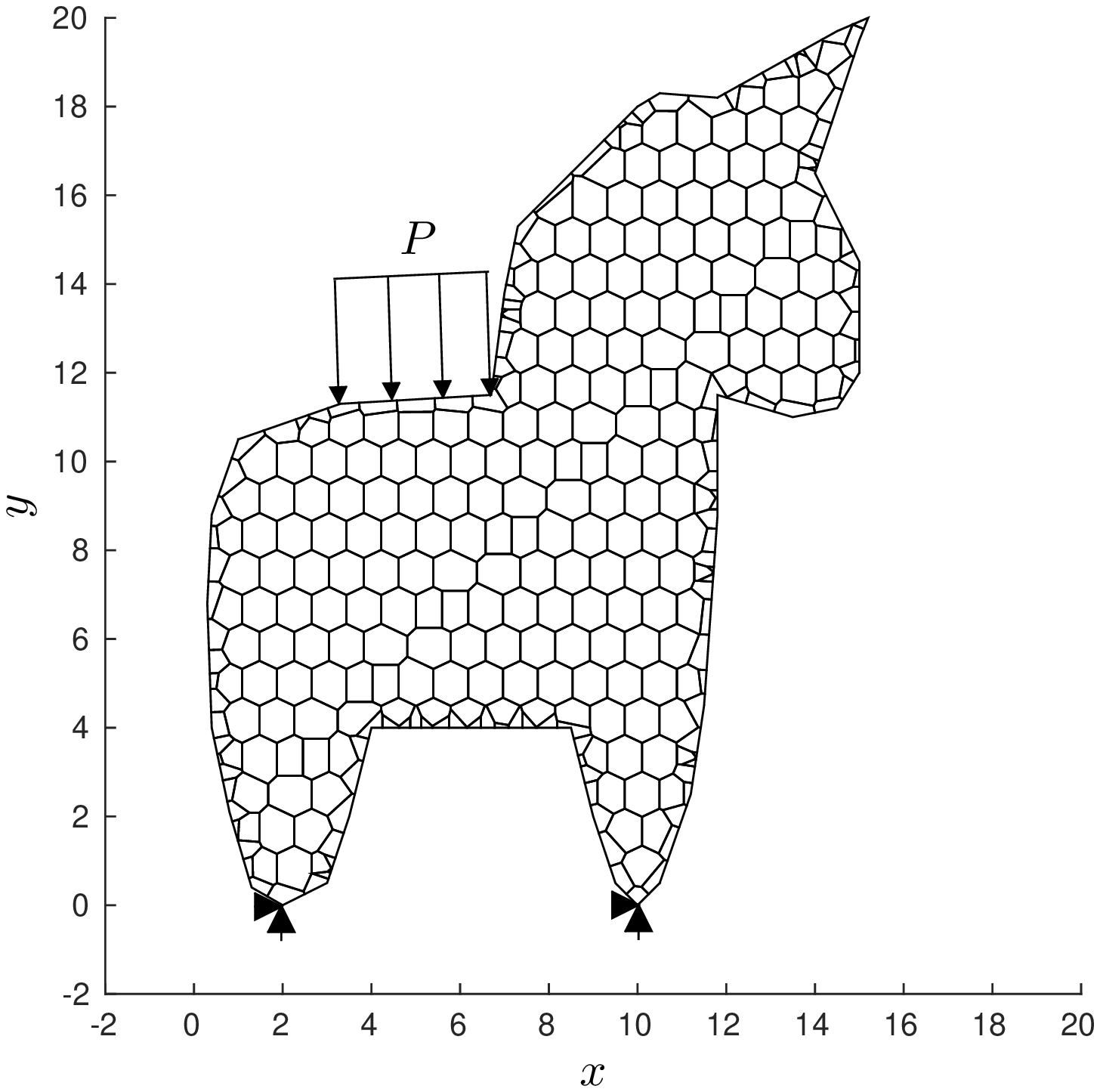, width = 0.35\textwidth}
}
\subfigure[]{\label{fig:17b}
\epsfig{file = 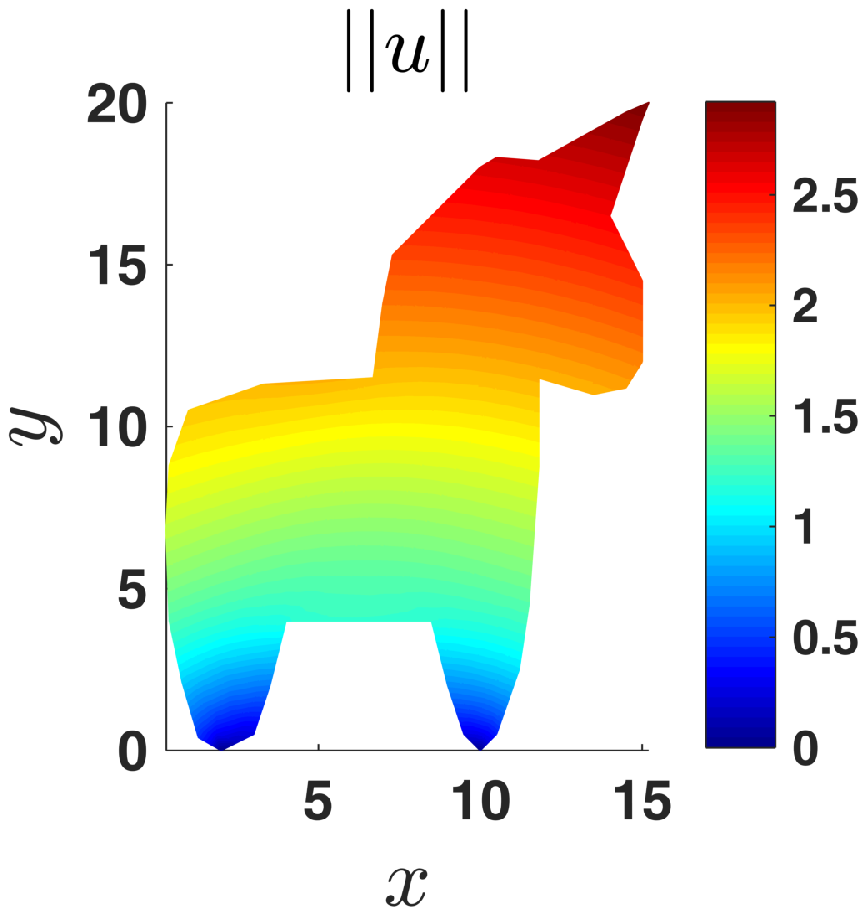, width = 0.43\textwidth}
}
}
\caption{Solution for the toy problem using \texttt{Veamy}. (a) Model geometry, polygonal mesh and boundary conditions, and (b) norm of the VEM displacements}
\label{fig:17}
\end{figure}

A toy problem consisting of a unicorn loaded on its back and fixed at its feet is modeled and solved using \texttt{Veamy}. The objective of this problem is to show additional capabilities for domain definition and mesh generation that are available in \texttt{Veamy}. The complete setup instructions for this problem are provided in the file ``UnicornTestMain.cpp'' that is located in the folder ``test/.'' The following material parameters are considered: $E_Y = 1\times 10^4$ psi, $\nu = 0.25$ and plane strain condition is assumed. The model geometry, the polygonal mesh and boundary conditions, and the VEM solution are shown in \fref{fig:17}.

\subsection{Poisson problem with a manufactured solution}
\label{sec:poissonproblem}
\begin{figure}[!tbhp]
\centering
\mbox{
\subfigure[]{\label{fig:18a}
\epsfig{file = 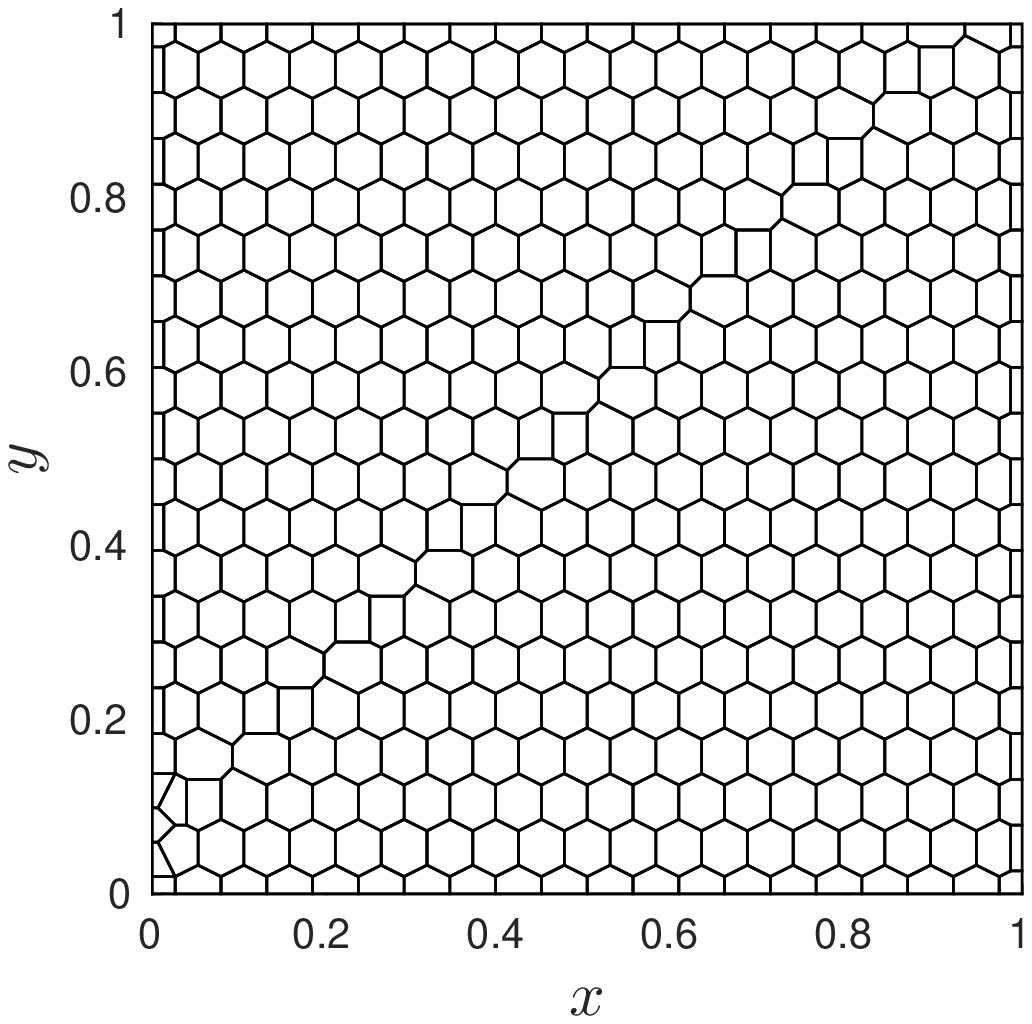, width = 0.42\textwidth}
}
\subfigure[]{\label{fig:18b}
\epsfig{file = 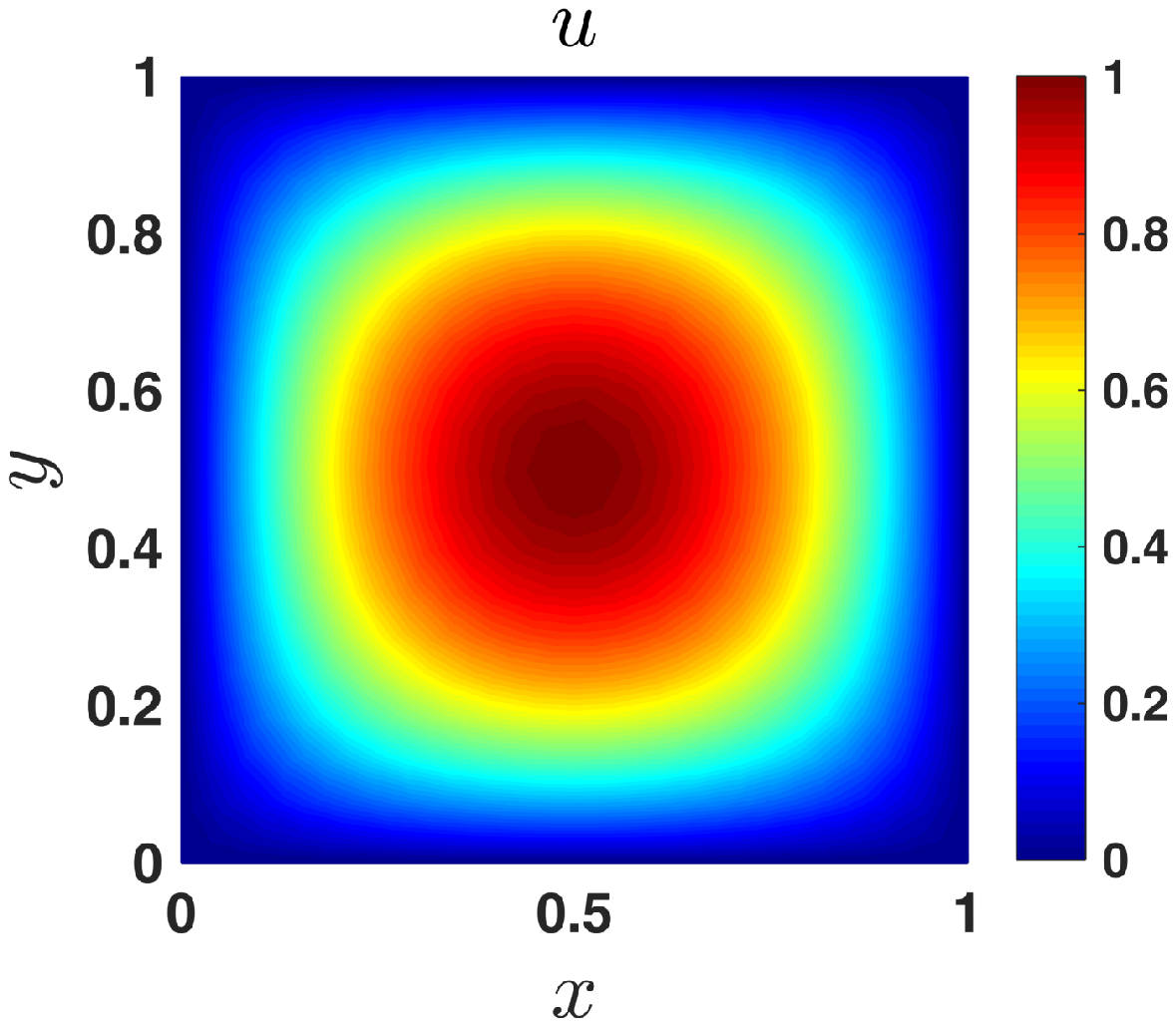, width = 0.45\textwidth}
}
}
\caption{Solution for the Poisson problem using \texttt{Veamy}. (a) Polygonal mesh and (b) VEM solution. The relative $L^2$-norm of the error is $2.6695\times 10^{-3}$ and the relative $H^1$-seminorm of the error is $6.7834\times 10^{-2}$}
\label{fig:18}
\end{figure}

We conclude the examples by solving a Poisson problem with a source term given by $f(\vm{x})= 32y(1-y)+32x(1-x)$, which is the outcome of letting the solution field be $u(\vx)=16xy(1-x)(1-y)$. A unit square domain is considered and $u(\vx)$ is imposed along the entire boundary of the domain resulting in the essential (Dirichlet) boundary condition $g(\vx)=0$. The complete setup instructions for this problem are provided in the file ``PoissonSourceTestMain.cpp'' that is located in the folder ``test/.'' The polygonal mesh and the VEM solution are shown in \fref{fig:18}. The relative $L^2$-norm of the error and the relative $H^1$-seminorm of the error obtained for the mesh shown in \fref{fig:18a} are $2.6695\times 10^{-3}$ and $6.7834\times 10^{-2}$, respectively.

\section{Concluding remarks}
\label{sec:conclusions}

In this paper, an object-oriented C++ library for the virtual element method
was presented for the linear elastostatic and Poisson problems in two-dimensions.
The usage of the library, named \texttt{Veamy},
was demonstrated through several examples. Possible extensions
of this library that are of interest include three-dimensional linear elastostatics, where
an interaction with the polyhedral mesh generator Voro++~\cite{rycroft:VORO:2009} seems very appealing,
and nonlinear solid mechanics~\cite{BeiraodaVeiga-Lovadina-Mora:2015,chi:VEMFD:2017,artioli:AOVEMIN:2017,wriggers:VEMCIFD:2017}.
\texttt{Veamy} is free and open source software.

%

\begin{acknowledgements}
AOB acknowledges the support provided by Universidad de Chile through the ``Programa VID Ayuda de Viaje 2017'' and
the Chilean National Fund for Scientific and Technological Development (FONDECYT) through grant CONICYT/FONDECYT No. 1181192.
The work of CA is supported by CONICYT-PCHA/Mag\'ister Nacional/2016-22161437. NHK is grateful for the support provided by
the Chilean National Fund for Scientific and Technological Development (FONDECYT) through grant CONICYT/FONDECYT No. 1181506.
\end{acknowledgements}



\end{document}